\documentclass[11pt,a4paper]{article} 
\pdfoutput=1
\usepackage{jheppub}
\usepackage{url}
\usepackage[latin1]{inputenc}
\usepackage{amsfonts}
\usepackage{subfigure}
\usepackage{soul}
\usepackage{mciteplus}
\usepackage{bm}
\usepackage{bbm}
\usepackage{scalefnt}
\usepackage{booktabs}
\usepackage{threeparttable}

\newbox\charbox
\newbox\slabox
\def\s#1{{      
 \setbox\charbox=\hbox{$#1$}
 \setbox\slabox=\hbox{$/$}
 \dimen\charbox=\ht\slabox
 \advance\dimen\charbox by -\dp\slabox
 \advance\dimen\charbox by -\ht\charbox
 \advance\dimen\charbox by \dp\charbox
 \divide\dimen\charbox by 2
 \raise-\dimen\charbox\hbox to \wd\charbox{\hss/\hss}
 \llap{$#1$}
}}

\graphicspath{%
{./}%
{figs/}%
}

\newcommand{\newc}{\newcommand}

\makeatletter

\newcommand{\Rmnum}[1]{\expandafter\@slowromancap\romannumeral #1@}
\makeatother

\newc{\squark}{\tilde{q}}
\newc{\ssup}{\tilde{u}}
\newc{\ssdown}{\tilde{d}}
\newc{\ssstrange}{\tilde{s}}
\newc{\sscharm}{\tilde{c}}
\newc{\sstop}{\tilde{t}}
\newc{\ssbottom}{\tilde{b}}
\newc{\sse}{\tilde{e}}
\newc{\ssmu}{\tilde{\mu}}
\newc{\sstau}{\tilde{\tau}}
\newc{\ssnu}{\tilde{\nu}}
\newc{\ssnue}{\tilde{\nu}_{e}}
\newc{\ssnumu}{{\tilde{\nu}_{\mu}}}
\newc{\ssnutau}{{\tilde{\nu}_{\tau}}}
\newc{\ssbnue}{\bar{\tilde{\nu}}_{e}}
\newc{\ssbnumu}{\bar{\tilde{\nu}}_{\mu}}
\newc{\ssbnutau}{\bar{\tilde{\nu}}_{\tau}}
\newc{\neut}{{\tilde{\chi}}^0}
\newc{\charge}{\tilde{\chi}}
\newc{\gluino}{\tilde{g}}
\newc{\MSUSY}{M_\text{SUSY}}
\newc{\CP}{\mathcal{CP}}
\newc{\MStop}{M_S}

\newc{\SMU}[1]{M_{\tilde{U}_#1}}
\newc{\SMD}[1]{M_{\tilde{D}_#1}}
\newc{\SMQ}[1]{M_{\tilde{Q}_#1}}
\newc{\SML}[1]{M_{\tilde{L}_#1}}
\newc{\SME}[1]{M_{\tilde{E}_#1}}

\newc{\brhnp}{\text{BR}(h\to \text{NP})}
\newc{\brhinv}{\text{BR}(h\to \text{inv.})}
\newc{\brhnpsusy}{\text{BR}(h\to \text{NP}+\text{SUSY})}
\newc{\brhsusy}{\text{BR}(h\to \text{SUSY})}

\newc{\htogaga}{h\to\gamma\gamma}
\newc{\htobb}{h\to b\bar{b}}
\newc{\htotautau}{h\to \tau^+\tau^-}

\newc{\gev}{~\mathrm{GeV}}
\newc{\tev}{~\mathrm{TeV}}
\newc{\CL}[1]{#1\%~\mathrm{C.L.}}

\newc{\FH}{\texttt{FeynHiggs}}
\newc{\FHv}[1]{\texttt{FeynHiggs-#1}}
\newc{\SH}{\texttt{SusHi}}
\newc{\SHv}[1]{\texttt{SusHi-#1}}
\newc{\HB}{\texttt{HiggsBounds}}
\newc{\HBv}[1]{\texttt{HiggsBounds-#1}}
\newc{\HS}{\texttt{HiggsSignals}}
\newc{\HSv}[1]{\texttt{HiggsSignals-#1}}

\newc{\htg}[1]{{{\color{green}  #1}}  \color{black}}
\newc{\htb}[1]{{{\color{blue}  #1}}  \color{black}}
\newc{\htr}[1]{{{\color{red}  #1}} \color{black}}

\newc{\abbrev}{\rm\scalefont{.9}}
\newc{\nnlo}{{\abbrev NNLO}}
\newc{\nnnlo}{{\abbrev NNNLO}}
\newc{\nlo}{{\abbrev NLO}}
\newc{\lo}{{\abbrev LO}}
\newc{\qcd}{{\abbrev QCD}}
\newc{\smallr}{{\scriptscriptstyle R}} %
\newc{\smallf}{{\scriptscriptstyle F}} %
\newc{\muF}{\mu_\smallf}
\newc{\muR}{\mu_\smallr}

\newc{\eqn}[1]{Eq.\,(\ref{#1})}
\newc{\fig}[1]{Fig.\,\ref{#1}}
\newc{\figs}[1]{Figs.\,\ref{#1}}
\newc{\citere}[1]{Ref.~\cite{#1}}
\newc{\citeres}[1]{Refs.~\cite{#1}}

\title{Light Stop Mass Limits from Higgs Rate Measurements in the MSSM: Is MSSM Electroweak Baryogenesis Still Alive After All?}

\author[a]{Stefan~Liebler,}
\affiliation[a]{Deutsches Elektronen-Synchrotron (DESY), Notkestra{\ss}e 85, 22607 Hamburg, Germany}
\emailAdd{stefan.liebler@desy.de}

\author[b]{Stefano~Profumo}
\emailAdd{profumo@scipp.ucsc.edu}
\affiliation[b]{Department of Physics and Santa Cruz Institute for Particle Physics (SCIPP),\\ University of California Santa Cruz, CA 95064, USA}

\author[b]{and Tim~Stefaniak}
\emailAdd{tistefan@ucsc.edu}

\abstract{
\noindent We investigate the implications of the Higgs rate measurements from Run~1 of the LHC for the mass of the light scalar top partner (stop) in the Minimal Supersymmetric Standard Model (MSSM). We focus on light stop masses, and we decouple the second, heavy stop and the gluino to the multi-TeV range in order to obtain a Higgs mass of $\sim 125\gev$. We derive lower mass limits for the light stop within various scenarios, taking into account the effects of a possibly light scalar tau partner (stau) or chargino on the Higgs rates, of additional Higgs decays to undetectable ``new physics'', as well as of non-decoupling of the heavy Higgs sector. Under conservative assumptions, the stop can be as light as $123\gev$. Relaxing certain theoretical and experimental constraints, such as vacuum stability and model-dependent bounds on sparticle masses from LEP, we find that the light stop mass can be as light as $116\gev$. Our indirect limits are complementary to direct limits on the light stop mass from collider searches and have important implications for electroweak baryogenesis in the MSSM as a possible explanation for the observed matter-antimatter asymmetry of the Universe.
}

\arxivnumber{1512.09172}


\begin{document}
\maketitle


\section{Introduction}

A long-standing problem in particle physics and cosmology is the question of the origin of the matter-antimatter asymmetry of the Universe. A well-motivated framework for the dynamical generation of this baryon-antibaryon asymmetry is electroweak baryogenesis (EWBG; for a recent review see \citere{Morrissey:2012db}). In this framework, the electroweak phase transition is strongly first-order, with expanding bubbles of broken electroweak phase providing out-of-equilibrium regions, one of the three Sakharov conditions \cite{Sakharov:1967dj} for the dynamical generation of a baryon-antibaryon asymmetry. If sufficiently large sources of $\CP$ violation exist and are associated with degrees of freedom light enough to be abundant at the time of the phase transition, sphaleron transitions can convert a net left-handed chiral current into a net baryon-antibaryon asymmetry in the regions around the expanding bubble walls. The asymmetry can then diffuse inside the regions of broken electroweak phase, ``freeze in'' inside the bubbles, and survive to date, as long as sphaleron processes are suppressed enough in the broken phase. This latter condition implies, in turn, a specific quantitative requirement on the properties of the electroweak phase transition.

While electroweak baryogenesis demands ingredients beyond the particle content of the Standard Model (SM)  \cite{Morrissey:2012db}, it is simple enough to construct working examples for such a scenario in the context of well-motivated extensions of the SM, such as the minimal supersymmetric extension to the Standard Model (MSSM). The general requirements on the specific incarnation of the MSSM that leads to successful baryogenesis at the electroweak scale always involve (i) a light scalar top (stop) quark, with a mass much below the top quark mass, in order to have an enhanced strongly-first order phase transition, as well as (ii) new sources of $\CP$ violation, typically either in the electroweak-ino sector or in the sfermion sector.

The electroweak baryogenesis framework additionally possesses the important feature of being an {\em eminently testable framework} for the production of the baryon asymmetry in the Universe. Several studies have pointed out a variety of tests that would probe in different ways, but rather conclusively, this route for the generation of a baryon asymmetry in the Universe, including collider studies, searches for the electric dipole moment of elementary particles or atoms, gravity waves, dark matter searches etc.~(an incomplete list of references includes e.g.~\citeres{Carena:1996wj,Delepine:1996vn,Carena:1997gx,Carena:1997ki,Carena:2002ss,Balazs:2004bu,Cirigliano:2006dg,Li:2008ez,Carena:2008rt,Carena:2008vj,Cirigliano:2009yd,Carena:2012np,Kozaczuk:2011vr,Kozaczuk:2012xv}).

As mentioned above, a common denominator to any MSSM electroweak baryogenesis model is a light right-handed stop.\footnote{The lighter stop state is generally considered to be mostly right-handed in order to avoid constraints from electroweak precision observables~\cite{Espinosa:2012in,Djouadi:1998sq}.} Such a light, SU$(3)_c$ triplet degree of freedom with an $\mathcal{O}(1)$ coupling to the Higgs has evidently important implications for the Higgs sector, which data from the Large Hadron Collider (LHC) is putting under closer and closer scrutiny. In particular, the mass and signal rate measurements of the SM-like Higgs boson that has been discovered by the LHC experiments ATLAS~\cite{Aad:2012tfa} and CMS~\cite{Chatrchyan:2012xdj} yield stringent constraints on the MSSM light stop scenario (see e.g.~\citeres{Dermisek:2007fi,Arvanitaki:2011ck,Carena:2011aa,Cohen:2012zza,Curtin:2012aa,D'Agnolo:2012mj,Farina:2013ssa,Gori:2013mia,Fan:2014txa,Katz:2015uja}).

Important constraints on MSSM electroweak baryogenesis also arise from collider searches for the direct production of light stops, see e.g.~\citeres{Ferretti:2015dea,Eifert:2014kpa,Aad:2015pfx} for recent discussions of the current status. The experimental constraints on the light stop scenario are quite strong under the simplifying assumption that the stop decays purely to a charm quark and a stable lightest neutralino, $\sstop_1 \to c \neut_1$~\cite{Aad:2014nra,CMS:2014yma}.\footnote{Recently it was shown in \citere{Tackmann:2016jyb} that jet veto resummation impacts the uncertainties on direct bounds of slepton searches. In case additional jets are vetoed, see e.g. the monojet-like searches in \citere{Aad:2014nra}, also direct squark searches are similarly affected.} In the case of 4-body stop decays to the neutralino via virtual top quarks and $W$ bosons, $\sstop_1 \to b f\bar{f} \neut_1$, or admixtures of these decay modes, the constraints are weaker~\cite{Grober:2014aha,Belyaev:2015gna}. Nevertheless, both possibilities (and arbitrary admixtures thereof) are by now excluded by the latest ATLAS results~\cite{Aad:2015pfx} for light stop masses below the top mass. Alternatively, if charginos are light, the light stop can undergo the decay chain $\sstop_1\to b\tilde{\chi}_1^\pm$, with successive decay of the chargino to a neutralino and, e.g., a (virtual) $W$ boson, $\tilde{\chi}^\pm_1 \to W^{(*)\,\pm} \neut_1$. The limit depends on the masses of all involved particles, i.e.~the light stop, chargino and neutralino mass, and unexcluded parameter regions still exist for stop masses below the top mass~\cite{Aad:2015pfx}. Such a scenario can e.g.~be realized if the light chargino and lightest neutralinos have a large Higgsino component. Potential admixtures of light stop decay modes with and without intermediate charginos may further weaken the current exclusion limits. We conclude that these searches highly depend on the assumed decay mode(s) of the stop and on the mass spectrum of the involved supersymmetric particles. Furthermore, small new physics effects \emph{beyond} the MSSM, such as a small $R$-parity violating coupling, may drastically change the exclusion limits obtained from direct searches. While direct searches are very important, the model dependence of the derived limits emphasizes the need of the complementary approach for obtaining constraints on the light stop scenario in the MSSM that we follow here --- the study of the implications from Higgs precision measurements.

At tree-level the light Higgs boson mass in the MSSM is bounded from above by the $Z$ boson mass, $m_h^\text{tree} \le M_Z$. In order to lift the Higgs mass to its observed value of $m_h \sim 125\gev$~\cite{Aad:2015zhl} large radiative mass corrections are needed. The dominant contributions to the Higgs mass come from the stop sector~\cite{Haber:1990aw,Okada:1990vk,Ellis:1990nz,Haber:1996fp,Carena:2000dp}, with logarithmic sensitivity on the stop masses and quadratic and quartic sensitivity on the stop mixing parameter. Subdominant contributions arise from the sbottom and the gluino sector (see \citeres{Heinemeyer:2004ms, Djouadi:2005gj} for reviews). Under the assumption that the masses of the supersymmetric partner particles, in particular those of the stops and the gluino, are not too far above the TeV-scale, the Higgs boson mass restricts the light stop to be heavier than $200\gev$~\cite{Draper:2011aa,Bechtle:2012jw} (see also Refs.~\cite{Han:2013kga,Kobakhidze:2015scd}), thus ruling out, at face value, the possibility of successful EWBG in the MSSM. Moreover, under these circumstances, light stop masses of $\sim 200\gev$ are viable only for a large stop mixing parameter, which would suppress the coupling of the light stop to the light Higgs and thus prevent it from having a significant effect on the electroweak phase transition~\cite{Carena:2008rt,Menon:2009mz}. Consequently, in order to allow for light stop masses \emph{below} the top mass which exhibit a substantial influence on the strength of the electroweak phase transition, a large mass splitting in the stop sector with a multi-TeV heavy stop and a small to moderate stop mixing parameter is needed.\footnote{New physics \emph{beyond} the MSSM can provide new tree-level contributions to the Higgs mass, thus reducing the need for large radiative corrections. A prominent example is the Next-to-Minimal Supersymmetric Standard Model (NMSSM), where the MSSM Higgs sector is extended by a complex scalar singlet, see e.g. \citeres{Huber:2006wf,Carena:2011jy,Kozaczuk:2013fga,Kozaczuk:2014kva}.}

The implications of a light stop scenario compatible with MSSM electroweak baryogenesis for the light Higgs phenomenology have been discussed in \citeres{Menon:2009mz,Cohen:2012zza,Curtin:2012aa} (see also \citere{Djouadi:2005gj}): A light stop with a small stop mixing parameter generically leads to a strong enhancement of the loop-induced light Higgs coupling to gluons and a moderate suppression of its coupling to photons. Since the main Higgs production mode at the LHC in the inclusive channels is the gluon fusion process, $gg\to h$, a light stop thus leads typically to too high Higgs signal rates,  in contradiction with the current LHC measurements. Moreover, Higgs search channels targetting the Higgs production in the vector boson fusion (VBF) process or in association with a $W$ or $Z$ boson ($Wh$/$Zh$), do not feature a rate enhancement from a light stop. These tensions can be exploited to constrain the light stop scenario of the MSSM with Higgs signal rate measurements from the LHC.

In 2012, the authors of \citere{Curtin:2012aa} performed a global fit to very early measurements of the Higgs signal rates and claimed that EWBG in the MSSM had been ruled out. Their analysis, however, neglects the possibility of  other light charged sparticles influencing the Higgs rates. In particular, light scalar tau partners (staus) and charginos are known to substantially affect the Higgs decay rate to photons~\cite{Carena:2013iba,Belyaev:2013rza}. Moreover, it has been argued in \citere{Carena:2012np} that a light neutralino $\neut_1$ with a mass lower than around $60\gev$ allowing for the invisible\footnote{Due to the assumption of conserved $R$-parity and a neutralino LSP in \citere{Carena:2012np} the decay $h\to\neut_1\neut_1$ is considered to be an invisible Higgs decay. However, the mechanism would also work for a sucessively decaying neutralino.} Higgs decay $h\to\neut_1\neut_1$  globally reduces the Higgs rates of channels with SM particles in the final state. This mechanism can substantially reduce the tension between the predicted Higgs rates and the measurements. However, the tension between the rates of gluon fusion dominated Higgs channels and VBF/$Wh$/$Zh$ dominated Higgs channels still remains.

In the present study, in part motivated by probing MSSM electroweak baryogenesis, we embark on an update on mass limits for the light stop from Higgs rate measurements. We consider four distinct scenarios: one where we decouple every particle in the MSSM with the exception of a light right-handed stop, and three where we allow for additional light degrees of freedom. We focus on cases of relevance for MSSM electroweak baryogenesis, specifically with a light chargino or light stau and/or a non-decoupled heavy Higgs sector. We perform a detailed global $\chi^2$ analysis in each case, using $85$ signal strength measurements in various Higgs signal channels from the LHC and Tevatron experiments by employing the dedicated computer program \HS~\cite{Bechtle:2013xfa,HSreleasenote11,Bechtle:2014ewa} (version {\tt 1.4.0}) (based on the computer program~\HB~\cite{Bechtle:2008jh,Bechtle:2011sb,Bechtle:2013gu,Bechtle:2013wla,Bechtle:2015pma}). We discuss the preferred parameter regions and correlations and derive lower limits on the stop mass for each scenario.
We find that in all scenarios the light stop mass can be lower than $155\gev$ at 95$\%$ confidence level --- a mass value, where recent lattice studies have demonstrated that the conditions needed for electroweak baryogenesis can still be fulfilled~\cite{Laine:2012jy}. In particular, in the presence of light staus or charginos we find that the light stop mass can be substantially lower, down to around $\sim 123\gev$, while vacuum stability requirements and model-dependent collider bounds from LEP on sparticle masses are still satisfied. Further relaxing these model-dependent constraints we find that the light stop mass can be as light as $116\gev$. The latter lower limit is obtained in the case of a light stau and a non-decoupled heavy Higgs sector, where the light Higgs boson Yukawa and gauge boson couplings deviate at the percent level from the couplings of a SM Higgs boson.

Since our study deals with a large splitting between the electroweak scale and the masses of some colored SUSY particles, we need to discuss the effect of large logarithms on our results. For the prediction of the Higgs boson masses and branching ratios we employ the code \FH~\cite{Heinemeyer:1998yj,Heinemeyer:1998np,Degrassi:2002fi,Frank:2006yh} (version {\tt 2.11.0}). The \FH\ package has recently implemented the re-summation of large logarithms involving the masses of heavy SUSY particles and the top quark mass~\cite{Hahn:2013ria}. Similarly, logarithms of the masses of the gluino and the heavy second stop with respect to the electroweak scale occur in higher-order predictions of the gluon fusion cross section, $\sigma(gg\to h)$, which we calculate using the code \SH{}~\cite{Harlander:2012pb,Bagnaschi:2014zla,Liebler:2015bka} (version {\tt 1.4.1}). We argue below that the parameter space we consider still allows for a perturbative treatment of the gluon fusion cross section, as the dependence of $\sigma(gg\rightarrow h)$ on the masses of the heavy colored SUSY particles is observed to be small. Therefore the precise numerical values of the gluino mass and the second heavy stop mass are of minor relevance for our study and can thus be adjusted to tune the light Higgs boson mass $m_h$ to $\sim 125\gev$.
In contrast, the smallness of the light stop mass, $m_{\sstop_1}$, induces an additional uncertainty for both the gluon fusion cross section and the partial widths for Higgs decay to gluons, $\Gamma(h\rightarrow gg)$, since the next-to-leading order terms to $\sigma(gg\rightarrow h)$ within \SH\ and to $\Gamma(h\to gg)$ in \FH\ are implemented in expansions of either a vanishing Higgs mass or a heavy SUSY spectrum, which both require $m_h\ll 2m_{\sstop_1}$. Other theoretical and parametric uncertainties to $\sigma(gg\rightarrow h)$ are included in a similar way as in \citere{Bagnaschi:2014zla}. All these uncertainties are carefully implemented in our statistical analysis.

The remainder of our study is organized as follows: We first provide an introduction to the theoretical framework in Section~\ref{Sect:framework}, which includes a description of the stop sector with a large mass splitting. We then comment on the Higgs mass calculation in such a scenario, and give a detailed discussion of light Higgs boson production. We also include the description of our prescription for implementing SUSY corrections to the decays of the light Higgs boson into photons. Subsequently, we discuss theoretical uncertainties for gluon fusion, emphasizing in particular the contributions from the light stop, which -- due to employed approximations at higher order -- come with an additional uncertainty as a function of the light stop mass. In Section~\ref{Sect:numerics} we discuss our numerical procedure, which includes our parameter space choices and a prescription for the employed codes used for the fitting procedure. We then present in Section~\ref{Sect:Results} our results for four distinct scenarios, which include a light stop and possibly a light stau, a light chargino and non-decoupling effects. All scenarios are motivated in the context of electroweak baryogenesis by arguments we provide in Section~\ref{Sect:EWBG}. Finally, we conclude in Section~\ref{Sect:conclusions}. In Appendix~\ref{App:measurements} we list the Higgs rate measurements that are used in our analysis.

\section{Theoretical framework}
\label{Sect:framework}

In this section we introduce the theoretical framework on which our study is based. After a discussion of the stop and  Higgs sector in the $\CP$-conserving MSSM we elaborate on the calculation of higher-order corrections to light Higgs boson production and decay rates, focussing in particular on the contributions arising from a light stop and, potentially, a light chargino or stau. At the end of this section we provide a thorough discussion of the theoretical uncertainties for light Higgs boson production and decay rates which will be incorporated in our numerical analysis.

\subsection{MSSM Higgs sector and stop sector with large mass splitting}

We consider the phenomenological $\CP$-conserving MSSM, where SUSY soft-breaking parameters are real and defined at the low scale, i.e.~not too far above the electroweak (EW) scale. In the following we briefly introduce the parameters  relevant for our study.

\begin{figure*}
\centering
\includegraphics[width=0.65\textwidth]{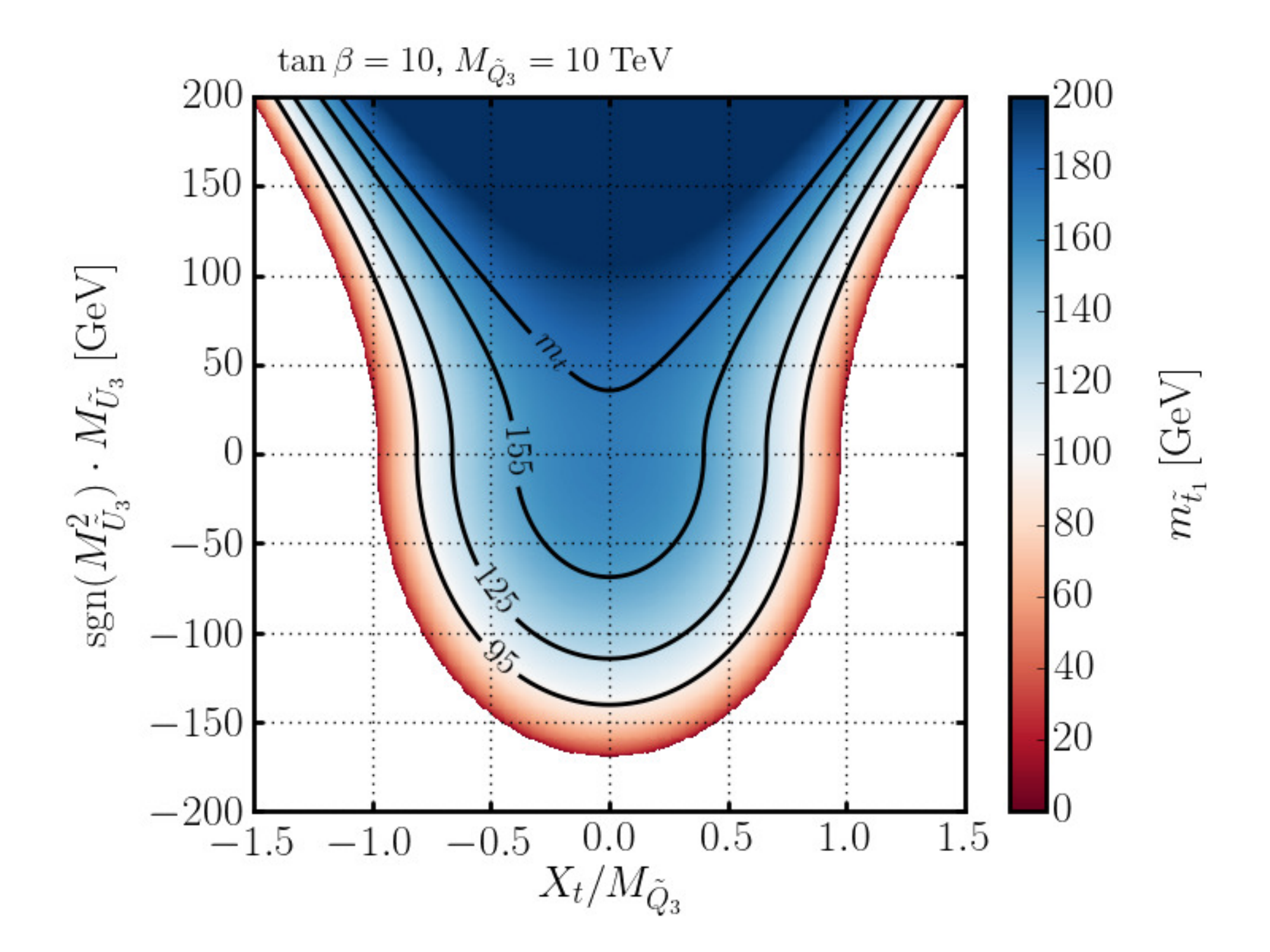}
\vspace{-0.4cm}
\caption{Tree-level mass for the light stop $m_{\sstop_1}$ as a function of the right-handed stop soft-breaking mass parameter, $\SMU{3}$, and the stop mixing parameter, $X_t/\SMQ{3}$, for  $\SMU{3} \ll \SMQ{3} = 10\tev$ and $\tan\beta = 10$.}
\label{Fig:stopmass}
\end{figure*}

At tree-level, the MSSM Higgs sector depends on only two parameters:  $\tan\beta \equiv v_u/v_d$, the ratio of the vacuum expectations values (vevs) of the two Higgs doublets, and $M_A$, the mass of the $\CP$-odd (or pseudoscalar) Higgs boson $A$.
Beyond tree-level, Higgs masses and couplings receive important radiative corrections, especially from the top/stop sector, as well as, for large values of $\tan\beta$,  from the bottom/sbottom and tau/stau sector. The tree-level mass spectrum of the third generation sfermions is determined by the following mass matrices (in the basis of the current eigenstates $\tilde{f}_L$, $\tilde{f}_R$ with $f=t,b,\tau$):
\begin{align}
\mathcal{M}_{\sstop}^2 = \begin{pmatrix} \SMQ{3}^2 +m_t^2 + (\tfrac{1}{2} -\tfrac{2}{3} s_w^2) M_Z^2\cos2\beta & m_t X_t \\ m_t X_t & \SMU{3}^2 + m_t^2 + \tfrac{2}{3} s_w^2 M_Z^2 \cos2\beta \end{pmatrix}\,,
\label{Eq:Mstop}
\end{align}
\begin{align}
\mathcal{M}_{\ssbottom}^2 = \begin{pmatrix} \SMQ{3}^2 +m_b^2 + (-\tfrac{1}{2} +\tfrac{1}{3} s_w^2) M_Z^2\cos2\beta & m_b X_b \\ m_b X_b & \SMD{3}^2 + m_b^2 - \tfrac{1}{3} s_w^2 M_Z^2 \cos2\beta \end{pmatrix}\,,
\end{align}
\begin{align}
\mathcal{M}_{\sstau}^2 = \begin{pmatrix} \SML{3}^2 +m_\tau^2 + (-\tfrac{1}{2} +s_w^2) M_Z^2\cos2\beta & m_\tau X_\tau \\ m_\tau X_\tau & \SME{3}^2 + m_\tau^2 - s_w^2 M_Z^2 \cos2\beta \end{pmatrix}\,,
\end{align}
where $\SMQ{3},\SML{3}$ and $\SMU{3},\SMD{3},\SME{3}$ are the left- and right-handed soft-supersymmetry-breaking sfermion masses
of the third sfermion generation, respectively, $m_t$, $m_b$, $m_\tau$ and $M_Z$ the top and bottom quark, $\tau$ lepton and $Z$ boson mass, respectively, and $s_w \equiv \sin\theta_w$ with the weak mixing angle $\theta_w$. The stop, sbottom and stau mixing parameters, $X_{t,b,\tau}$, are defined as
\begin{align}
X_t = A_t - \mu/\tan\beta,\\
X_{b,\tau} = A_{b,\tau} - \mu \tan\beta\,.
\end{align}
Here, $A_{f}$ ($f=t,b,\tau$) are the trilinear Higgs-sfermion couplings, and $\mu$ is the Higgsino mass parameter. Diagonalizing the stop mass matrix, \eqn{Eq:Mstop}, yields the following tree-level mass for the lighter stop state:
\begin{align}
\label{Eq:treelevelstopmass}
m_{\sstop_{1}}^2 =  m_t^2 + \frac{1}{2}\Big[ & \SMQ{3}^2 + \SMU{3}^2 + \tfrac{1}{2} M_Z^2\cos2\beta  \nonumber  \\
& - \sqrt{\left(\SMU{3}^2- \SMD{3}^2 +(\tfrac{1}{2} - \tfrac{4}{3}s_w^2 )M_Z^2 \cos2\beta \right)^2 + 4m_t^2 X_t^2} \, \Big]\,.
\end{align}
A light stop mass below the top mass, $m_{\sstop_1} < m_t$, can easily be obtained for negative $\SMU{3}^2$ and/or large $X_t$ values of the order of $\SMQ{3}$. The terms $\propto \cos2\beta$ also lower the light stop mass (for the physically relevant case of $\tan\beta > 1)$. \eqn{Eq:treelevelstopmass} is illustrated in \fig{Fig:stopmass} where we show $m_{\sstop_1}$ as a function of $\SMU{3}$ and $X_t/\SMQ{3}$ for fixed values of $\SMQ{3} = 10\tev$ and $\tan\beta = 10$.  In both codes that we employ, \FH{} and \SH{}, the stop sector is renormalized on-shell, such that the tree-level mass in \eqn{Eq:treelevelstopmass} coincides with the on-shell stop mass.

The mass of the light Higgs~$h$ in the MSSM is lifted above its maximal tree-level value~$M_Z$ by higher order corrections,
mainly driven by stop corrections. For illustration, the one-loop correction to the light Higgs boson mass in the decoupling limit
can be written in the form~\cite{Haber:1990aw,Okada:1990vk,Ellis:1990nz,Haber:1996fp,Carena:2000dp}
\begin{align}
 (\Delta m_h^2)^{t,\sstop}_{\rm one-loop}\approx \frac{3m_t^4}{2\pi^2v^2}\left(\log\left(\frac{\MStop^2}{m_t^2}\right)+\frac{X_t^2}{\MStop^2}-\frac{X_t^4}{12\MStop^4}\right),
\label{Eq:higgsstopcorr}
\end{align}
with $\MStop\equiv\sqrt{m_{\sstop_1}m_{\sstop_2}}$ and the Higgs vacuum expectation value $v = \sqrt{v_u^2 + v_d^2} \approx 246.2\gev$.
In our case we are interested in a very light stop, and we thus focus on a large mass splitting in the stop sector by setting $\SMU{3}\ll \SMQ{3}$, i.e.~$m_{\sstop_1}\ll m_{\sstop_2}$,
so that the Higgs mass $m_h$ is lifted to the measured value $\sim 125\gev$. 
As can be seen from \eqn{Eq:higgsstopcorr} a very large stop mixing parameter $X_t\gg \mathcal{O}(\MStop)$ induces a negative contribution to the light Higgs  mass, $m_h$.
On the other hand, $X_t\lesssim \mathcal{O}(\MStop)$ results in a very small stop mixing, since the relevant parameter $X_t$ remains small compared to $\SMQ{3}$.
We therefore set the stop mixing parameter to $X_t=0$ throughout our analysis,
which is also preferred by electroweak baryogenesis. The absence of a large stop mixing is crucial for
our findings since in this case the gluon fusion cross section and the $h\rightarrow gg$ rate
are enhanced through the contributions from a light stop.
We note that a vanishing stop mixing reduces the effects of complex phases of $A_t$ and $\mu$
in the stop sector. Assuming no $\CP$ violation in the stop sector is therefore well-motivated, and warranted from two-loop constraints from electric dipole moments in the context of electroweak baryogensis \cite{Cirigliano:2009yd}.
In principle, the alternative ansatz $\SMQ{3}\ll \SMU{3}$, which results in
a light sbottom in addition to the light stop might also be of interest. A light left-handed stop is however incompatible with electroweak precision
observables, see e.g.~\citere{Espinosa:2012in}. Similarly, a non-zero stop mixing in combination with a large mass splitting between the stops can induce large contributions to those observables~\cite{Djouadi:1998sq}, which also motivates the choice of vanishing stop mixing, $X_t=0$.

For the precise calculation of the Higgs boson mass~$m_h$ and the branching ratios of the light Higgs~$h$
we make use of the public code \FH~\cite{Heinemeyer:1998yj,Heinemeyer:1998np,Degrassi:2002fi,Frank:2006yh},
which also incorporates the re-summation of logarithms arising from a heavy SUSY spectrum~\cite{Hahn:2013ria}.
Our specific case of one light stop together with a heavy stop and a heavy gluino was discussed in an
effective field theory approach in \citere{Espinosa:2001mm}, see their Eq.~(30).
However, we point out that our findings are
mostly insensitive to the heavy stop and the heavy gluino mass. Their actual numerical values
can be adjusted such that the light Higgs boson mass $m_h$ fits $\sim 125\gev$. We only demand a small stop mixing, which has a direct influence
on light Higgs boson production and decay.

\subsection[Higher-order and SUSY corrections to Higgs rates]{Higher-order and SUSY corrections to Higgs production and decay rates}
\label{Sect:higherorder}

In this subsection we discuss the implementation of the relevant light Higgs production and decay processes within the two codes \FH{} and \SH{}, and include a description of higher-order contributions with a focus on the stop contributions.
Although we include the full squark sector including squark mixing for what concerns Higgs production and decay in the two codes,
we derive simplified formulae in order to describe the observed behavior as a function of the light stop mass.
As explained above, we assume that the stop mixing term $X_t$ is small compared to $m_{\tilde{t}_2}$, such that the mixing in the stop sector remains small.
Using the notation of the \SH{} manual~\cite{Harlander:2012pb}, the coupling of the light Higgs $h$ to the lightest stop in the decoupling limit can  be written in the form
\begin{align} 
g_{\tilde{t},11}^h= 2+\frac{M_Z^2}{m_t^2}\left(1-\frac{4}{3}s_w^2\right) \cos 2\beta
\label{Eq:couplingstop11}
\end{align}
and is thus independent of the lightest stop mass $m_{\tilde{t}_1}$. The contribution of the lightest stop to the \lo{} gluon fusion
cross section at the amplitude level is given by the simple expression
\begin{align}
\mathcal{A}_{\tilde{t}_1}^{\lo{}} = -\frac{6m_t^2}{4m_h^2}g_{\tilde{t},11}^h(1-\tau_{\tilde{t}1}f(\tau_{\tilde{t}1}))\quad\xrightarrow{\tau_{\sstop_1} \gg 1}\quad
g_{\tilde{t},11}^h\frac{m_t^2}{m_{\tilde{t}_1}^2}\left(\frac{1}{8}+\frac{m_h^2}{m_{\tilde{t}_1}^2}+\ldots\right)
\end{align}
with $f(\tau)=\text{arcsin}^2(1/\sqrt{\tau})$ and $\tau_{\tilde{t}1}=4m_{\tilde{t}_1}^2/m_h^2\gg 1$,
which -- if expanded in inverse powers of $\tau_{\tilde{t}1}$ -- scales like $1/m_{\tilde{t}_1}^2$.
The contribution from the heavy stop $\tilde{t}_2$ can be consistently neglected. 
A simple formula at \lo{} \qcd{} for the stop contributions to gluon fusion can also be
taken from Eq. (4.18) and (4.19) of \citere{Grojean:2013nya},
which confirms the trivial scaling.
Higher order corrections, remarkably, follow the same behaviour $1/m_{\tilde{t}_1}^2$ as
 can be analytically deduced from \citere{Degrassi:2008zj} at \nlo{} \qcd{}.
If we consistently expand the two-loop light stop contributions presented in Eq. (22) - (24)
of \citere{Degrassi:2008zj} in the heavy masses~$m_{\tilde{t}_2}$ and $m_{\tilde{g}}$ we obtain
\begin{align}\nonumber
G^{2l}_t&=-\frac{1}{12m_{\tilde{t}_1}^2}\left(C_A+C_F\frac{11}{2}\right)
+\frac{1}{2m_{\tilde{t}_1}^2}C_F\left[
\frac{1}{6}\log\left(\frac{m_{\tilde{t}_1}^2}{m_t^2}\right)
+\frac{1}{12}\log\left(\frac{m_{\tilde{t}_2}^2}{m_t^2}\right)
+\frac{1}{4}\log\left(\frac{m_{\tilde{g}}^2}{m_t^2}\right)
\right]\\
\tilde{G}^{2l}_t&=\tilde{F}^{2l}_t=
-\frac{1}{4m_{\tilde{t}_1}^2}\left(\frac{1}{3}C_A+\frac{5}{2}C_F\right)
-\frac{1}{12m_{\tilde{t}_1}^2}C_F\log\left(\frac{m_{\tilde{g}}^2}{m_{\tilde{t}_1}^2}\right)
\label{eq:twoloopcontr}
\end{align}
with $C_F=4/3$ and $C_A=3$. They enter the two-loop amplitude to gluon fusion in the form
\begin{align}
 \mathcal{A}_{\tilde{t}_1}^{\nlo{}}=-\frac{3}{4}\left[2m_t^2G^{2l}_t + M_Z^2\cos 2\beta \left(1-\frac{4}{3}s_w^2\right)\tilde{G}^{2l}_t\right]\,,
\end{align}
where $\mathcal{A}_{\tilde{t}_1}^{\nlo{}}$ is normalized as in the \SH{} manual~\cite{Harlander:2012pb}.
Due to the small mixing in the stop sector the contribution $F^{2l}_t$
yields a vanishing contribution. Note that terms proportional to 
$\mu/m_{\tilde{t}_2}$ and $\mu/m_{\tilde{g}}$ are set to zero.
Apart from the overall suppression by
$1/m_{\tilde{t}_1}^2$ the \nlo{} \qcd{} contributions are also logarithmically
dependent on the heavy SUSY masses. This behaviour was also discussed
in \citere{Harlander:2003bb} in the context of the
Appelquist-Carazzone decoupling theorem~\cite{Appelquist:1974tg}.
\citere{Muhlleitner:2008yw} worked out the decoupling of heavy gluinos
by means of an effective theory, where Higgs-stop and Higgs-top couplings
are independent of each other below the gluino mass scale.
We consider a large splitting between the electroweak scale
and in particular the gluino mass. Nevertheless the contributions given in
\eqn{eq:twoloopcontr} are still well within the perturbative regime. We checked explicitly that
the application of the resummed gluino contributions according to
Eq.~(27) of \citere{Muhlleitner:2008yw} at the level of the \lo{} \qcd{}
cross section leads to a very similar dependence on the gluino mass
than the one of the two-loop contribution provided in \eqn{eq:twoloopcontr}, which
justifies our perturbative treatment even though large logarithms are present.
If however the effective coupling in
Eq.~(27) of \citere{Muhlleitner:2008yw} gets below $0.6\cdot 2g_Q^{\mathcal{H}}m_Q^2/v$,
which can be translated into $m_{\tilde{g}}\gg 300\tev$,
the perturbative treatment becomes questionable.
Our perturbative treatment of the gluon fusion cross section leaves
a small dependence on the actual value of the gluino mass and the heavy
second stop mass, which however only enters first at \nlo{} \qcd{}.
As we emphasized the previous formulas are based on the absence of
a large mixing in the stop sector. In this case the contribution of
a light stop to the gluon fusion cross section is positive in contrast
to SUSY scenarios with two relatively light stops with sizable mixing, which
mostly yield a negative contribution.

The \SH{} code includes both top and bottom quark contributions with full massive quarks at \nlo{} \qcd{}~\cite{Spira:1995rr,Harlander:2005rq},
and \nnlo{} \qcd{} contributions in the effective theory of a
heavy top quark~\cite{Harlander:2002wh,Anastasiou:2002yz,Ravindran:2003um,Harlander:2002vv,Anastasiou:2002wq}.
Stop contributions are added at \nlo{} \qcd{} in the so-called vanishing Higgs mass limit (VHML)~\cite{Harlander:2004tp,Degrassi:2008zj}.
These contributions go beyond the illustrative discussion above, and allow for arbitrary stop mixing.
In the VHML limit they are even known at \nnlo{} \qcd{}~\cite{Pak:2010cu,Pak:2012xr}.
\SH{} includes approximate \nnlo{} stop contributions following Eq.~(15) of \citere{Bagnaschi:2014zla}.
Moreover \SH{} takes into account electroweak contributions induced by light quarks~\cite{Aglietti:2004nj,Bonciani:2010ms}.

For what concerns the Higgs decay into gluons, $h\rightarrow gg$,
\FH{} includes, in addition to the bottom and top-quark contributions at \nlo{} \qcd{},
also the \nlo{} \qcd{} stop contributions, according to \citeres{Lee:2003nta,Spira:1995rr}.
These are based on the assumption of a heavy loop mass or, alternatively, a small Higgs mass. This allows us to associate the VHML uncertainty, that we will deduce in the next subsection for the gluon fusion process, $gg\rightarrow h$, with the decay $h\rightarrow gg$.

Similarly, also \nlo{} \qcd{} contributions for the Higgs decay into photons, $h\rightarrow \gamma\gamma$,
are taken into account in \FH{},
both for quark and for the stop contributions, although the influence of the latter is smaller
compared to $h\rightarrow gg$.
We also include a discussion of the influence of light staus and charginos on our results, which in particular affect the
decay of the light Higgs into photons. In order to understand the observed behavior we subsequently
provide formulas for the Higgs to diphoton decay width at \lo{}.
Apart from the dominant SM-like contributions via the top-quark and
$W$-boson loop, it is given by~\cite{Kalyniak:1985ct,Bates:1986zv,Weiler:1988xn,Gunion:1988mf,Djouadi:1996pb}
\begin{align}
 \Gamma(h\rightarrow \gamma\gamma) =
 \frac{G_Fm_h^3\alpha_s^2}{128\sqrt{2}\pi^3}\left|\sin(\beta-\alpha)A_1(\tau_W)+\frac{4}{3}\frac{\cos\alpha}{\sin\beta}A_{1/2}(\tau_t)
 +\mathcal{A}^{\rm{SM}}+\mathcal{A}^{\rm{SUSY}}\right|^2
\end{align}
with $\tau_i=4m_i^2/m_h^2$ and the functions~(see e.g. \citere{Carena:2012xa})
\begin{align}
A_0(\tau)&=-(1-\tau f(\tau))\tau,\qquad A_{1/2}(\tau)=2(1+(1-\tau)f(\tau))\tau,\\ A_1(\tau)&=-(2/\tau+3+3(2-\tau)f(\tau))\tau\,.
\end{align}
Notice that the amplitude~$\mathcal{A}^{\rm{SM}}$ contains other minor SM contributions,  e.g.~from light quarks, and the amplitude~$\mathcal{A}^{\rm{SUSY}}$ contains all possible SUSY contributions. The latter are predominantly induced through the charged Higgs $H^\pm$, a light stop, a light stau and/or a light chargino.
In particular the contributions from $\tilde{t}_1$, $\tilde{\tau}_1$ and $\tilde{\chi}^\pm_1$ read
\begin{align}
\mathcal{A}^{\rm{SUSY}}= c^h_{\tilde{t}_1}A_0(\tau_{\tilde{t}_1}) + c^h_{\tilde{\tau}_1}A_0(\tau_{\tilde{\tau}_1})+
c^h_{\tilde{\chi}^\pm_1}A_{1/2}(\tau_{\tilde{\chi}^\pm_1})\,,
\end{align}
where we have employed the following abbreviations
{\allowdisplaybreaks
\begin{align}\label{eq:staucoupling}
 c^h_{\tilde{t}_1}&=
 \frac{2}{9m_{{\tilde{t}_1}}^2}\left[6m_t^2\frac{\cos\alpha}{\sin\beta} + 
 6m_t\sin\theta_{\tilde{t}}\cos\theta_{\tilde{t}}\left(\mu \frac{\sin\alpha}{\sin\beta}+ A_{t}\frac{\cos\alpha}{\sin\beta}\right)\right.\\\nonumber
 \qquad &\left.+M_Z^2\sin(\alpha+\beta)(-4s_w^2+\sin^2\theta_{\tilde{t}}(-3+8s_w^2)\right]\,,\\
 c^h_{\tilde{\tau}_1}&=
 -\frac{1}{4m_{{\tilde{\tau}_1}}^2}\left[2m_\tau^2\frac{\sin\alpha}{\cos\beta}+2m_\tau\sin\theta_{\tilde{\tau}}\cos\theta_{\tilde{\tau}}
 \left(\mu\frac{\cos\alpha}{\cos\beta} +A_{\tau}\frac{\sin\alpha}{\cos\beta}\right)\right.\\\nonumber
 \qquad &\left.+2M_Z^2\sin(\alpha+\beta)\left(-2s_w^2+\sin^2\theta_{\tilde{\tau}}(-1+4s_w^2)\right)\right]\,,\\
 c^h_{\tilde{\chi}^\pm_1}&=
 -\frac{\sqrt{2}M_W}{m_{\tilde{\chi}^\pm_1}}(U_{12}V_{11}\sin\alpha-U_{11}V_{12}\cos\alpha)\,.
\end{align}}
In the expressions above we assumed the chargino mixing matrices $U$ and $V$ to be real.
In the decoupling limit, $\sin(\beta-\alpha)\rightarrow 1$, and assuming no stop mixing the light Higgs coupling to the light stop yields, similar to the expression in \eqn{Eq:couplingstop11}:
\begin{align}
 c^h_{\tilde{t}_1}=\frac{2m_t^2}{3m_{\sstop_1}^2} \left[2+\frac{M_Z^2}{m_t^2}\left(1-\frac{4}{3}s_w^2\right) \cos 2\beta\right]\,.
\end{align}
In the decoupling limit the light Higgs coupling to the light stau is given by 
\begin{align}
  c^h_{\tilde{\tau}_1}&=
 -\frac{1}{4m_{{\tilde{\tau}_1}}^2}\left[-2m_\tau^2-m_\tau\sin2\theta_{\tilde{\tau}}
 X_\tau-2M_Z^2\cos2\beta\left(-2s_w^2+\sin^2\theta_{\tilde{\tau}}(-1+4s_w^2)\right)\right]\,,
\end{align}
where, in turn, $\sin(2\theta_{\tilde{\tau}})\propto m_\tau X_\tau$.
The coupling is thus enhanced by large values of $\mu\tan\beta$. In contrast, the chargino contribution is
proportional to the mixing in the chargino sector, which is enhanced for small values of $\tan\beta$.

The other branching ratios as well as the other production modes are much less affected by light stops
(and other light SUSY particles and a charged Higgs boson)
and are all taken from \FH{}:
Light stops enter bottom-quark annihilation~\cite{Harlander:2003ai}/bottom-quark
associated production~\cite{Dittmaier:2003ej,Dawson:2003kb} (denoted here as $gg\to b\bar{b}h$) only through the resummation
encoded in $\Delta_b$~\cite{Hempfling:1993kv,Hall:1993gn,Carena:1994bv,Carena:1999py,Carena:2002bb,Guasch:2003cv,Noth:2008tw,Noth:2010jy,Mihaila:2010mp}.
However, for a SM-like light Higgs $h$ bottom-quark
annihilation is irrelevant, even for large values of $\tan\beta$. Also the effect of light stops
on the partial width of the decay $h\rightarrow b\bar b$ through $\Delta_b$ is small.
Vector boson fusion (VBF) is only mildly affected by \qcd{} corrections. The same is true for the effect of light stops on VBF processes~\cite{Hollik:2008xn,Rauch:2010mi,Figy:2010ct}, particularly in the decoupling limit.
The Higgsstrahlung process, $pp\to Zh$, can be affected by stop contributions, especially for the subprocess $gg\rightarrow ZH$~\cite{Kniehl:2011aa},
which however only contributes at the level of $\mathcal{O}(10\%)$ to the inclusive Higgsstrahlung cross section
and only gains in relative size in the region of large transverse Higgs momenta~\cite{Harlander:2013mla}.
We take both the VBF and the Higgsstrahlung cross section from \FH{}, which reweights
the SM Higgs cross sections with the effective coupling of the light Higgs boson to heavy gauge bosons.
The same is true for the decay modes $h\rightarrow VV~(V=W^\pm,Z)$, where \FH{} approximates the MSSM results
through a reweighting of SM predictions obtained from {\tt Prophecy4f}~\cite{Bredenstein:2006ha,Bredenstein:2006rh}.
Therefore our predictions of the VBF and the Higgsstrahlung cross sections
as well as the decay width $\Gamma(h\rightarrow VV)$ are not affected
by a light stop (and other light SUSY particles and the charged Higgs boson). However, the branching ratio $\text{BR}(h\rightarrow VV)$ still indirectly depends on the light stop mass through the partial widths $\Gamma(h\rightarrow gg)$ and $\Gamma(h\to\gamma\gamma)$, which modify the total width.

\begin{figure}[t]
\subfigure[~(SM normalized) production rates of the light Higgs.]{\includegraphics[width=0.49\textwidth]{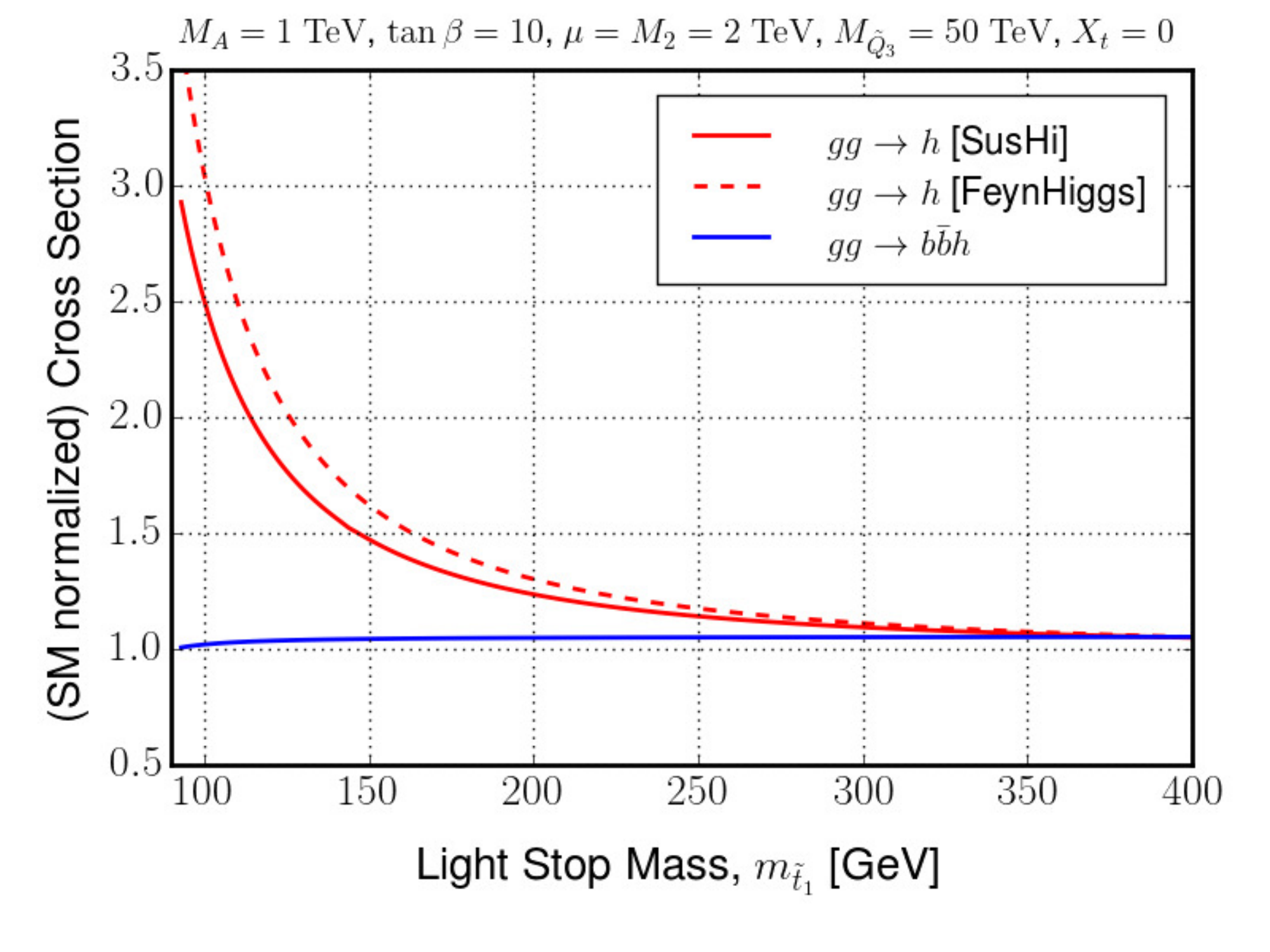}}
\subfigure[~Decay rates of the light Higgs.]{\includegraphics[width=0.49\textwidth]{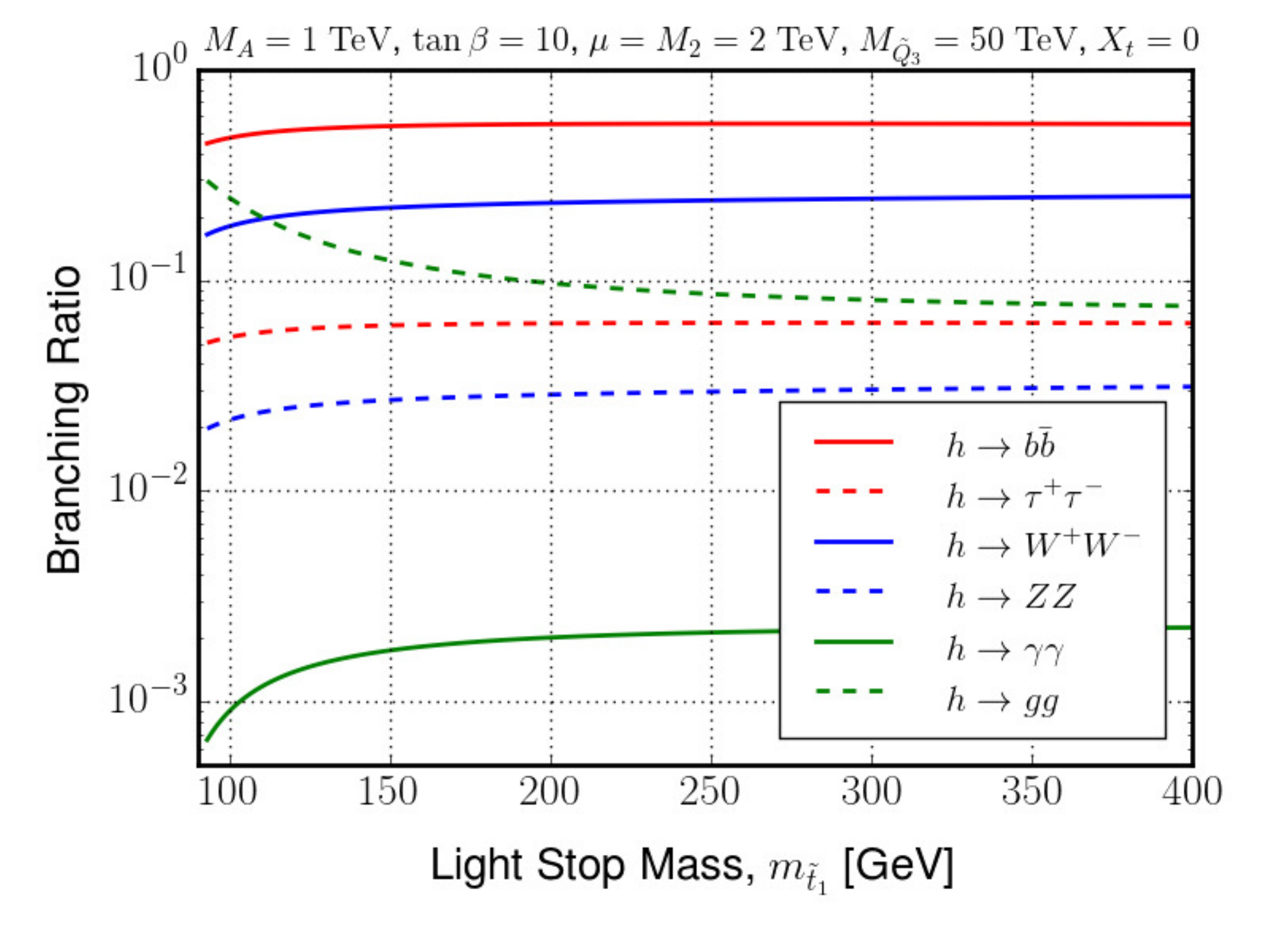}}
\caption{Rates for the production (a) and decays (b) of the light Higgs as a function of the light stop mass. We set $M_A=1\tev$, $\tan\beta=10$, $\mu=M_2 = 2\tev$, $M_{\tilde{Q}_3} = 50\tev$, $X_t = 0$, $M_{\tilde{D}_3} = 40\tev$ and $M_3 = 75\tev$.}
\label{Fig:nomix_CL}
\end{figure}

An illustration of the effect of a light stop on the gluon fusion cross section and the branching ratios is given in \fig{Fig:nomix_CL}. We choose the parameters $M_A=1\tev$, $\tan\beta=10$, $\mu=M_2 = 2\tev$, $M_{\tilde{Q}_3} = 50\tev$, $X_t = 0$, $M_{\tilde{D}_3} = 40\tev$ and $M_3 = 75\tev$. At smaller light stop masses, $m_{\sstop}$, we observe a strong enhancement of the the gluon fusion cross section, $\sigma(gg\to h)$, and the partial width $\Gamma(h\to gg)$. The latter leads to a suppression of all other decay modes through its influence on the total Higgs decay width. For large $m_{\sstop_1}$ the cross sections and branching ratios approach the predictions for the SM Higgs boson, as expected in the decoupling limit of the MSSM. 
The slight enhancement of bottom-quark annihilation, $gg\rightarrow b\bar bh$, is consistent with the delayed decoupling of the bottom-quark Yukawa coupling of the light Higgs boson.

\subsection{Discussion of theoretical uncertainties for $gg\rightarrow h$ and $h\rightarrow gg$}
\label{Sect:uncertainties}

We closely follow \citere{Bagnaschi:2014zla} for our procedure to estimate the theoretical uncertainties for the gluon fusion cross section.
Apart from the renormalization and
factorization scale uncertainty and the PDF$+\alpha_s$ uncertainty, we have to
consider two additional sources of theoretical uncertainties of particular relevance for light
stop scenarios. First we estimate the uncertainty from the fact that stop contributions at \nlo{}
and \nnlo{} are implemented in the vanishing Higgs mass limit (VHML).
Secondly, our implementation of \nnlo{} stop contributions is non-exact, since we miss three-loop
contributions in the Wilson coefficient $C^2$, see again \citere{Bagnaschi:2014zla}.

\begin{figure}[t]
\centering
\includegraphics[width=0.6\textwidth]{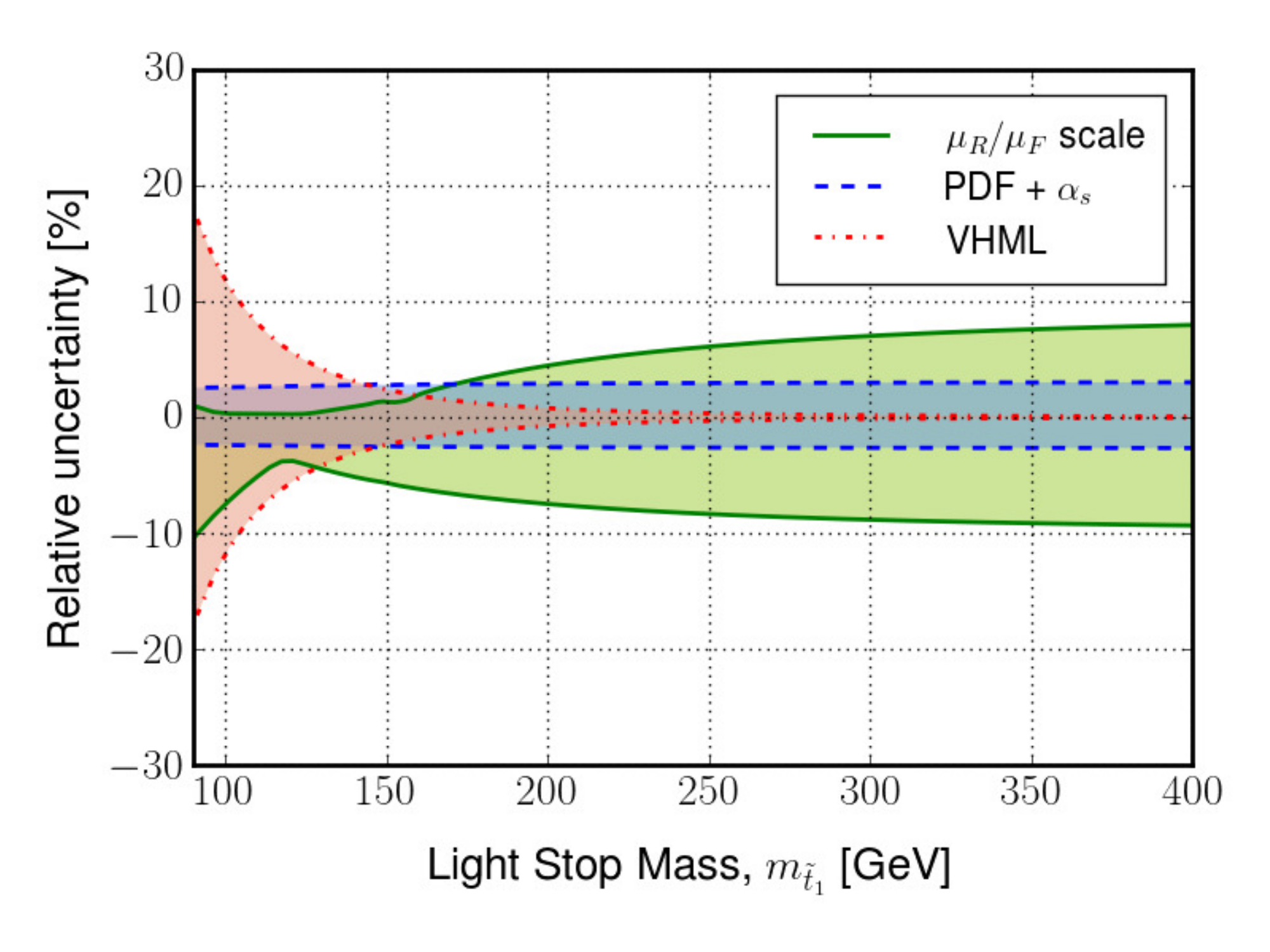} 
\vspace{-0.6cm}
\caption{Relative uncertainties (in \%) for the gluon fusion cross section
as a function of $m_{\tilde{t}_1}$. The renormalization
and factorization scale uncertainty corresponds to the green, solid curves,
the PDF$+\alpha_s$ uncertainty (using the MSTW2008 prescription) to the
blue, dashed curves and the VHML expansion uncertainty to the red, dotted-dashed
curves. The uncertainty due to non-exact NNLO stop contributions is far less than $1\%$ and thus not shown.
}
\label{fig:gghuncertainties} 
\end{figure}

Our subsequent discussion of theoretical uncertainties is performed for a typical SUSY scenario used in our analysis, i.e.~a scenario in the decoupling limit with a right-handed light stop. For this purpose we link \SH{} to \FH{} and choose similar MSSM parameters as in the previous example, namely $M_A=1\tev, \, \tan\beta=10,\, M_3=75\tev, X_t\approx 0$.
Again, the soft-breaking masses are fixed to $50\tev$, except for the right-handed soft breaking
masses in the sbottom sector, which is set to $\SMD{3}=40\tev$. We vary 
\begin{align}
 \text{sgn}(\SMU{3}^2)\SMU{3}\in [ -150,400]\gev
\end{align}
and thus obtain values of $m_{\tilde{t}_1}$ between $80\gev$ and $430\gev$.
For our purposes it is crucial that the Higgs mass is close to $125\gev$. The following uncertainty discussion is mostly insensitive to the other SUSY parameters as long as the stop mixing remains small. We list the relevant theoretical uncertainties for the gluon fusion cross section $\sigma(gg\rightarrow h)$:

\begin{itemize}
\item The PDF$+\alpha_s$ uncertainty is obtained following the MSTW2008~\cite{Martin:2009iq} prescription as provided
by the PDF4LHC group~\cite{Alekhin:2011sk,Botje:2011sn}. The result, as a function of the light stop mass, is shown with the blue, dashed curve in \fig{fig:gghuncertainties}. As expected (see e.g.~\citeres{Bagnaschi:2014zla,Liebler:2015bka}), the PDF$+\alpha_s$ error is mainly dependent on the Higgs mass and mostly insensitive to the SUSY scenario. In the following
we therefore pick the full PDF$+\alpha_s$ uncertainty (combining the results of different PDF fitting groups)
for a SM Higgs as it is provided by the LHC Higgs Cross Section Working Group (LHCHXSWG), given by $+7.5\%,-6.9\%$ for a Higgs mass of $125\gev$.
\item In order to estimate the uncertainty associated with the choice of the renormalization scale, $\muR$, and factorization scale, $\muF$,
we pick a nine-point combination for $\muR$ and $\muF$.
Similar to \citere{Bagnaschi:2014zla} we pick the pairs $(\muR,\muF)$
out of $\muR=\lbrace m_h/4,m_h/2,m_h\rbrace$ and $\muF=\lbrace m_h/4,m_h/2,m_h\rbrace$ with
the constraint $1/2\leq \muR/\muF\leq 2$. Since we observe a cancellation of the scale uncertainties
between top and stop contributions for light stop masses, we add two more scale choices: $(\muR,\muF)=(3m_h/8,3m_h/8)$ and $(\muR,\muF)=(3m_h/4,3m_h/4)$. Out of the nine pairs
we identify the minimal and maximal cross section as lower and upper bound for the scale uncertainty.
The green, solid curves in \fig{fig:gghuncertainties} reflect the obtained renormalization and
factorization scale uncertainties with respect to our central choice $\muR=\muF=m_h/2$.
Since the cancellation effect of top and stop scale uncertainties can be considered a coincidence,
we may use the scale uncertainties of a SM Higgs as provided by the LHCHXSWG as a conservative approach,
giving $+7.2\%,-7.8\%$ for a Higgs mass of $125\gev$. This is close to our scale uncertainty
without stop contributions.
\item Our implementation of two-loop and approximate three-loop stop contributions 
is based on the VHML, which assumes a massless Higgs boson, and is analogous to the
heavy-top limit known for the top contribution in the SM. Strictly speaking, the parameter
$\tau_{\sstop_1}=4m_{\tilde{t}_1}^2/m_h^2\gg 1$ is assumed to be large.
The VHML thus becomes invalid if $m_{\tilde{t}_1}\lesssim 62.5\gev$ for a Higgs mass of $m_h\sim 125\gev$. 
Even for light stop masses in the range $m_h/2 \lesssim m_{\sstop_1}\lesssim m_h$ the approximation seems questionable and potentially inaccurate.
We therefore assign an uncertainty to the VHML expansion as follows: At LO we know the exact amplitude,
$\mathcal{A}_{\tilde{t}_1}^{\lo{}}$, as well as its VHML expansion,
$\mathcal{A}_{\tilde{t}_1}^{\lo{},\text{VHML}} = g_{\tilde{t},11}^h m_t^2/(8m_{\tilde{t}_1}^2)$.
We multiply all occurrences of two-loop stop amplitudes including the
two-loop amplitude entering the approximate \nnlo{} stop contributions by the test factor
$t=\mathcal{A}_{\tilde{t}_1}^{\lo{}}/\mathcal{A}_{\tilde{t}_1}^{\lo{},\text{VHML}}$
and use the relative difference of the obtained cross section to the cross section with $t=1$ as symmetric expansion uncertainty.
The result is shown with the red, dotted-dashed curves in \fig{fig:gghuncertainties}. Unsurprisingly,
the uncertainty becomes large when we approach the very light stop mass region, indicating the increasing invalidity of the VHML expansion.
\item We estimate the uncertainty due to the fact that we neglect three-loop contributions from
the variation of the Wilson coefficient $C^2$ in the interval $[0,2C^2]$ as it was done in \citere{Bagnaschi:2014zla}.
Therein $C^2$ includes the top-induced contribution to gluon fusion only.
The error we obtain is, however, rather small and far below $\pm 1\%$. We can thus neglect it here, given the comparably large VHML expansion uncertainty.
\item As argued before, we are left with a mild logarithmic dependence on the gluino mass through
higher order corrections to gluon fusion, which we keep in our calculation. In principle
we could assign an additional uncertainty to make our statements independent of the heavy SUSY spectrum. However,
we deem that we are taking already a conservative enough stance with the uncertainties described above and we neglect this last source.
\end{itemize}

In summary, for our purposes we can take over the PDF$+\alpha_s$ uncertainties as well
as the renormalization and factorization scale uncertainties from the SM Higgs case as provided by
the LHCHXSWG. However, we add a third source of uncertainty for gluon fusion that reflects the
uncertainty in the expansion in a vanishing Higgs mass for the \nlo{} and \nnlo{} stop contributions.
This uncertainty becomes dominant for a light stop mass $m_{\tilde{t}_1}$ below $m_h$.
It is a function of $m_{\tilde{t}_1}$ and the remaining SUSY parameter dependence can be neglected.
Above $m_{\tilde{t}_1}>430\gev$ we can set the VHML uncertainty to zero.

We also apply the VHML expansion uncertainty to the partial width of the Higgs decay into gluon, $\Gamma(h\rightarrow gg)$,
since the \nlo{} \qcd{} corrections encoded in \FH{} are based on an expansion in a heavy loop mass
or small Higgs mass as well.
Due to our conservative approach for what concerns
Higgs production and the decay into gluons we do not add an additional uncertainty
to the Higgs decay into photons, which is affected by light stop contributions at the level
of up to $\mathcal{O}(10-30\%)$. In the next section we describe how these cross section
and partial width uncertainties are incorporated into our numerical analysis with \HS.

\section{Numerical procedure}
\label{Sect:numerics}

Our numerical scans proceed as follows: For a scan point given by specific values of the scanned SUSY parameters\footnote{The exact selection of scan parameters will be specified for each scan in Section~\ref{Sect:Results}.}, e.g.
\begin{align}
\SMU{3},~\SML{3},~\SME{3},~\tan\beta,~M_A,~\mu,~M_2,~A_t,~A_b,~A_\tau, \dots\,,
\end{align}
we evaluate the Higgs sector predictions with \FHv{2.11.0}, starting with a first guess for the decoupled SUSY mass scale, $\MSUSY$, between $50$ and $300\tev$ (exact value depends on $\tan\beta$). For numerical stability reasons, we choose the following configuration for the decoupled SUSY masses:
For the first and second generation sleptons and squarks, we choose
\begin{align} 
\SML{j} = \SME{j} =  \SMU{j} = \SMD{j} = \SMQ{j}  \equiv \MSUSY\quad(j=1,2)\,;
\end{align}
For the decoupled third-generation squark soft-breaking mass parameters, we choose
\begin{align}
\SMQ{3} &= \MSUSY\,, \label{Eq:MQ3}\\
\SMD{3} &= 0.8 \cdot \MSUSY\,; \label{Eq:MD3}
\end{align}
The gluino mass is fixed to
\begin{align} 
M_3 = 1.5 \cdot \MSUSY\,; \label{Eq:M3}
\end{align}
In the case where the third generation sleptons are also decoupled, we additionally set
\begin{align} 
\SML{3} = \SME{3} = \MSUSY\,. \label{Eq:ML3}
\end{align}
After the first guess evaluation, $\MSUSY$ is adjusted through an iterative procedure, until the light Higgs mass is predicted to be in the vicinity of its observed value,
\begin{align}
124\gev \le m_h \le 126\gev\,.
\end{align}
The evaluation of the light Higgs partial decay widths and production cross sections then proceeds with the use of \FH\ (version \text{2.11.0}) and \SH\ (version \text{1.4.1}). The latter is used for the $gg\to h$ cross section only. These predictions are fed to \HS\ (version \text{1.4.0}) for the evaluation of the $\chi^2$ compatibility with the Higgs rate measurements. In fact, since the runtime of \HS\ is very short, we perform at this point an additional scan over the branching fraction of a (not further specified) light Higgs decay mode to ``new physics'' (NP), $\brhnp$, ranging between $0\%$ and $50\%$ in steps of $0.5\%$. This allows for an overall reduction of the known Higgs decay rates to SM particles, which may partially compensate for a possible enhancement in the Higgs production cross section. In the MSSM, such a decay could be represented by the Higgs decaying into a pair of stable neutralinos, thus leading to a missing energy signature.  Another example would be an unexpectedly large decay rate to SM particles that can hardly be detected at the LHC, e.g.~light flavored quarks such as charm quarks. Examples for novel, yet undetectable Higgs decay modes can be found beyond the MSSM, e.g.~decays to light supersymmetric particles (e.g.~neutralinos), which successively decay hadronically via $R$-parity violating interactions.

In the \HS\ evaluation we compare the light Higgs predictions against the latest Higgs rate measurements from ATLAS and CMS from LHC Run~1. For completeness, we also include the available measurements from the Tevatron experiments. A detailed listing of all included Higgs rate observables including references is given in Appendix~\ref{App:measurements}. We do not include the $\chi^2$ contribution from the Higgs mass, which can also be obtained from \HS, as we are only interested in the rate constraints, and in our scenario the Higgs mass is adjustable via the tuning of $\MSUSY$. In the $\chi^2$ test, \HS\ takes into account the correlations of some of the most important systematic uncertainties, in particular, of the theoretical uncertainties of the production cross sections and branching ratios. Moreover, \HS\ allows the theoretical rate uncertainties of the tested model to be different than in the SM. Internally, this is done by first subtracting (in a Gaussian way) the SM theoretical uncertainty from the measurement's uncertainty, and then adding back in the model's theoretical uncertainty. The correlations among the production cross sections and branching ratios, as e.g.~induced by common parametric dependences on the top and bottom masses, the strong coupling, etc., or, in the latter case, through the dependence on the total decay width, are fully taken into account in \HS\ through two covariance matrices, for both the SM and the supersymmetric model: The covariance matrix $C_\sigma$ is given in the basis of the five Higgs production modes at the LHC, ($gg\to h$, $pp\to qqh$ (VBF), $pp\to WH$, $pp\to ZH$, $pp\to t\bar{t}h$), and the covariance matrix $C_\text{BR}$ is specified in the basis of the SM Higgs decay modes. For more information we refer to \citeres{Bechtle:2013xfa,HSreleasenote11,Bechtle:2014ewa}.

As discussed in Section~\ref{Sect:uncertainties}, there is one additional theoretical uncertainty in the calculation of the gluon fusion cross section, $\sigma(gg\to h)$, as well as of the partial width of the Higgs decay to gluons, $\Gamma(h\to gg)$, arising from the VHML expansion. This uncertainty is incorporated in the $\chi^2$ evaluation as a function of the light stop mass (cf.~\fig{fig:gghuncertainties}) in the following way: For each light stop mass between $80\gev$ and $430\gev$, in steps of $1\gev$, we re-evaluate $C_\sigma$ and $C_\text{BR}$, taking into account the additional VHML uncertainty. Since this uncertainty is inherent to only the $gg\to h$ and $h\to gg$ processes it does not introduce any additional correlation between the cross section and the partial width uncertainties of the various production and decay modes, respectively. However, it should be noted that the additional $\Gamma(h\to gg)$ uncertainty propagates into the \emph{branching ratio} uncertainties of \emph{all} decay modes, since it increases the uncertainty on the total decay width. In the scan, the covariance matrices are selected for each point according to the predicted $m_{\sstop_1}$ value and fed into \HS.

\HS\ provides a $\chi^2$ value for each scan point in the parameter space. In every scan we determine the best fit (BF) point, given by the point of least $\chi^2$ value, $\chi^2_\text{BF}$. We define the \emph{most favored} and \emph{favored} points in the parameter space by their $\chi^2$ difference to the BF point being
\begin{align}
\chi^2 - \chi^2_\text{BF} \le 2.30 \qquad \text{and} \qquad \chi^2 - \chi^2_\text{BF} \le 5.99\,,
\end{align}
respectively. Given linearity in the mapping of model parameters to statistical observables, as well as the validity of the Gaussian approximation in the treatment of all uncertainties, these values correspond to the $68\%$ and $95\%$ confidence level (C.L.) regions for two statistical degrees of freedom. While these circumstances may not be completely  fulfilled in the MSSM for the given observables, this simple statistical treatment is useful to determine the statistically preferred parameter regions and has found wide applicability in related studies (For more detailed discussions and a demonstration of a more thorough $P$--value estimation in supersymmetric models, see \citeres{Bechtle:2013mda,Bechtle:2014yna,Bechtle:2014ewa,Bechtle:2015nua}).

\section{Results}
\label{Sect:Results}

In this section we explore the constraints from the measured Higgs signal rates on models with a light stop mass in four different scenarios: In the first three cases we assume the decoupling limit, $M_A \gg M_Z$, and consider: (\textbf{A}) only a light stop, (\textbf{B}) a light stop and a light stau, and (\textbf{C}) a light stop and a light chargino. In all three scenarios we allow for an additional unobservable ``new physics'' Higgs decay mode, parametrized by the scan parameter $\mathrm{BR}(h\to \text{NP})$. The currently strongest $\CL{95}$ limits on the branching fraction of an invisible Higgs decay, $\brhinv$, under the assumption of the Higgs production cross sections being as predicted in the SM, are
\begin{align}
\brhinv \le 28\%, \quad &\mbox{from~ATLAS~\cite{Aad:2015txa}},\label{Eq:brhinvlimit_ATLAS}\\
\brhinv \le 36\%, \quad &\mbox{from~CMS~\cite{CMS:2015naa}}.\label{Eq:brhinvlimit_CMS}
\end{align} 
As will be seen below, even if we assumed the ``new physics'' Higgs decay mode entirely led to invisible final states, as would arise in the MSSM if the Higgs decays into a pair of light stable neutralinos, these limits would only marginally affect the allowed parameter space. 

In the last scenario~\textbf{D} we explore the constraints on the light stop mass in the non-decoupled parameter space region, and treat $M_A$ as a free parameter. In addition, we also consider the presence of light staus in this scenario. At the end of this section we present a summary of our findings.
We provide some theoretical motivation in the context of supersymmetric electroweak baryogenesis for the four scenarios under consideration in Section~\ref{Sect:EWBG}.

\subsection{Decoupling Limit with a light stop (scenario {\bf A})}
\label{Sect:Case1}

We start our discussion with the simple scenario of a light right-handed stop and the remaining SUSY particles being decoupled, cf.~Eqs.~\eqref{Eq:MQ3}--\eqref{Eq:ML3}. We furthermore assume the MSSM Higgs decoupling limit by setting $M_A = 1\tev$. The numerical scan is performed over the right-handed soft-breaking stop mass, $\text{sgn}(\SMU{3}^2)\SMU{3} \in [-150, 500]\gev$, and for each parameter point different values of $\brhnp$ are tested. The remaining MSSM parameters are fixed to
\begin{align}
\mu = M_2 = 1\tev,~\tan\beta = 10,~A_t = A_b = A_{\tau} = 100\gev\,, \label{Eq:scan1_fixedpars}
\end{align}
which yields a vanishing stop mixing parameter,
\begin{align}
X_{t} = A_{t} - \mu/\tan\beta = 0\,.
\end{align}
However, since everything except the right-handed stop is well decoupled, the exact choice of the parameters in \eqn{Eq:scan1_fixedpars} is to a good approximation irrelevant for the light Higgs phenomenology.

\begin{figure*}
\subfigure[~Contours indicate the Higgs signal rates for vector boson final states, $h\to VV$ ($V=W^\pm,Z$).]{\includegraphics[width=0.44\textwidth]{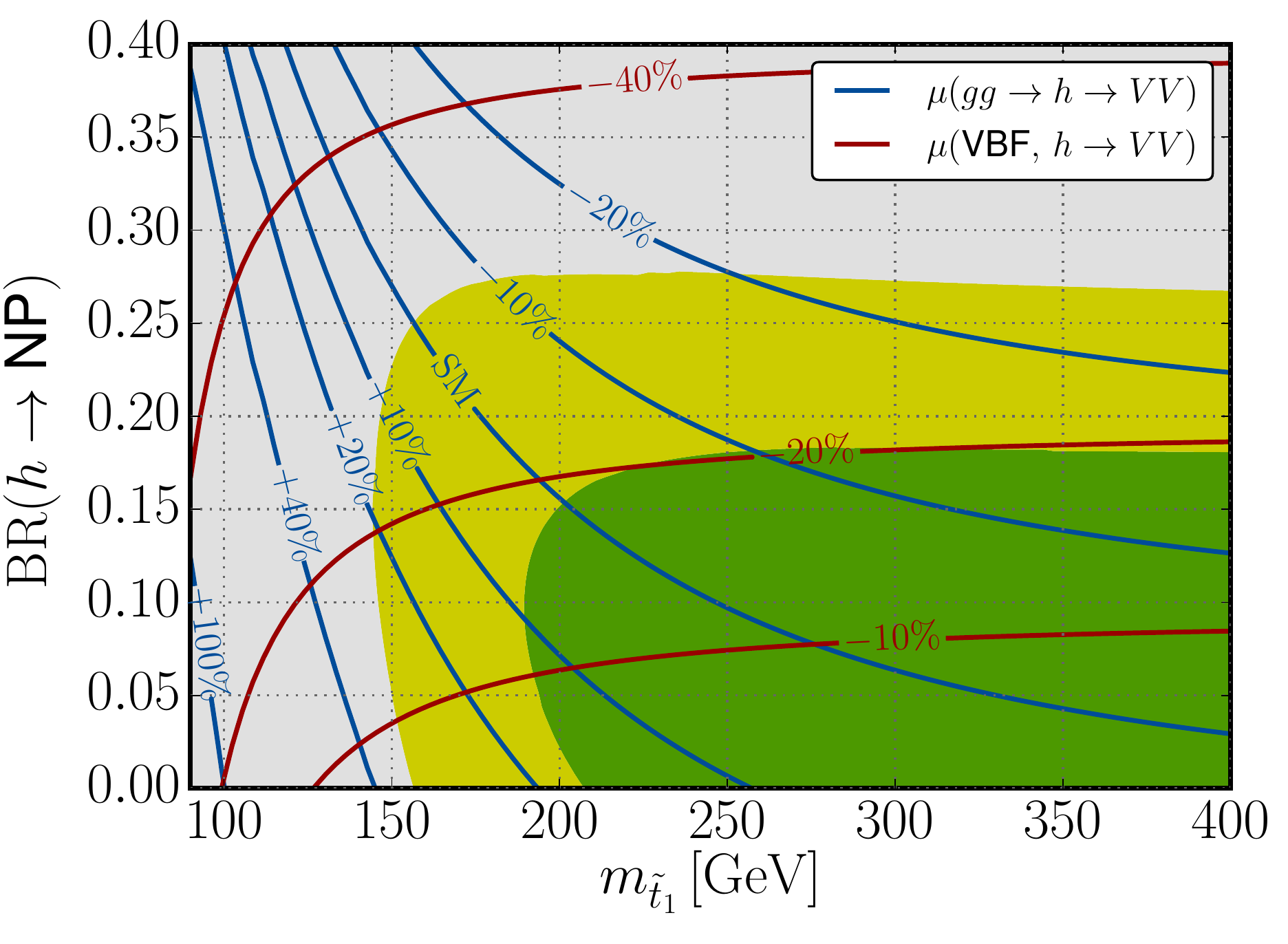}}\hfill
\subfigure[~Contours indicate the Higgs signal rates for di-photon final states, $h\to \gamma\gamma$.]{\includegraphics[width=0.44\textwidth]{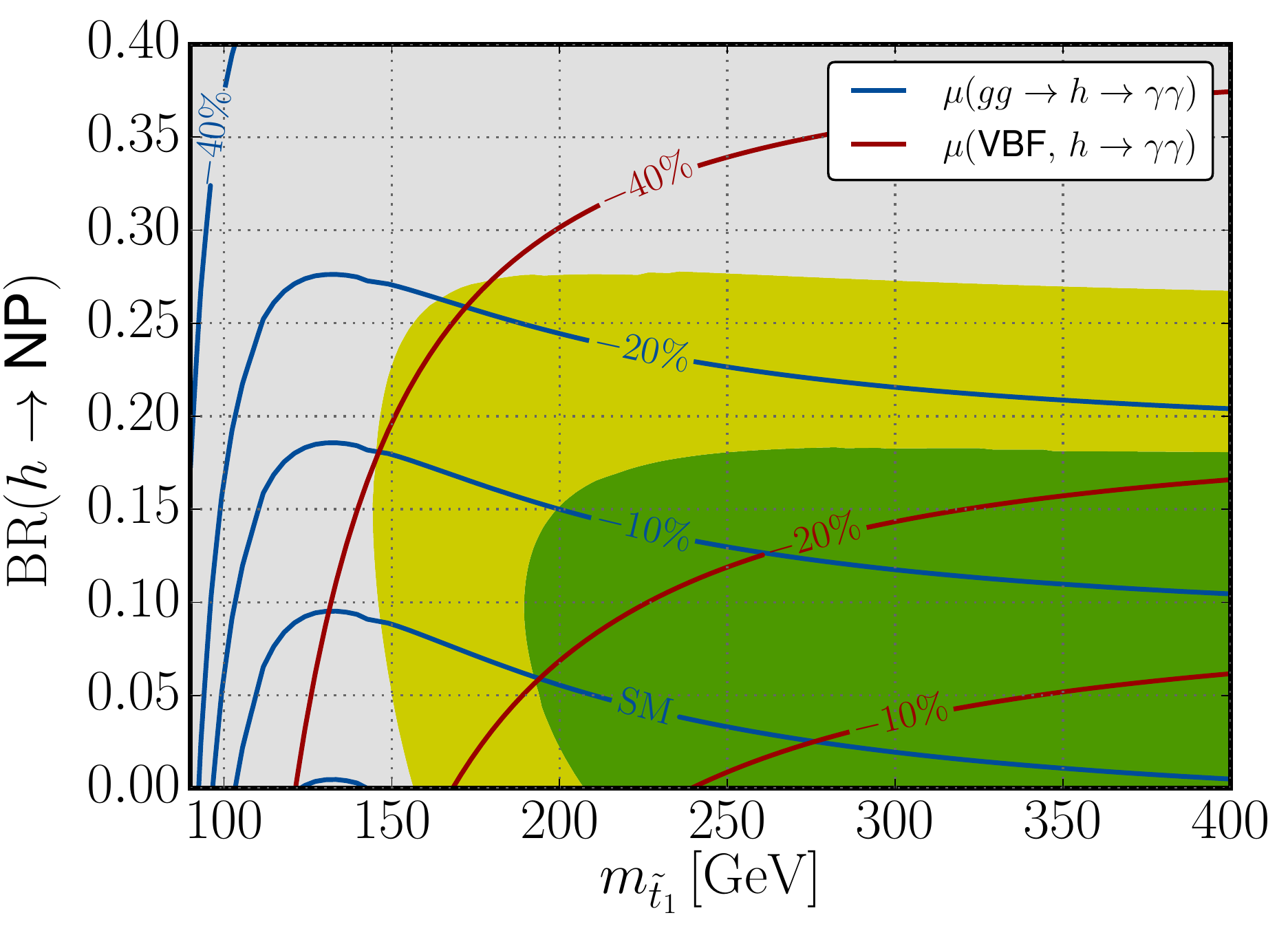}}
\caption{Results for scenario~\textbf{A} `\emph{Decoupling Limit with a Light Stop}' in the $(m_{\sstop_1},~\brhnp)$ plane. The green [yellow] region give the $68\%~\mathrm{C.L.}$ [$95\%~\mathrm{C.L.}$] allowed region from the Higgs signal rates. 
 The contour lines give the deviations in the Higgs signal rates from the SM prediction for vector boson final states, $h\to VV$ ($V=W^\pm,Z$), [{a}], and di-photon final states, $h\to \gamma\gamma$, [{b}].}
\label{Fig:Case1_brhnp}
\end{figure*}

The results are shown in \fig{Fig:Case1_brhnp}. The green (yellow) area indicates the parameter region allowed at $\CL{68}$ ($ \CL{95}$). The BF point lies outside the shown region at 
\begin{align}
(m_{\sstop_1}, \brhnp) = (527\gev, 3.0\%)
\end{align}
with a fit quality of $\chi^2/\mathrm{ndf} = 68.3/83$. Within the $\CL{95}$, the light stop mass can be as low as $144\gev$ in the presence of an additional ``new physics'' decay mode with a branching fraction of $\sim15\%$. In contrast, if no additional Higgs decay mode is allowed, $\brhnp \equiv 0$, the $\CL{95}$ lower mass limit on the lightest stop is $154\gev$. This illustrates that allowing an additional Higgs decay mode to suppress the Higgs decays to SM particles still has an effect on the light stop mass limit. The new physics decay rate can maximally be $\brhnp \lesssim 28\%$ in this scenario. If this decay mode is leading entirely to an invisible final state, $\brhnp \equiv \brhinv$, this is just at the current exclusion limit from ATLAS, see \eqn{Eq:brhinvlimit_ATLAS}.

The impact of the light stop on the Higgs signal rates can be seen in \fig{Fig:Case1_brhnp}(a) [(b)], where the blue and red contour lines indicate the Higgs signal rates for search channels with Higgs production in gluon fusion and vector boson fusion\footnote{The predictions for the SM normalized rates for VBF and associated Higgs production with a vector boson ($Vh$, with $V=W^\pm, Z$) are to a good approximation identical.} (VBF), respectively, and the Higgs decaying into a pair of vector bosons ($V=W^\pm, Z$) [photons]. The Higgs signal rate $\mu$ for a Higgs production mode $P(h)$ and decay mode $D(h)$ is defined as
\begin{align}
\mu ( P(h), D(h) ) \equiv \frac{\sigma(P(h)) \times \mathrm{BR} (D(h))}{\sigma_\text{SM}(P(h)) \times \mathrm{BR}_\text{SM} (D(h))}\,,
\label{Eq:mu_definition}
\end{align}
where $\sigma$ is the inclusive cross section at the LHC with a center-of-mass energy of $8\tev$, $\mathrm{BR}$ denotes the branching ratio, and the subscript `SM' indicates that the quantity is the prediction in the SM. The contours in \fig{Fig:Case1_brhnp} denote the deviation from the SM prediction, $\mu = 1$, as given by the contour labels. The VBF cross section as well as the partial width for the decay $h\to VV$ $(V=W^\pm, Z)$ are independent of the light stop mass. The red contours in \fig{Fig:Case1_brhnp}(a) thus indicate how much the branching ratio $\mathrm{BR}(h\to VV)$ is reduced when the light stop affects the partial widths $\Gamma(h\to gg)$ and $\Gamma(h\to \gamma\gamma)$, or when a new physics Higgs decay mode, $h\to\mathrm{NP}$, is added. As can be seen, a $20\%$ reduction of the Higgs signal rate in the VBF channels with vector boson final states is still tolerable within $\CL{68}$ in this scenario. In contrast, the signal rate for the channel $gg\to h\to VV$, given by the blue contours in \fig{Fig:Case1_brhnp}(a), increases significantly for a decreasing light stop mass due to the enhanced gluon fusion cross section, cf.~Section~\ref{Sect:higherorder}. In the $\CL{68}$ region, this rate is constrained to be within $\sim 15-18\%$ of its SM prediction in this scenario. At lighter stop masses the enhancement in the gluon fusion cross section needs to be compensated by a reduction of the SM Higgs decay rates through the increase of the new physics decay rate, $\brhnp$. However, this leads to a suppression of the signal rates in the VBF initiated Higgs channels. Eventually, the splitting between the signal rates of the gluon fusion and VBF initiated Higgs channels becomes too large to be consistent with the LHC measurements.

The Higgs rates for the di-photon channels, $h\to \gamma\gamma$, are also affected by the direct influence of the light stop on the partial width $\Gamma(h\to \gamma\gamma)$. As discussed in Section~\ref{Sect:higherorder}, a light stop leads to a reduction of $\Gamma(h\to \gamma\gamma)$, thus the SM normalized Higgs signal rates for the di-photon channels are in general smaller than those in the channels with vector boson final states. This also means that, for $\brhnp = 0$, the $gg\to h\to \gamma\gamma$ rate is closer to its SM prediction than the $gg\to h\to VV$ rate at light stop masses, due to the compensation between increasing gluon fusion cross section and decreasing partial width $\Gamma(h\to\gamma\gamma)$. Nevertheless, the splitting between the rates of gluon fusion and VBF initiated channels is the same as before, and penalizes the fit at light stop masses.

\begin{figure}
\centering
\includegraphics[width=0.6\textwidth]{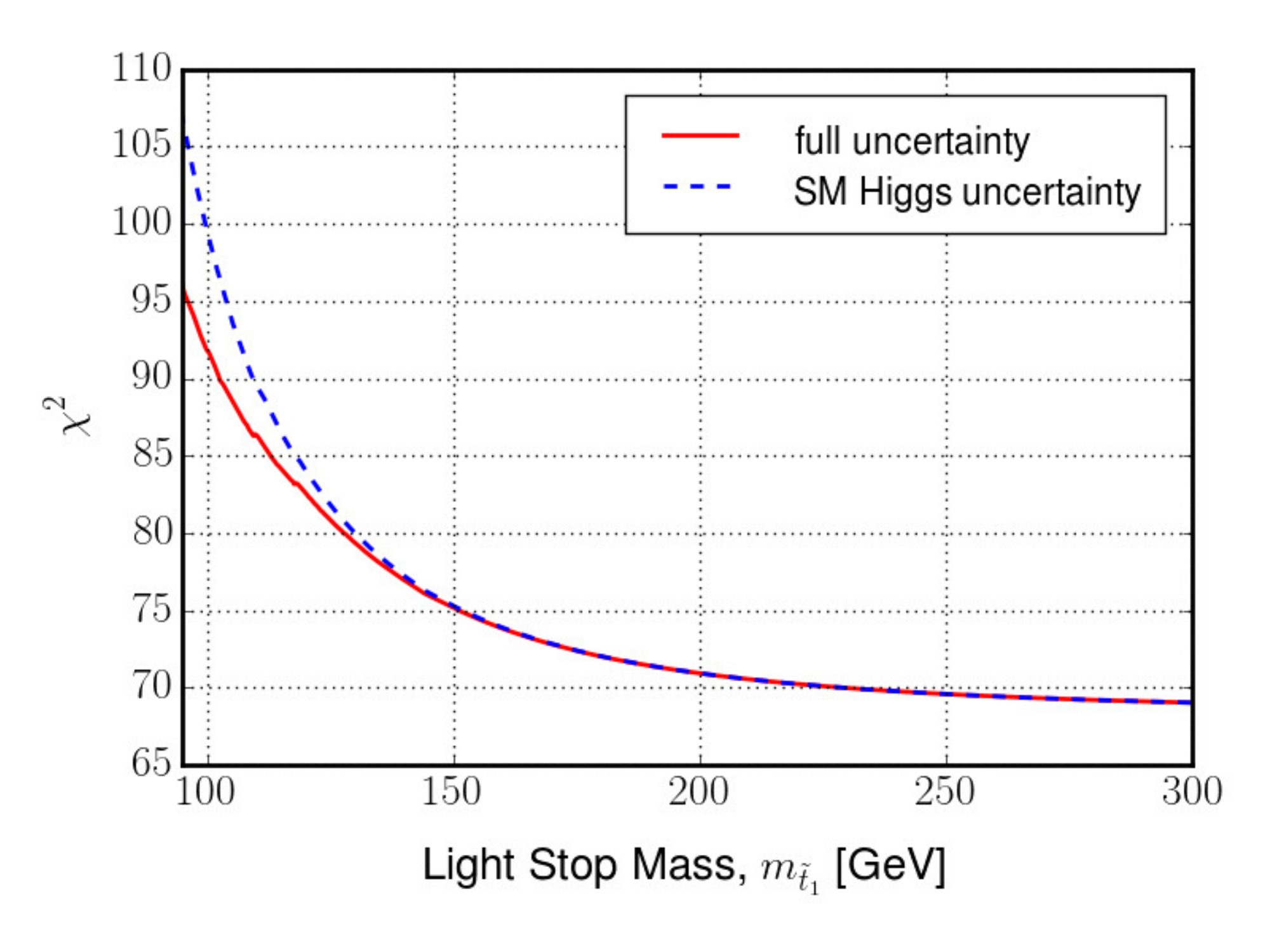}
\vspace{-0.6cm}
\caption{Impact of the rate uncertainties on the total $\chi^2$ as a function of the light stop mass for the case of no additional non-standard Higgs decays, $\brhnp=0$. The solid, red curve is obtained using the full rate uncertainties, cf.~Sect.~\ref{Sect:uncertainties}, whereas the dashed, blue curve is obtained using only the SM Higgs rate uncertainties.}
\label{Fig:chi2compare}
\end{figure}

Before moving on to the more general fits (\textbf{B}-\textbf{D}), we briefly come back to the discussion of theoretical rate uncertainties. In \fig{Fig:chi2compare} we show a comparison of the $\chi^2$ distribution over the light stop mass obtained when the full theoretical rate uncertainties (as discussed in Section~\ref{Sect:uncertainties}) are included versus the one obtained when only the rate uncertainties of a SM Higgs boson are included. Here we have set $\brhnp=0$. The two distributions are nearly identical for stop masses larger than $\gtrsim 140\gev$, while for smaller stop masses the $\chi^2$ distribution with full theoretical uncertainties has a shallower slope. In the above discussed scenario, the $\CL{95}$ region already ends at around $m_{\sstop_1} \gtrsim 144\gev$, thus the effect of the additional uncertainties on the allowed parameter space is negligible here. However, in the following scenarios, lower values of $m_{\sstop_1}$ are viable, and there the inclusion of the VHML rate uncertainty is important, leading to a slightly larger allowed parameter space than what would be obtained when only the SM Higgs rate uncertainties were considered.

\subsection{Decoupling Limit with a light stop and a light stau (scenario {\bf B})}
\label{Sect:Case2}

We now allow for light scalar tau partners (staus) in our fit. As discussed in Section~\ref{Sect:higherorder} these can substantially modify the decay width for the Higgs decay to two photons, $\Gamma(h\to \gamma\gamma)$. The implications of light staus on the signal rates of the discovered Higgs boson at $125\gev$ have recently been discussed in \citeres{Carena:2012mw,Carena:2012gp,Bechtle:2012jw,Carena:2013iba,Belyaev:2013rza,Kitahara:2013lfa}. The stau contributions to $\Gamma(h\to \gamma\gamma)$ are maximal for large values of $\mu\tan\beta$, cf.~\eqn{eq:staucoupling}. However, for small stau masses and very large values of $\mu\tan\beta$ charge-breaking minima can appear~\cite{Ratz:2008qh,Hisano:2010re,Carena:2012mw,Carena:2013iba,Kitahara:2013lfa}. Requiring metastability of the electroweak vacuum, an approximate upper bound on $\mu\tan\beta$ was presented in \citere{Hisano:2010re}, obtained from a fit to numerical results (where $\tan\beta = 70$ was fixed):
\begin{align}
\mu\tan\beta \; \lesssim & \;213.5 \cdot \sqrt{\SML{3}\SME{3}} - 17.0 \, (\SML{3} + \SME{3})\nonumber \\
& + 0.0452\gev^{-1} (\SML{3} - \SME{3})^2 - 1.3 \times 10^4\gev\,.
\label{Eq:CCBcondition}
\end{align}
Note that at large $\tan\beta$ there is a residual dependence on $\tan\beta$ induced by the radiatively corrected $\tau$-lepton Yukawa coupling~\cite{Carena:2012mw}. This dependence is not included in \eqn{Eq:CCBcondition}, and leads to the upper limit on $\mu\tan\beta$ becoming weaker at larger values of $\tan\beta$. Furthermore, the charge breaking vacuum constraint from the $\sstau$ sector, \eqn{Eq:CCBcondition}, should only be interpreted as an indicative constraint as dedicated numerical studies of the higher-order effective potential in certain MSSM scenarios~\cite{Camargo-Molina:2013sta,Camargo-Molina:2014pwa} have demonstrated that constraints on the parameter space from requiring vacuum (meta-)stability are oftentimes stronger.

We scan over the following three MSSM parameters and ranges:
\begin{align}
\text{sgn}(\SMU{3}^2)\SMU{3} &\in [-150, 500]\gev\,, \nonumber\\
\SML{3} \equiv \SME{3} &\in [70, 300]\gev\,, \nonumber\\
\tan\beta &\in [5,60]\,.
\end{align}
As before, we also allow for a new Higgs branching fraction to ``new physics'', $\brhnp$. We set the soft-breaking trilinear coupling for the stau sector to $A_\tau = 1000\gev$. In the case of $\CP$ violation in the stau sector, light quasi-degenerate staus with a non-zero trilinear coupling $A_\tau$ can be instrumental for successful electroweak baryogenesis while fulfilling all current constraints from electric dipole moments (EDMs)~\cite{Kozaczuk:2012xv}. 
The other parameters are left unchanged with respect to the previous fit, i.e.
\begin{align}
M_A = 1\tev,~\mu = M_2 = 1\tev,~X_t = X_b = 0\,. \label{Eq:scan2_fixedpars}
\end{align}

If $R$-parity is conserved, the lower $\CL{95}$ limit on the stau mass from LEP searches is around $m_{\tilde{\tau}_1} \gtrsim (87 - 93)\gev$, depending on the mass of the lightest neutralino, $m_{\neut_1}$, and assuming a mass splitting of $m_{{\sstau}_1} - m_{\neut_1} \ge 7\gev$ and the branching ratio $\mathrm{BR}(\sstau_1 \to \tau \neut_1) = 1$~\cite{LEPstaulimit}. In the case of $R$-parity violation, the limits from LEP searches for direct and indirectly decaying staus~\cite{Heister:2002jc,Abbiendi:2003rn,Abdallah:2003xc,Achard:2001ek} roughly range from $70\gev$ to $95\gev$ and strongly depend on the assumption of the dominant $R$-parity violating operator as well as, in some cases, on certain sparticle mass splittings, see \citeres{Barbier:2004ez,Agashe:2014kda} for an overview.

In our analysis we do not explicitly scan over the lightest neutralino mass, as its influence on the Higgs phenomenology is marginal (except in the case where the decay $h\to\neut_1\neut_1$ is kinematically accessible; in this case, however, the relevant effects are parametrized here by the additional branching fraction, $\brhnp$). Since the LEP stau mass limits mentioned above are model-dependent we discuss our results both with and without taking these constraints into account. In the first case, we simply require $m_{\sstau_1} \ge 90\gev$, given by the average lower mass limit in the $R$-parity conserving case. 

In the following numerical results, parameter points violating the criterion $m_{\sstau_1} \ge 90\gev$ will be shown in pale colors. In the figures we also introduce two new colors in order to illustrate the constraints arising from requiring metastability of the electroweak vacuum: The \emph{green} and \emph{yellow} [\emph{red} and \emph{blue}] parameter points indicate the $\CL{68}$ and $\CL{95}$ favored regions, respectively, and [do not] fulfill the vacuum metastability constraint, \eqn{Eq:CCBcondition}. Scan points that do not fall into any of these categories are disfavored by more than $\CL{95}$ from the Higgs signal rates and are shown in \emph{gray}. The points are plotted in the following order (from first to last, with the latter possibly overlapping the former): \emph{gray}, \emph{pale blue}, \emph{pale red}, \emph{blue}, \emph{red}, \emph{pale yellow}, \emph{pale green}, \emph{yellow}, \emph{green}.

The BF point has a fit quality of $\chi^2/\mathrm{ndf} = 67.8/81$ and features 
\begin{align}
(m_{\sstop_1}, m_{\sstau_1}, \brhnp) = (526\gev, 111\gev, 5.5\%)\,.
\end{align}
Adding the possibility of light staus to the model thus improves the minimal $\chi^2$ value by $\sim 0.5$ with respect to the previous fit in Section~\ref{Sect:Case1}, however, at the price of two additional free model parameters. Thus the number of statistical degrees of freedom, $\mathrm{ndf} = N_\text{observables} - N_\text{parameters}$, is reduced by two, and hence the overall fit quality is not improved by adding light staus to the fit. 

\begin{figure*}
 \subfigure[\label{Fig:Case2_brhnp_a}~$(m_{\sstop_1},~\brhnp)$ plane.]{\includegraphics[width=0.44\textwidth]{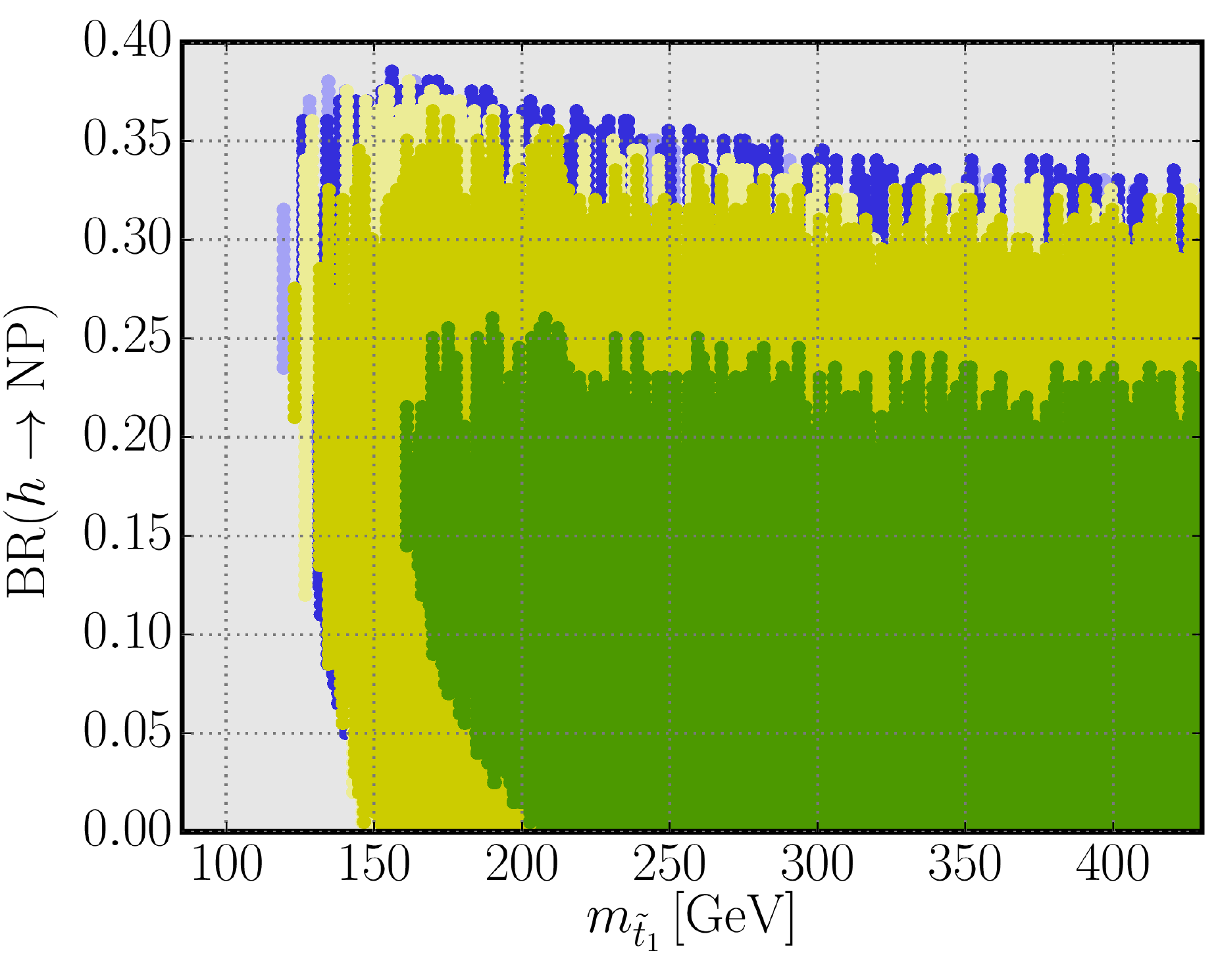}}\hfill
 \subfigure[~$(m_{\sstau_1},~\brhnp)$ plane.]{\includegraphics[width=0.44\textwidth]{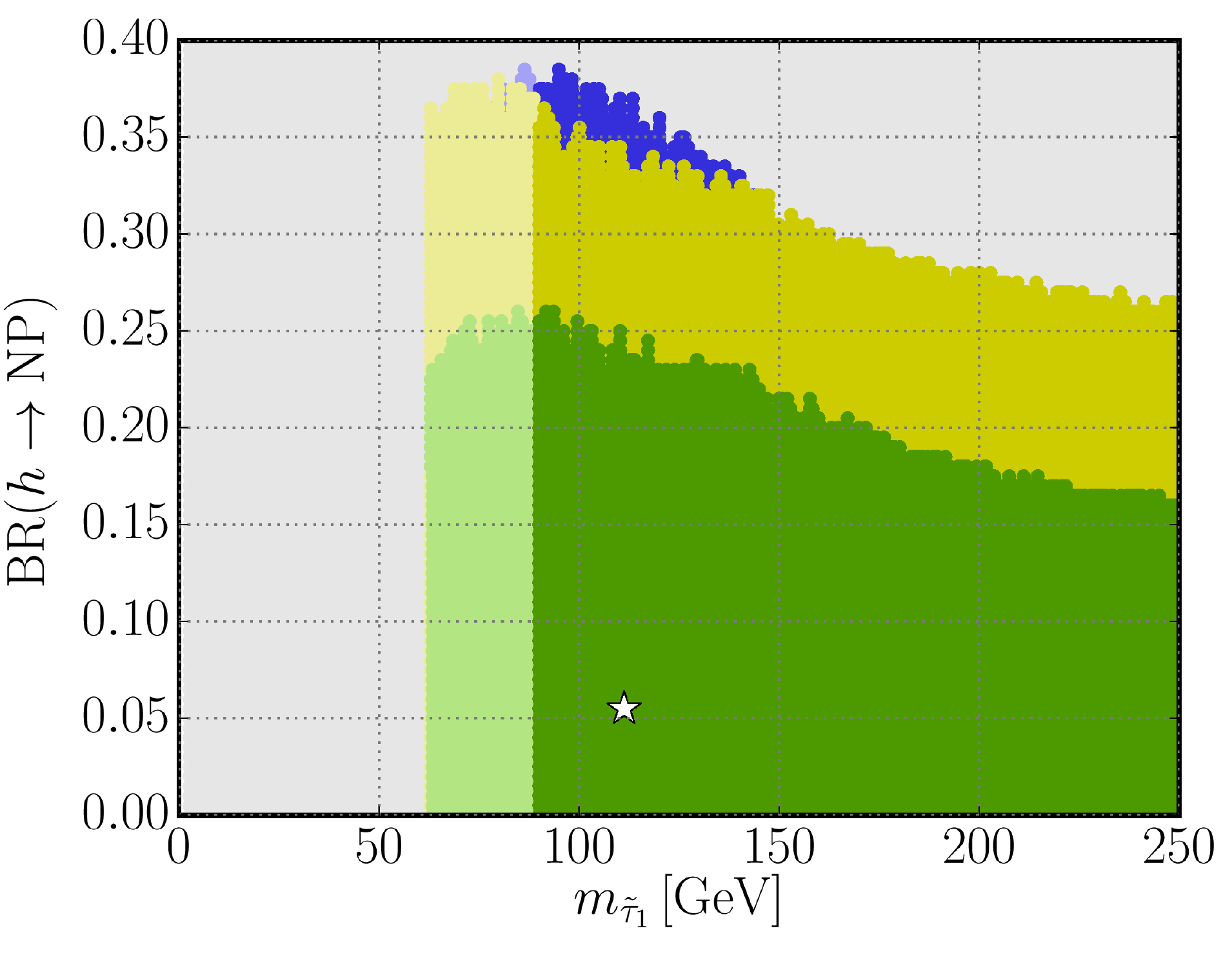}}
 \caption{Results for scenario~\textbf{B} `\emph{Decoupling Limit with a Light Stop and Light Stau}' in the $(m_{\sstop_1},~\brhnp)$ [a] and $(m_{\sstau_1},~\brhnp)$ [b] plane. The green (yellow) region is allowed from the Higgs signal rates at $68\%~\mathrm{C.L.}$ [$95\%~\mathrm{C.L.}$] and fulfills the metastability condition of the electroweak vacuum, \eqn{Eq:CCBcondition}, whereas the blue region is allowed at $\CL{95}$ but does not fulfill \eqn{Eq:CCBcondition}. The white star indicates the best fit point. The regions in paler colors violate the naive stau mass limit $m_{\sstau_1} \ge 90\gev$ from LEP. (See also text for a detailed discussion.)}
 \label{Fig:Case2_brhnp}
\end{figure*}

\begin{figure*}[t]
 \subfigure[~Dependence of $\Gamma(h\to\gamma\gamma)$ on $m_{\sstop_1}$.]{\includegraphics[width=0.44\textwidth]{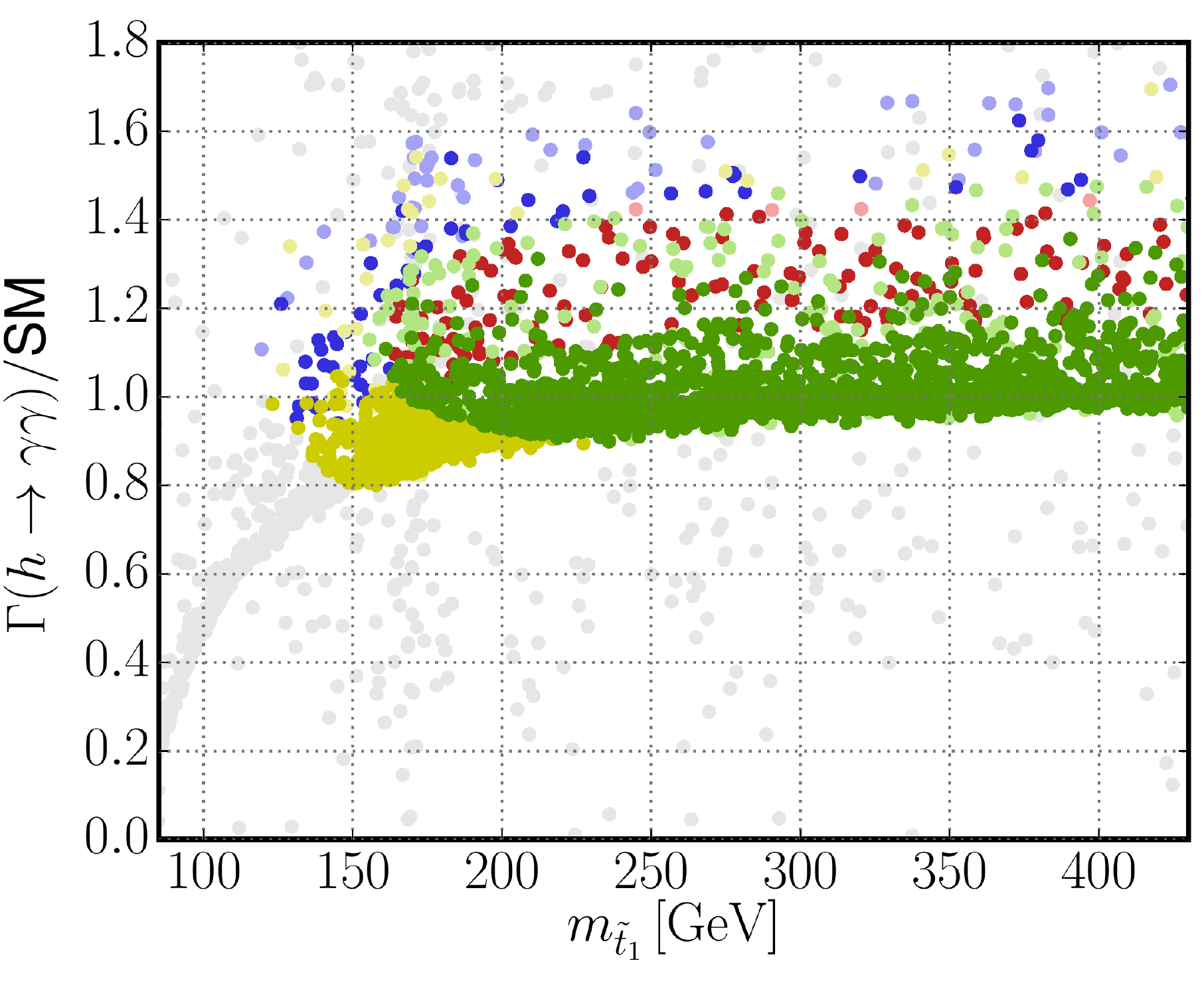}}\hfill
 \subfigure[~Correlation between $m_{\sstop_1}$ and $\mu\tan\beta$.]{\includegraphics[width=0.44\textwidth]{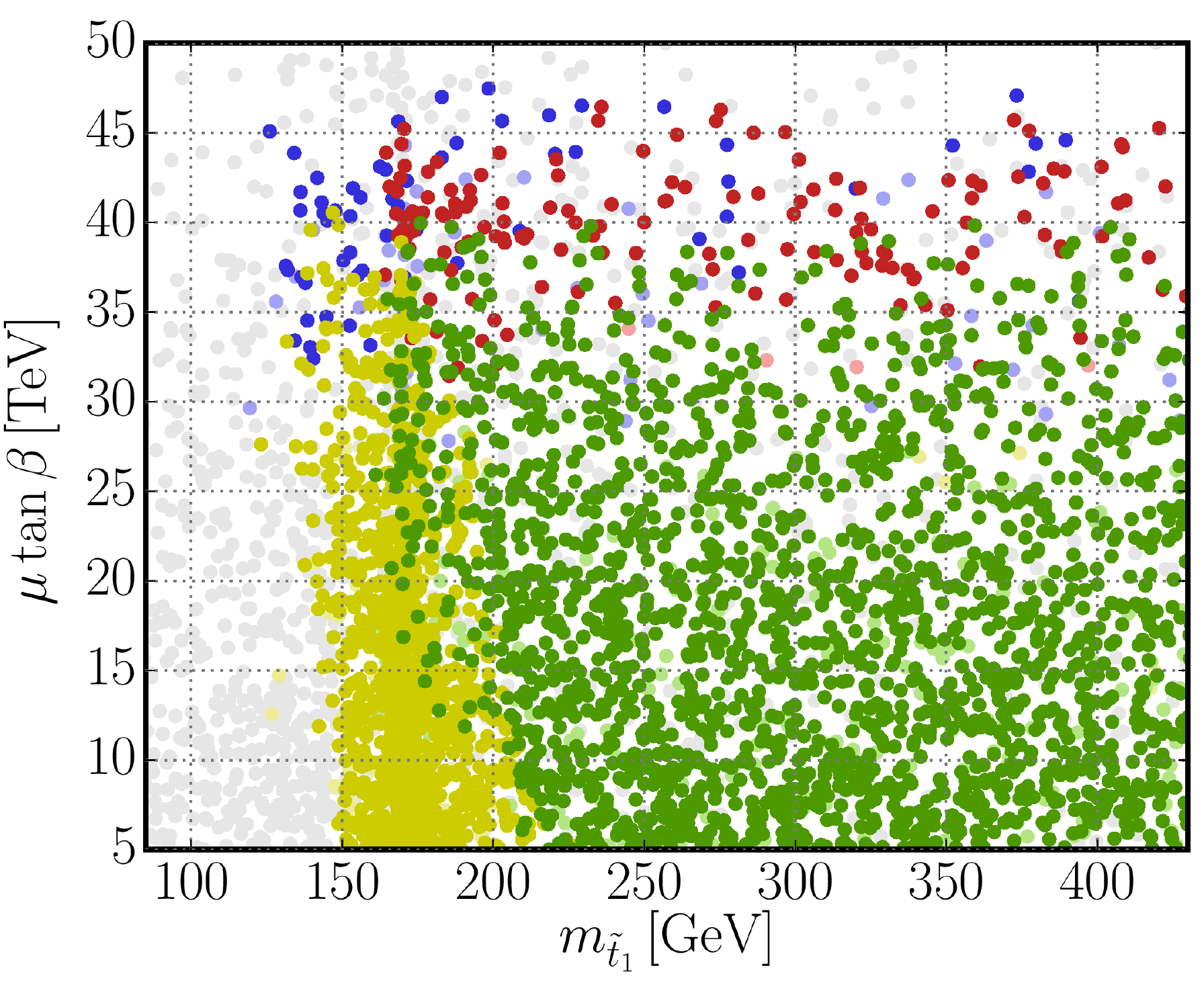}}
 \subfigure[~Dependence of $\Gamma(h\to\gamma\gamma)$ on $m_{\sstau_1}$.]{\includegraphics[width=0.44\textwidth]{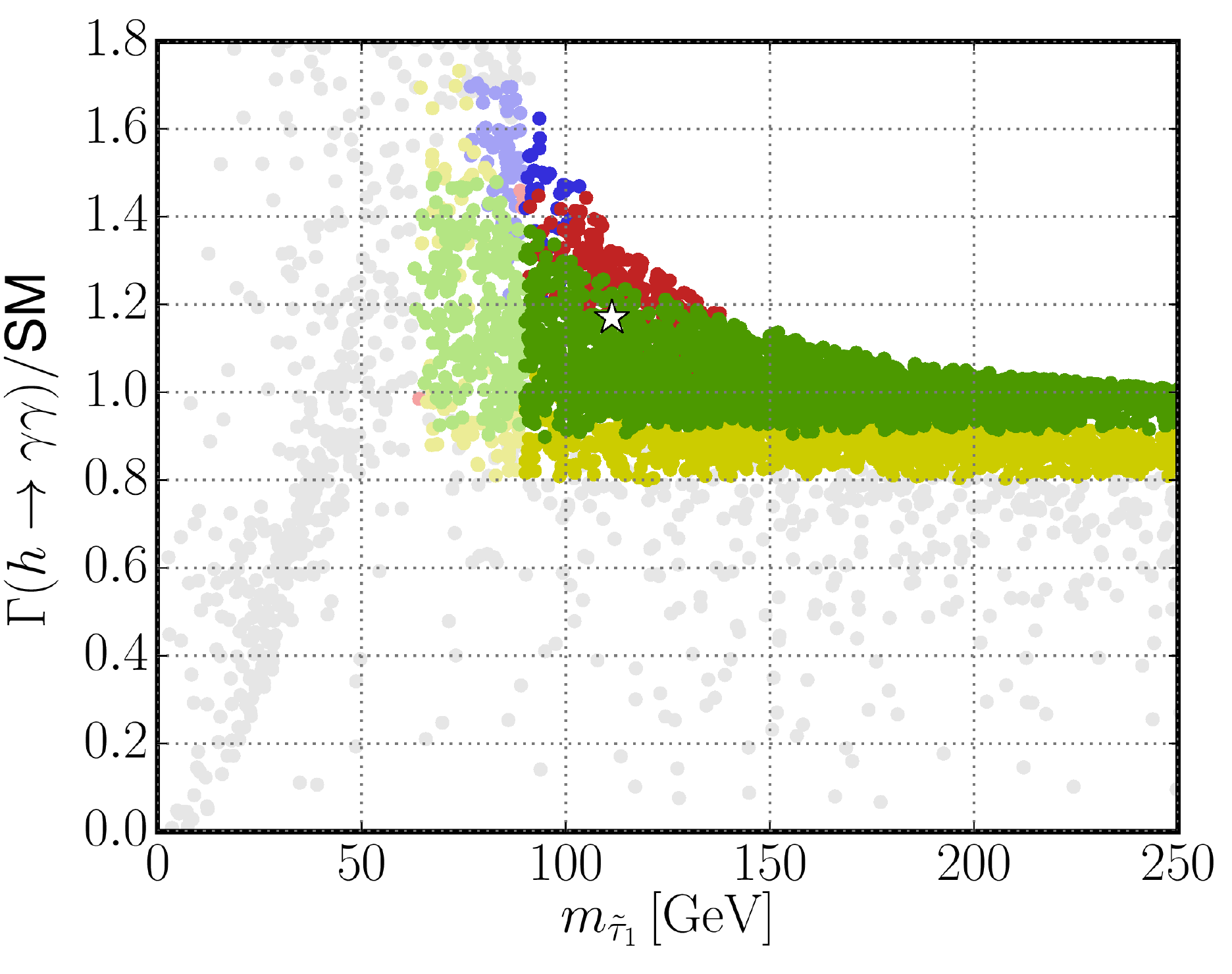}}\hfill
 \subfigure[~Correlation between $m_{\sstau_1}$ and $\mu\tan\beta$.]{\includegraphics[width=0.44\textwidth]{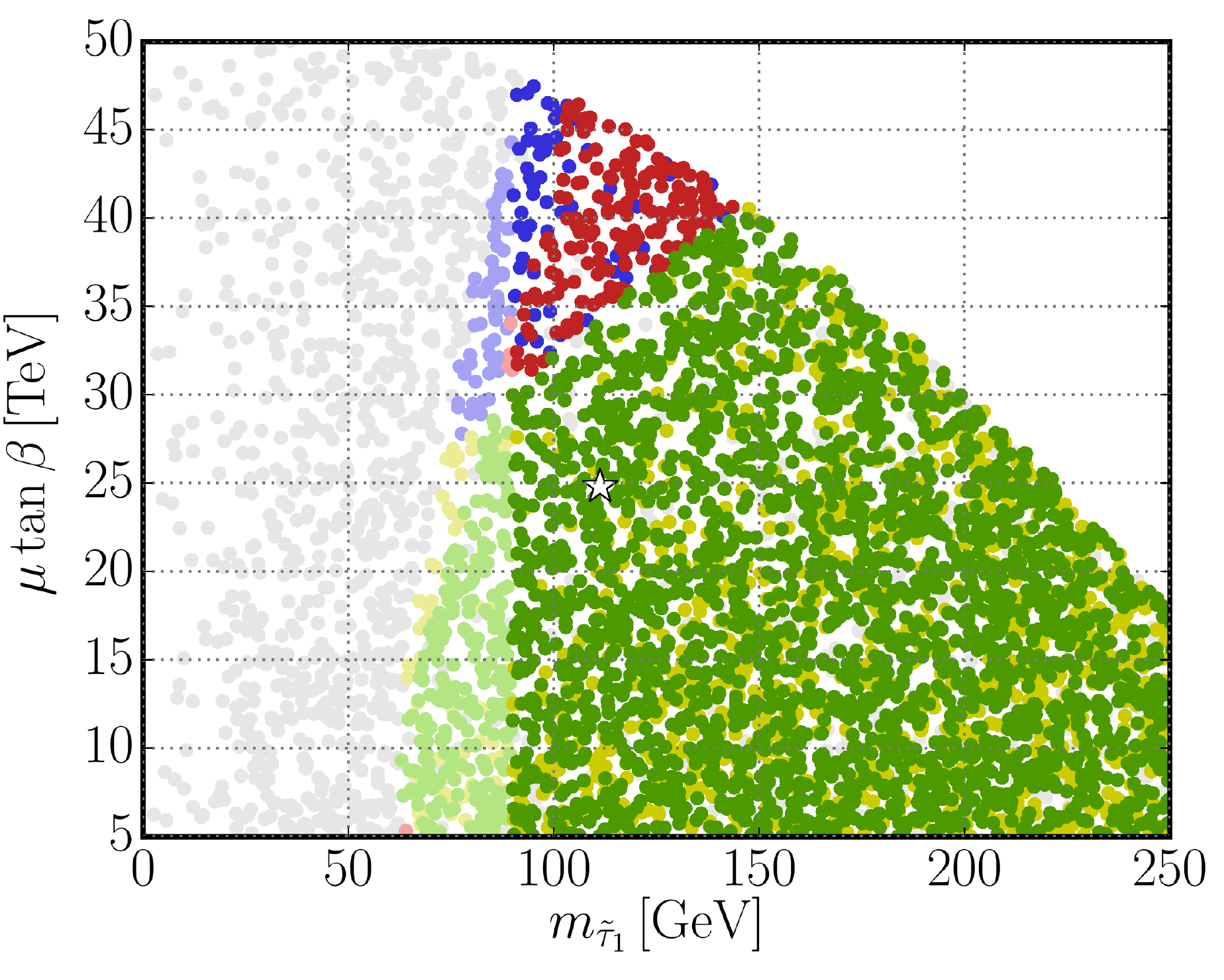}}
  \subfigure[~Dependence of $\Gamma(h\to\gamma\gamma)$ on $\mu\tan\beta$.]{\includegraphics[width=0.44\textwidth]{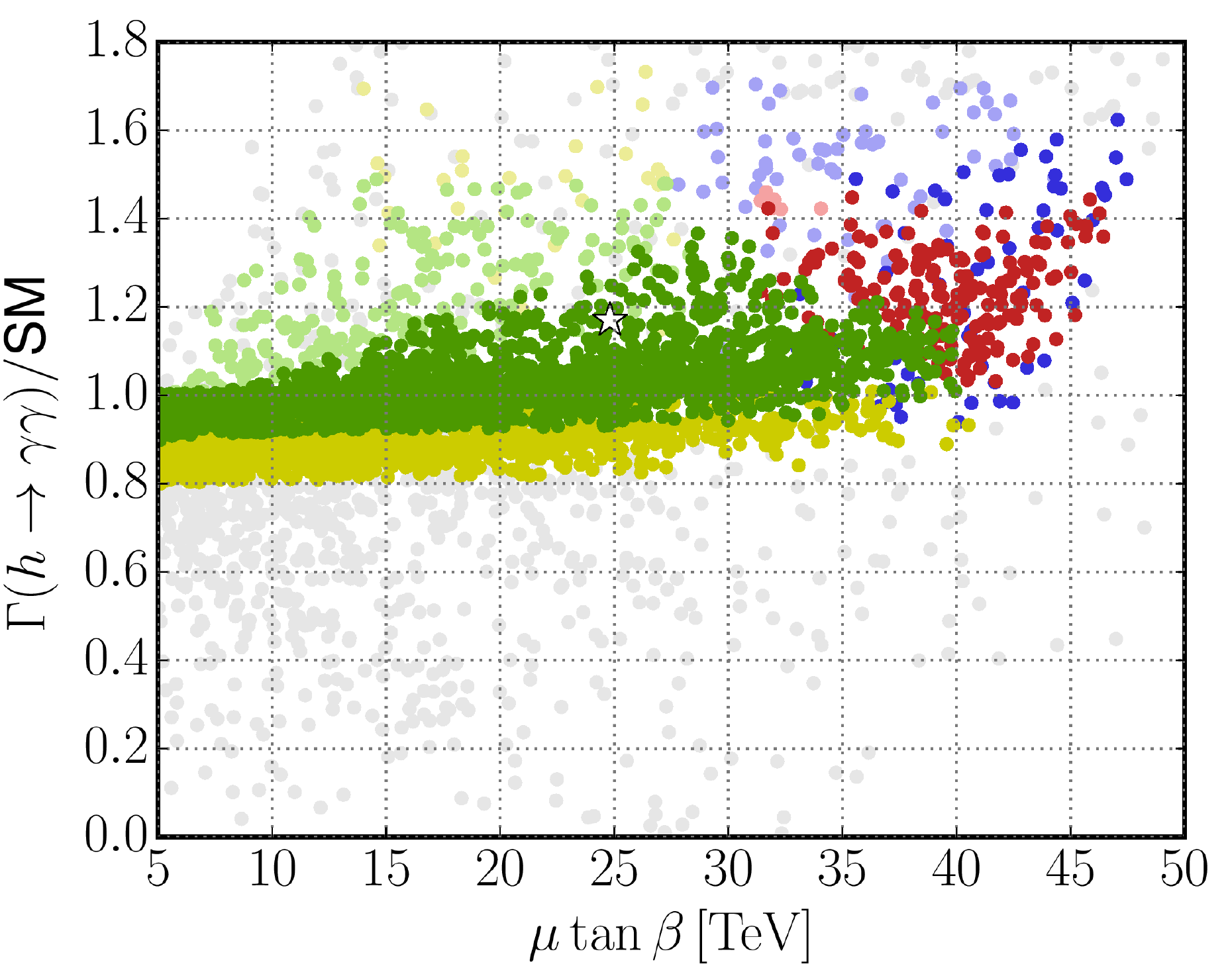}}\hfill
  \subfigure[~Correlation between $m_{\sstop_1}$ and $m_{\sstau_1}$.]{\includegraphics[width=0.44\textwidth]{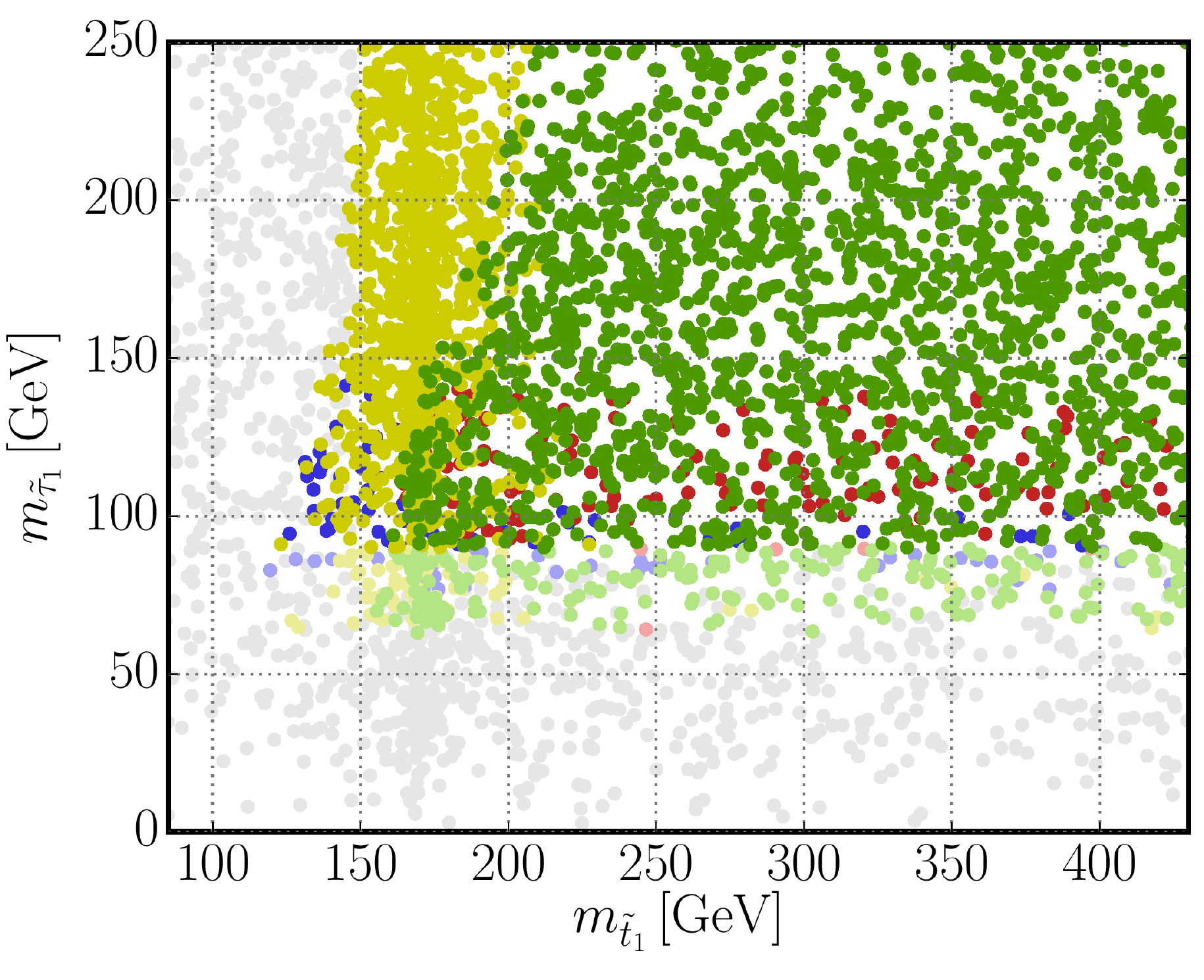}}
 \caption{Results for scenario~\textbf{B} `\emph{Decoupling Limit with a Light Stop and Light Stau}'. The color coding of the scan points is the same as in \fig{Fig:Case2_brhnp}, see also text for a detailed discussion.
 }
\label{Fig:Case2_other}
\end{figure*}
 
\fig{Fig:Case2_brhnp} shows the correlation of the light stop and light stau mass with the branching fraction of the new physics decay mode, $\brhnp$. Allowing for a sizable new physics branching fraction, $\brhnp \sim 25\%$, the light stop mass can be as low as $\sim 123\gev$ at $\CL{95}$, as can be seen in \fig{Fig:Case2_brhnp_a}. Disregarding the stau mass constraints from LEP and the constraints from the vacuum metastability condition weakens this lower limit only marginally, to $m_{\sstop_1} \gtrsim 119\gev$. In general, the $\CL{95}$ region features values of $\brhnp$ of up to $37\%$ at light stop and stau masses. If this new physics Higgs decay mode yields purely invisible final states, this region is already being probed and constrained by current LHC searches, cf.~\eqn{Eq:brhinvlimit_ATLAS}. Restricting to the case $\brhnp \equiv 0$, the $\CL{95}$ lower stop mass limit values $\sim 146\gev$.

The dependence of the partial width for the Higgs decay $h\to \gamma\gamma$ normalized to the SM prediction, denoted as $\Gamma(h\to\gamma\gamma)/\text{SM}$, on the light stop and stau masses, as well as on $\mu\tan\beta$, is shown in \figs{Fig:Case2_other}(a, c, e), respectively. The reduction and possible enhancement of $\Gamma(h\to\gamma\gamma)/\text{SM}$ at lighter stop and stau mass values, respectively, is evident from the slope of the point distribution in \figs{Fig:Case2_other}(a) and (c), respectively. The stau contribution to $\Gamma(h\to\gamma\gamma)/\text{SM}$ clearly grows with $\mu\tan\beta$, cf.~\fig{Fig:Case2_other}(e), and we can obtain an enhancement of the partial width by up to $\sim 40\%$ within the $\CL{95}$ preferred parameter region that is consistent with the naive LEP stau mass limit and vacuum metastability constraints. Disregarding the vacuum metastability or the stau mass constraint, the maximal enhancement  within the $\CL{95}$ region can be as large as $\sim 60\%$ or $\sim 70\%$, respectively. Note that these maximal values are obtained for large values of the light stop mass, where its influence on $\Gamma(h\to\gamma\gamma)$ is marginal.
The correlations between $m_{\sstop_1}$, $m_{\sstau_1}$ and $\mu\tan\beta$ are displayed in \figs{Fig:Case2_other}(b, d, f).
The lowest allowed stop mass values are obtained at large values of $\mu\tan\beta$ and low values of the light stau mass, where the stau contribution to $\Gamma(h\to\gamma\gamma)$ is sizable.

\subsection{Decoupling Limit with a light stop and a light chargino (scenario {\bf C})}

We now investigate the influence of a light chargino on the light stop mass limit. The light chargino contribution to the Higgs di-photon rate has first been calculated in \citeres{Kalyniak:1985ct,Bates:1986zv,Weiler:1988xn}, see also \citeres{Gunion:1988mf,Djouadi:1996pb,Djouadi:1996yq,Diaz:2004qt} for early studies of its implications and discovery potential. After the discovery of the Higgs boson, studies of the chargino contribution to $\Gamma(h\to \gamma\gamma)$ in various supersymmetric models~\cite{Casas:2013pta,Hemeda:2013hha,SchmidtHoberg:2012ip,Batell:2013bka} have revived due to a potential enhancement of the Higgs to di-photon rate seen awhile both in the ATLAS and CMS data. 

We elaborated upon the one-loop corrections to the di-photon partial decay width, $\Gamma(h\to\gamma\gamma)$, arising from the light chargino already in Section~\ref{Sect:higherorder}. The amplitude coefficient $c_{\charge_1^\pm}^h$ is proportional to the wino-Higgsino mixing in the chargino sector. We therefore choose $\mu \equiv M_2$ and allow for low values of $\tan\beta$ in the following numerical study in order to maximize the chargino contribution to $\Gamma(h\to \gamma\gamma)$. Note that this choice is well motivated by considerations in electroweak baryogenesis, see Section \ref{Sect:EWBG} for a discussion. We thus consider the following parameter space:
\begin{align}
\text{sgn}(\SMU{3}^2)\SMU{3} &\in [-150, 500]\gev\,,\nonumber\\
\mu \equiv M_2 &\in [50, 300]\gev\,,\nonumber\\
\tan\beta &\in [1, 20]\,. 
\end{align}
Again, we assume the decoupling limit, $M_1 = M_A = 1\tev$, and a vanishing stop mixing parameter, $X_t = 0$. Also, we allow for an additional ``new physics'' Higgs decay mode parametrized by $\brhnp$. However, due to the assumption $\mu \equiv M_2$ and allowing their values to be small, the masses of the three lightest neutralinos and the light chargino may be below $m_h/2$ such that Higgs boson decays to these SUSY particles become possible. We specify the sum of the Higgs decay branching fractions to these states as
\begin{align}
\brhsusy \equiv \sum_{i,j = 1,2,3} \text{BR}(h\to \neut_i \neut_j) + \text{BR}(h\to \charge_1^+ \charge_1^-)
\end{align}
in what follows. Thus, the total non-SM Higgs decay branching fraction is given by
\begin{align}
\brhnpsusy \equiv \brhnp + \brhsusy\,.
\end{align}

\citere{Batell:2013bka} scrutinized in detail the existing chargino mass limits from LEP, Tevatron and the LHC. The authors argue that the LEP chargino mass limit of $m_{\charge_1^+} \gtrsim 103.5\gev$~\cite{Agashe:2014kda} may be evaded in case of a sneutrino LSP decaying via a small R-parity violating coupling with a decay length around $10-100~\mathrm{cm}$. In such a case the chargino mass might be as low as $\gtrsim m_h/2$. We will indicate the parameter points that violate the naive LEP chargino mass limit $m_{\charge_1^+}\gtrsim 103.5\gev$ by pale colors in the following results.

The BF point is found at 
\begin{align}
(m_{\sstop_1}, m_{\charge_1^\pm}, \brhnp) = (523\gev, 117\gev, 6.0\%)
\end{align}
 with a fit quality of $\chi^2/\mathrm{ndf} = 68.0/81$. It features low values of $\tan\beta= 2.3$ and $\mu = M_2 = 190\gev$. We obtain a $\CL{95}$ mass limit on the light stop of $m_{\sstop_1} \gtrsim 123.3\gev$ in the region consistent with the LEP chargino mass constraint. Neither relaxing the LEP chargino mass limit nor imposing $\brhnp \equiv 0$ changes the picture. 
 
\begin{figure*}
 \subfigure[~($m_{\sstop_1}$, $\brhnpsusy$) plane.]{\includegraphics[width=0.44\textwidth]{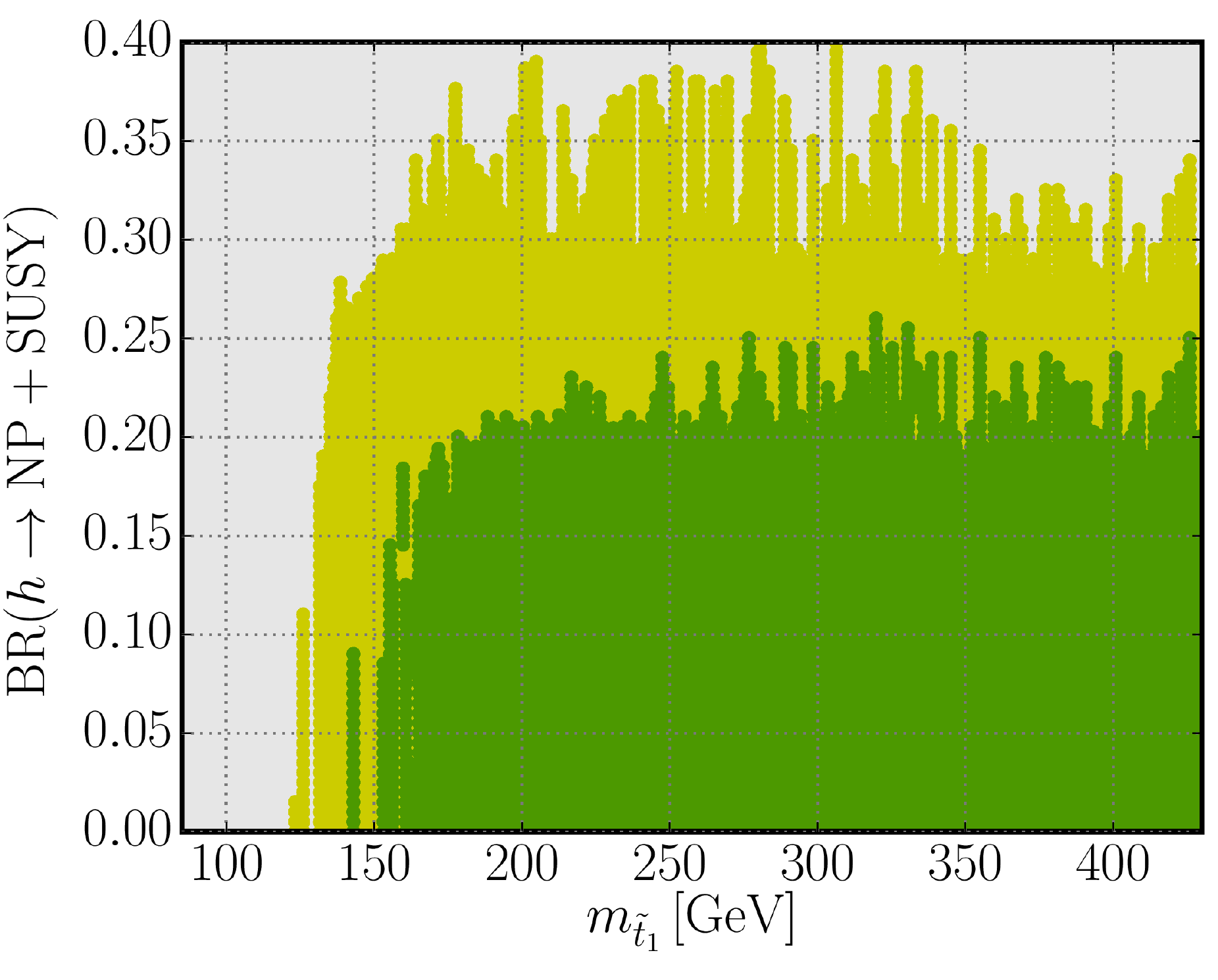}}
 \hfill
 \subfigure[~($m_{\charge^+_1}$, $\brhnpsusy$) plane.]{\includegraphics[width=0.44\textwidth]{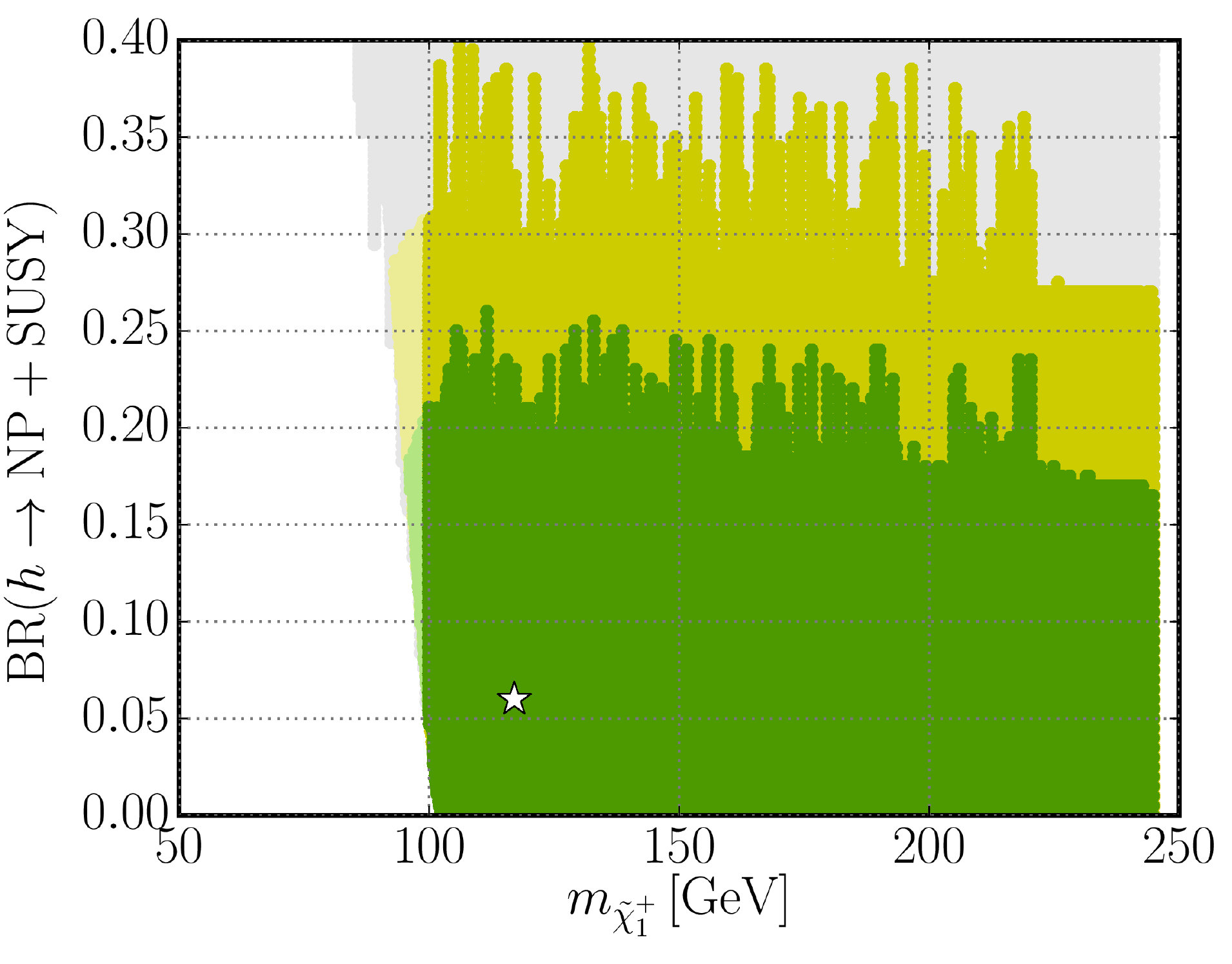}}
 \caption{Results for scenario~\textbf{C} `\emph{Decoupling Limit with a Light Stop and Light Chargino}' in the $(m_{\sstop_1},~\brhnpsusy)$ [a] and $(m_{\charge_1},~\brhnpsusy)$ [b] plane. The green (yellow) region is allowed from the Higgs signal rates at $68\%~\mathrm{C.L.}$ [$95\%~\mathrm{C.L.}$]. The white star indicates the best fit point. The regions in paler colors violate the naive chargino mass limit $m_{\charge^+_1} \ge 103.5\gev$ from LEP.}
 \label{Fig:Case3_brhnp}
 \end{figure*}

\begin{figure*}[t]
 \subfigure[~Dependence of $\Gamma(h\to\gamma\gamma)$ on $m_{\sstop_1}$.]{\includegraphics[width=0.44\textwidth]{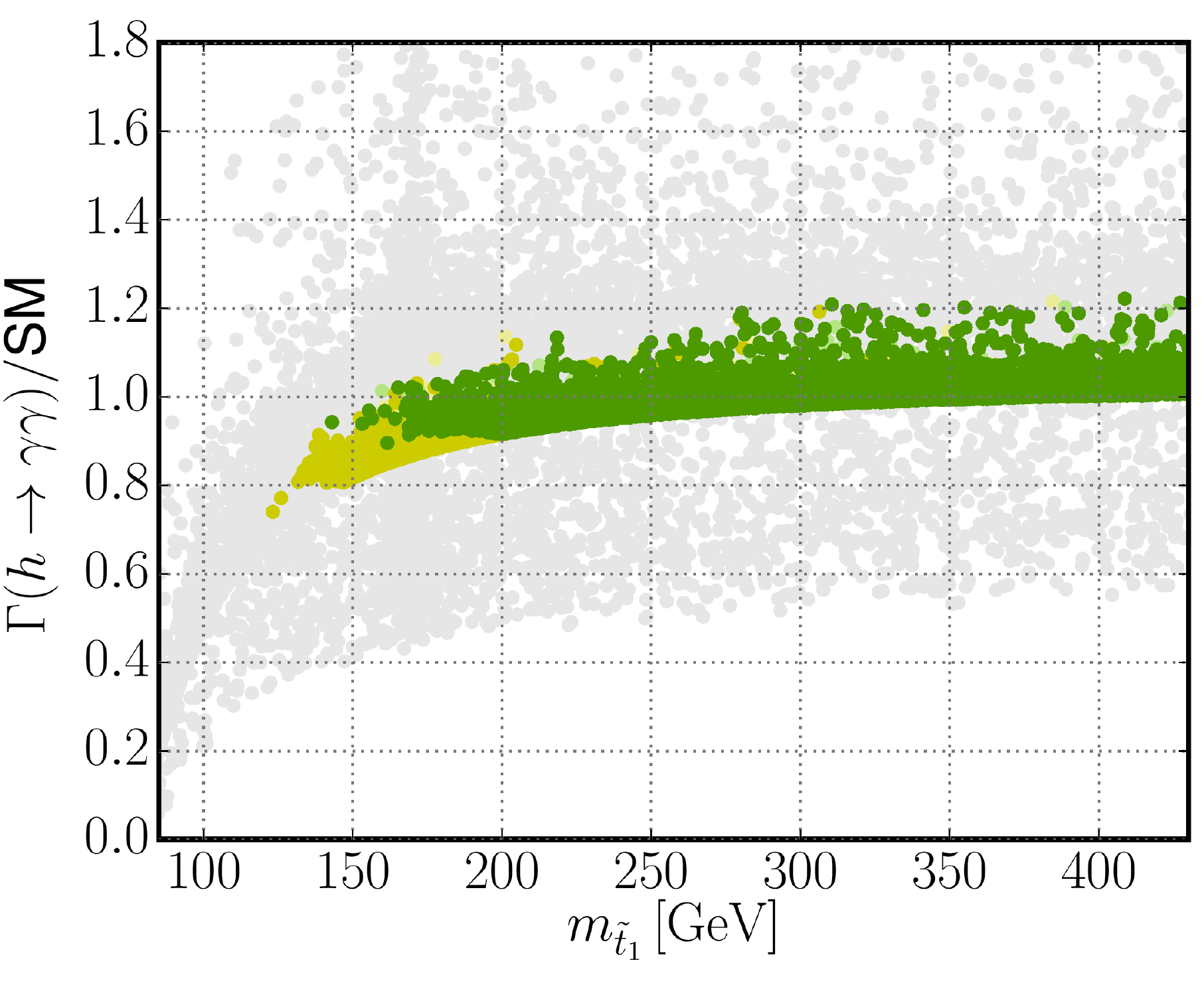}}\hfill
 \subfigure[~($\mu\tan\beta$, $\brhnpsusy$) plane.]{\includegraphics[width=0.44\textwidth]{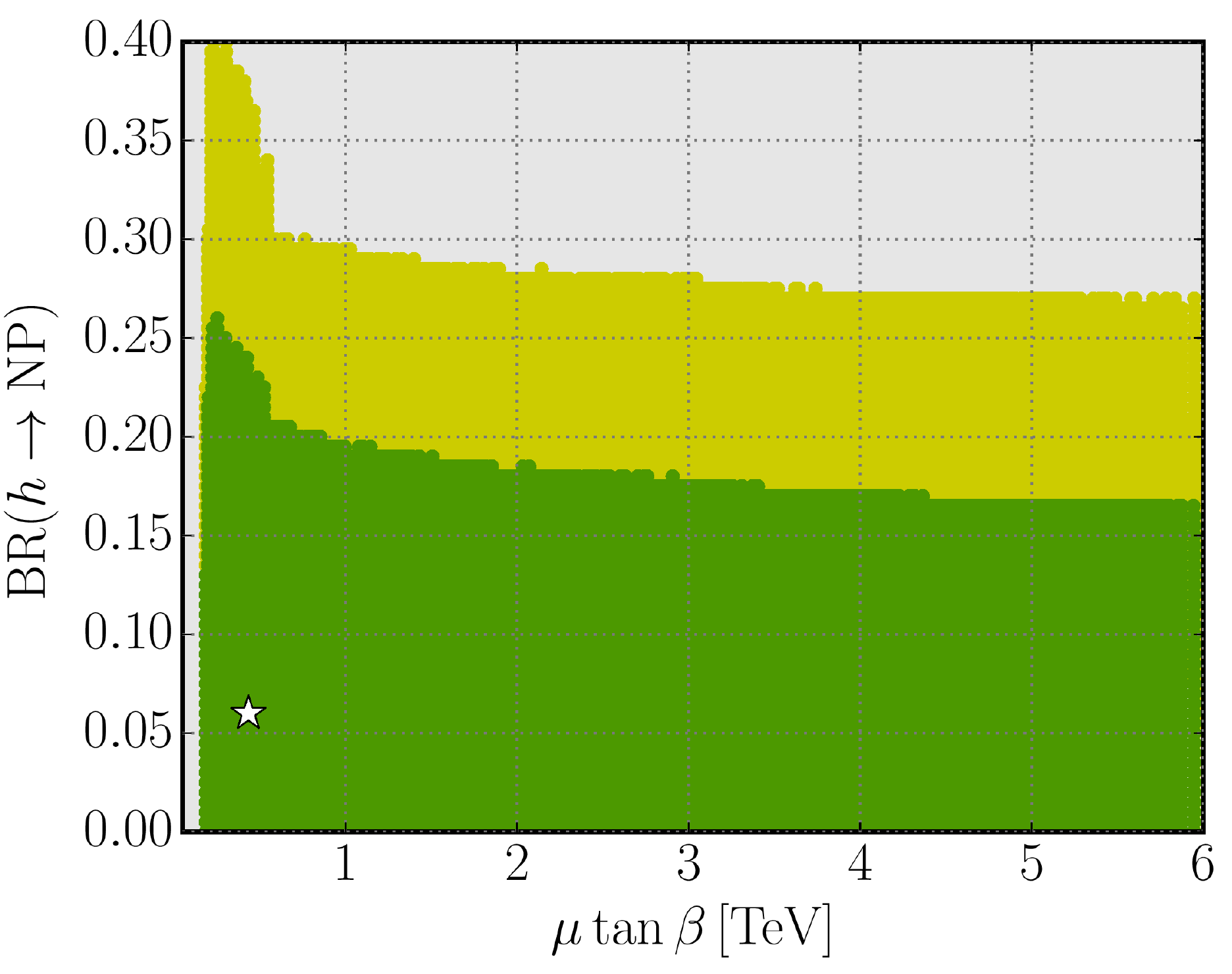}}\\
  \subfigure[~Dependence of $\Gamma(h\to\gamma\gamma)$ on $m_{\charge^+_1}$.]{\includegraphics[width=0.44\textwidth]{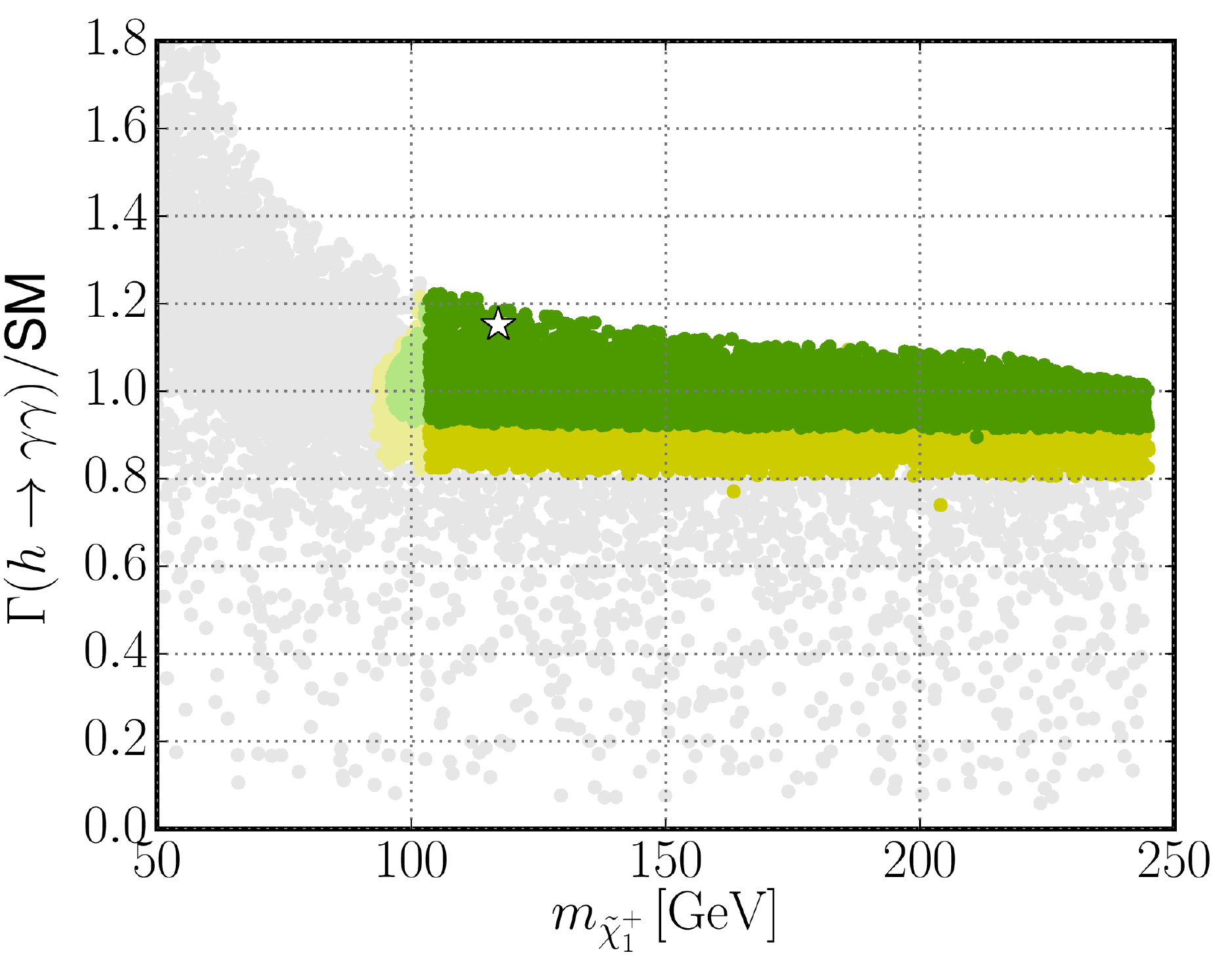}}\hfill
 \subfigure[~Correlation between $m_{\sstop_1}$ and $\mu\tan\beta$.]{\includegraphics[width=0.44\textwidth]{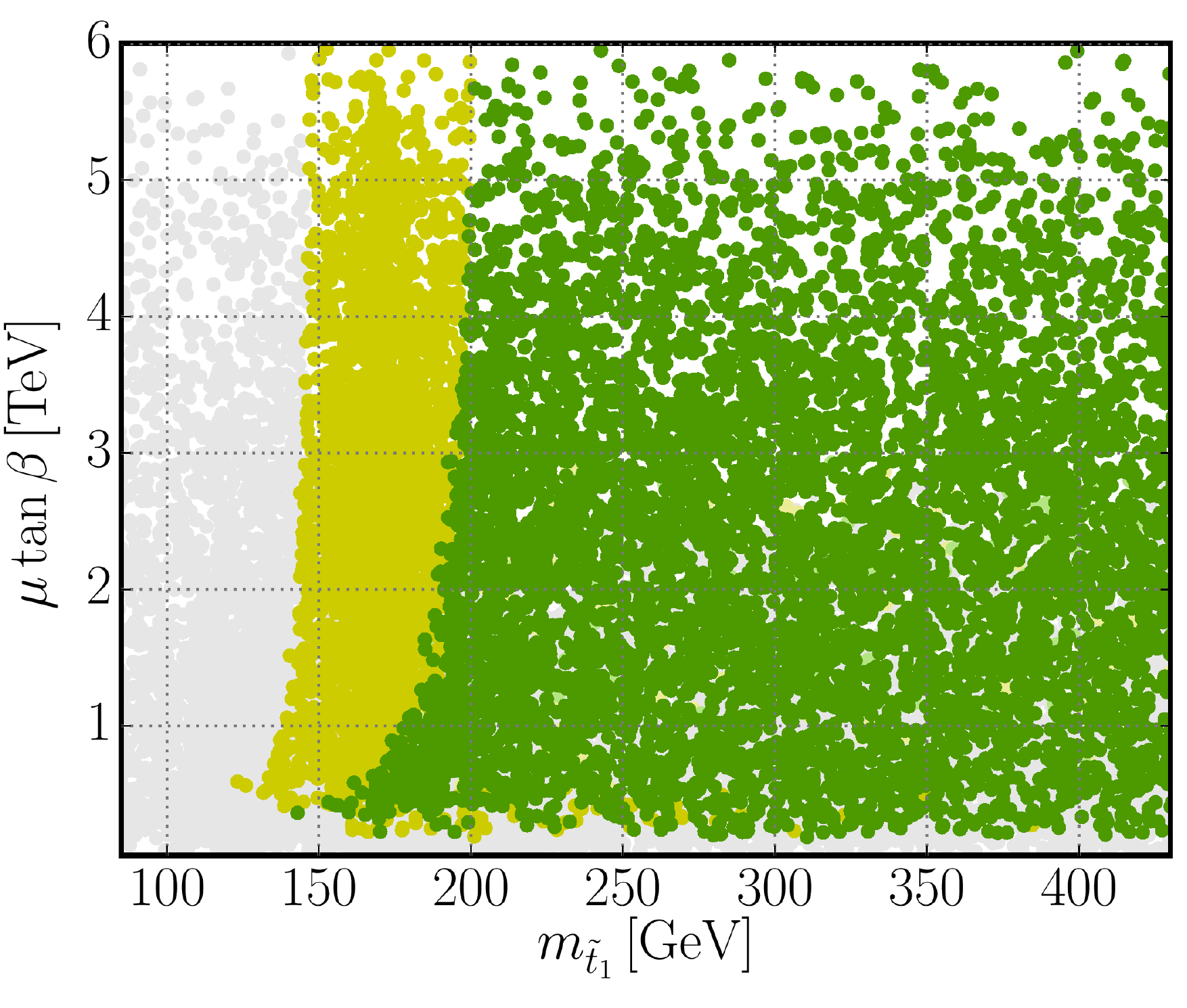}}\\
 \subfigure[~Dependence of $\Gamma(h\to\gamma\gamma)$ on $\mu\tan\beta$.]{\includegraphics[width=0.44\textwidth]{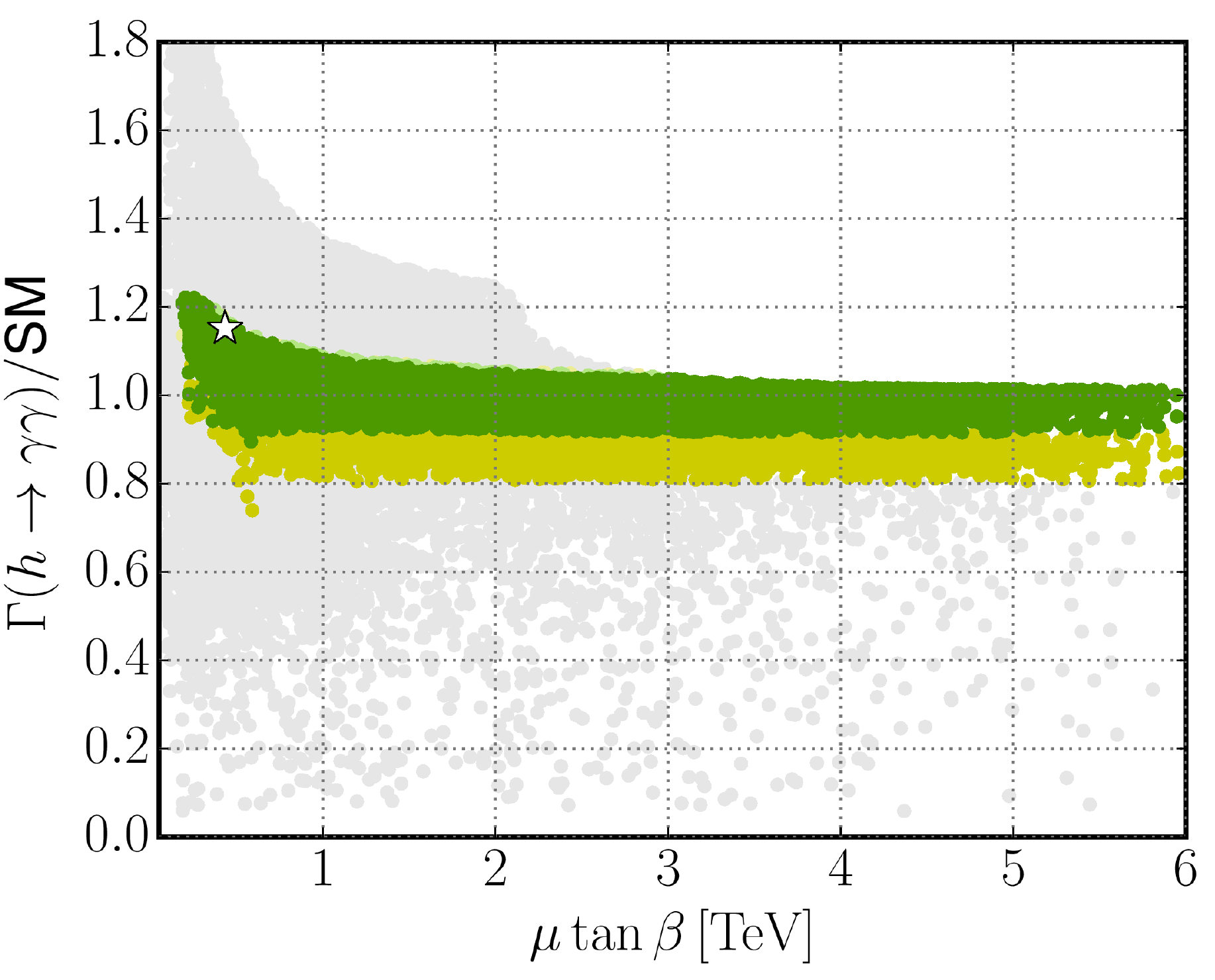}}\hfill
 \subfigure[~Correlation between $m_{\sstop_1}$ and $m_{\charge^+_1}$.]{\includegraphics[width=0.44\textwidth]{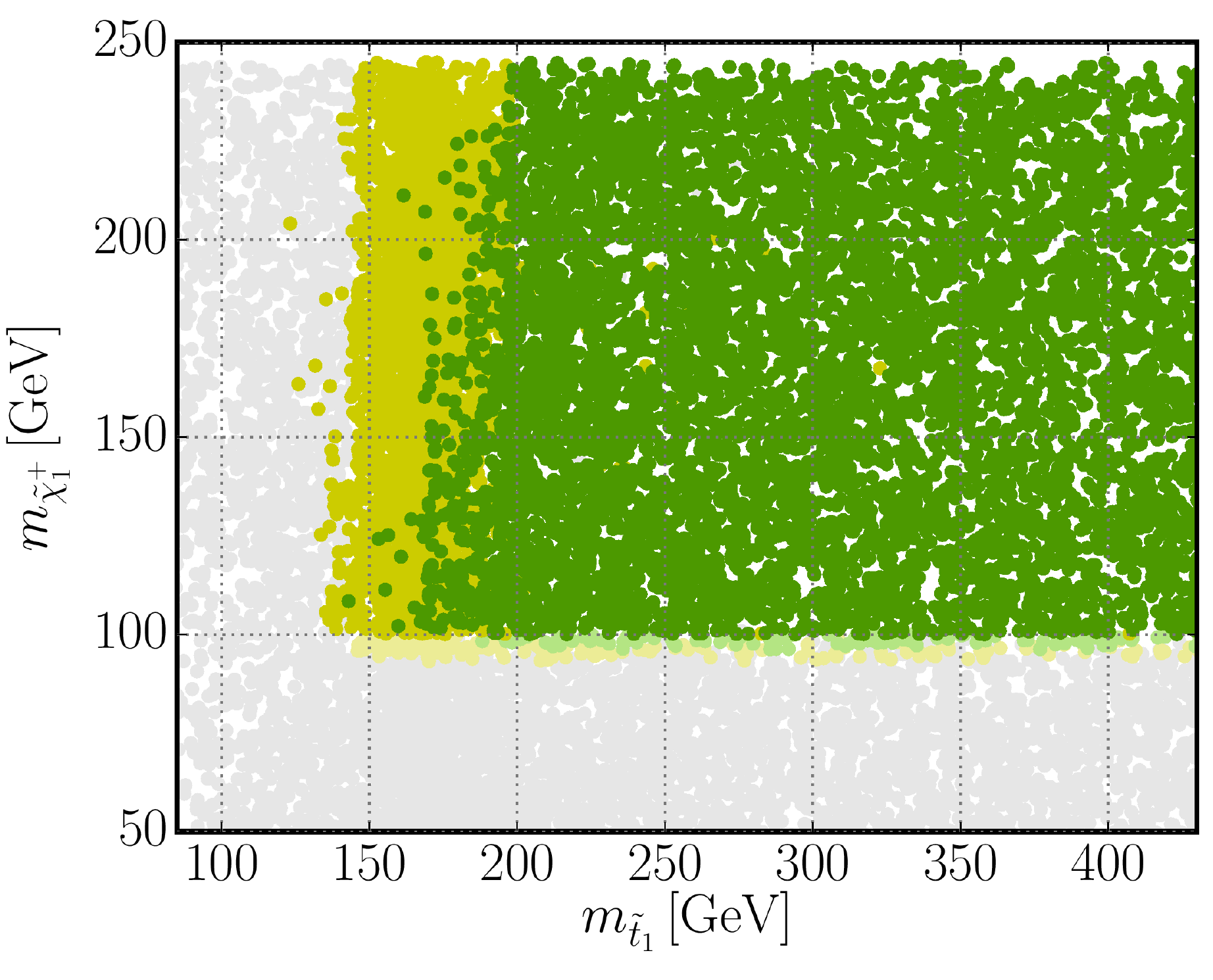}}
 \caption{Dependence of the SM normalized partial widths for $h\to\gamma\gamma$ and various parameter correlations for the scenario~\textbf{C} `\emph{Decoupling Limit with a Light Stop and Light Chargino}'. The color coding of the scan points is the same as in \fig{Fig:Case3_brhnp}.}
 \label{Fig:Case3}
 \end{figure*}

In \fig{Fig:Case3_brhnp} we show the Higgs decay branching fraction to non-SM final states, $\brhnpsusy$, as a function of the light stop and chargino mass. The maximal Higgs decay branching fraction to SUSY particles (dominantly to the lightest neutralino) is $\brhsusy \le 30\%$ at the $\CL{95}$, which can only be saturated if the light chargino mass is around the LEP limit, $m_{\charge_1^+} \sim (90 -105) \gev$. At lower chargino masses we inevitably obtain larger branching fractions, thus these points are disfavored, see \fig{Fig:Case3_brhnp}(b). The wiggly edges of the preferred parameter regions in \fig{Fig:Case3_brhnp} result from the fact that for most of the parameter space, $\brhnpsusy$ is constrained to be $\lesssim 17 - 20\%~[27 - 30\%]$ at the $\CL{68}$ [$\CL{95}$], except for the case that $\mu\tan\beta \lesssim 500\gev$. This can clearly be seen in \fig{Fig:Case3}(b), where $\brhnpsusy$ is shown as a function of $\mu\tan\beta$. In this parameter region, $\mu\tan\beta \lesssim 500\gev$, the chargino contribution to $\Gamma(h\to\gamma\gamma)$ is significant, as shown in \fig{Fig:Case3}(b), and can lead to an enhancement of the di-photon partial width of up to $\sim23\%$. Consequently, larger values of $\brhnpsusy$ are allowed here.

\figs{Fig:Case3}(a) and (c) show the influence of the light stop and light chargino mass, respectively, on the SM normalized partial width for the di-photon decay, $\Gamma(h\to\gamma\gamma)/{\rm SM}$. Moreover, we show in \figs{Fig:Case3}(d) and (f) how the allowed values for the light stop mass correlate with $\mu\tan\beta$ and with  the light chargino mass, $m_{\charge_1^+}$, respectively. It is evident from \fig{Fig:Case3}(d) that the light stop achieves its lowest allowed mass values in the low $\mu\tan\beta$ region, where the chargino contribution to $\Gamma(h\to\gamma\gamma)$ is substantial. The bulk of the parameter points featuring a very light stop, $m_{\sstop_1} \lesssim 150\gev$ tend to prefer a low chargino mass around $(100-150)\gev$. However, we also find a few points with larger values of $m_{\charge^+_1}$  around $\sim 200\gev$.


\subsection{Non-decoupling effects (scenario {\bf D})}

In the previous fits we investigated the impact of a non-zero branching fraction for the additional Higgs decay to ``new physics'' on the mass limits obtained from the Higgs signal rates. The desired suppression of  well-measured SM Higgs decay modes needed to partially compensate for the enhanced Higgs production rates may also, however, derive from an increase in the partial width of the dominant but relatively poorly measured Higgs decay mode to bottom quarks, $\Gamma(h\to b\bar{b})$. This is possible with an enhancement of the light Higgs coupling to bottom quarks if the decoupling limit is not quite realized. However, the combination of ATLAS and CMS results from the LHC Run~1 indicates a slight deficit both in the $hb\bar{b}$ and $h\tau^+\tau^-$ coupling determination with respect to the SM expectation at the level of roughly $2\sigma$ and $1\sigma$, respectively\footnote{Here we refer to a fit result that employs a very general parametrization of the Higgs production and decay rates in terms of $\kappa$ scale factors, see Section 6.1 in \citere{CMS:2015kwa}. The significance of the deviations may be different in other models (e.g.~the MSSM) which feature stronger correlations among the Higgs couplings/rates.}~\cite{CMS:2015kwa}. As a result, it remains to be seen to what extent the data allows for the presence of a Higgs-bottom quark Yukawa coupling enhancement that would compensate the increase of the gluon fusion cross section at small stop masses.

As another aspect of this study, we address the following question: Assuming the existence of a light stop with a mass below the top quark mass, $m_{\sstop_1} < m_t$, and possibly other light SUSY states such as a light stau, how low can the pseudoscalar Higgs mass, $M_A$, be? In other words, how large are the non-decoupling effects in the Higgs sector that are still allowed by the currently available Higgs data under these circumstances? Moreover, we will briefly address how large the branching fractions for heavy Higgs decays to light stops or staus can be in this scenario. We comment in Section~\ref{Sect:EWBG} on the physical significance of a non-decoupling value for $M_A$ for electroweak baryogenesis.

The tree-level Yukawa sector of the MSSM is that of a Type-II Two Higgs doublet model (2HDM). The Lagrangian is given by\footnote{In this discussion we focus on the Higgs boson couplings to third generation quarks and neglect the full generation structure of the Yukawa couplings.}
\begin{align}
-\mathcal{L}_\text{Yuk} = \epsilon_{ij} \left[h_b \bar{b}_R H_D^i Q_L^j + h_t \bar{t}_R Q_L^i H_U^j \right] + \text{h.c.}\,,
\end{align}
where $H_U$, $H_D$ are the hypercharge $+\tfrac{1}{2}$, $-\tfrac{1}{2}$ Higgs fields that couple to up- and down-type quarks, respectively, $Q_L$ the left-handed quark $\text{SU}(2)$ doublet field, and $t_R$, $b_R$ the right-handed top and bottom quark $\text{SU}(2)$ singlet fields.  We sum over the weak $\text{SU}(2)$ indices $i,j =1,2$, where $\epsilon_{ij}$ is the anti-symmetric tensor. $h_b$ and $h_t$ denote the bottom and top Yukawa couplings, respectively, that are related at tree-level to the bottom and top quark masses,
\begin{align}
m_b = h_b \frac{ v}{\sqrt{2}} \cos\beta, \qquad m_t = h_t \frac{v}{\sqrt{2}} \sin\beta\,,
\end{align}
with the Higgs vacuum expectation value $v \approx 246.2\gev$. The corresponding tree-level couplings of the light $\CP$-even Higgs  boson to the bottom and top quarks are given by
\begin{align}
g_{hb\bar{b}} = g_{hb\bar{b}}^\text{SM} \cdot ( \sin(\beta-\alpha) - \cos(\beta-\alpha) \tan\beta)\,,\\
g_{ht\bar{t}} = g_{ht\bar{t}}^\text{SM} \cdot ( \sin(\beta-\alpha) + \cos(\beta-\alpha) \cot\beta)\,,
\end{align}
respectively, where the SM Higgs boson coupling to fermion species $f$ is given by $g_{hf\bar{f}}^\text{SM} = m_f/v$, and $\alpha$ is the mixing angle in the $\CP$-even Higgs sector. In the decoupling limit, $M_A \gg M_Z$, we have $\sin(\beta-\alpha) \rightarrow 1$ and $\cos(\beta-\alpha) \rightarrow 0$, thus, in the complete decoupling, the SM Higgs Yukawa couplings to bottom and top quarks are recovered. However, in the absence of this complete decoupling, the deviation from the SM value for the down-type Yukawa coupling is $\tan\beta$ enhanced, such that the decoupling of the $hb\bar{b}$ coupling is delayed~\cite{Haber:2000kq,Carena:2001bg,Gunion:2002zf}. The same behavior is observed for the $\tau$-lepton Yukawa coupling.

Beyond tree-level, the Higgs-fermion Yukawa couplings receive important higher-order SUSY corrections that partly violate the Type-II 2HDM Yukawa structure. After integrating out the SUSY particles, the effective Lagrangian for the down-type quark Yukawa sector reads~\cite{Hempfling:1993kv,Hall:1993gn,Carena:1994bv,Carena:1999py,Carena:2002bb,Guasch:2003cv,Noth:2008tw,Noth:2010jy,Mihaila:2010mp}
\begin{align}
-\mathcal{L}_\text{Yuk} = \epsilon_{ij} h_b \bar{b}_R H_D^i Q_L^j - \tilde{h}_b \bar{b}_R Q_L^i H_U^{j\,*}  + \text{h.c.}\,,
\label{Eq:deltabdef}
\end{align}
We omit a correction term $\delta h_b$ to the first term of \eqn{Eq:deltabdef}, which is known to be small~\cite{Carena:2014nza}.
Conventionally, the ratio of the effective couplings is denoted as
\begin{align}
\Delta_b \equiv \frac{\tilde{h}_b \,v_u}{h_b \, v_d}\,,
\end{align}
such that the relation between the Higgs-bottom quark Yukawa coupling and the physical bottom mass is modified to
\begin{align}
m_b =h_b \frac{v}{\sqrt{2}} \cos\beta \,(1+\Delta_b)\,,
\end{align}
and the light Higgs coupling to bottom quarks can be expressed as
\begin{align}
g_{hb\bar{b}} = g_{hb\bar{b}}^\text{SM} \cdot \left(\sin(\beta-\alpha) - \cos(\beta-\alpha) \tan\beta\right)\frac{1}{1+\Delta_b}
\left[  1+\Delta_b\left(1-\frac{\cos(\beta-\alpha)}{\sin\beta\sin\alpha}\right)\right]\,.
\end{align}
It is important to note that in the complete decoupling limit we have $\cos(\beta-\alpha) =0$ and thus the influence of the radiative $\Delta_b$ corrections on the light Higgs-fermion Yukawa couplings vanishes.

For $\Delta_b$ we should take into account not only sbottom induced contributions, but also chargino induced contributions.
Thus, $\Delta_b$ is in the most general form given by~\cite{Hofer:2009xb}
\begin{align}
 \Delta_b = \frac{2}{3\pi}\alpha_sm_g\mu\tan\beta I(m_{\tilde{b}_1}^2,m_{\tilde{b}_1}^2,m_{\tilde g}^2)
 -\frac{y_t^2}{16\pi^2}A_t\mu(D_2-M_2^2D_0)+\frac{g^2}{16\pi^2}M_2\mu(D_2-m_{\tilde{t}_1}^2D_0)\,,
\end{align}
where we have assumed real parameters for $A_t$, $\mu$ and $M_2$ and no mixing in the stop sector.
The function $I(a,b,c)$ is given by $(a-b)(b-c)(a-c)I(a,b,c)=ab\ln(a/b)+bc\ln(b/c)+ca\ln(c/a)$,
whereas $D_2$ and $D_0$ are functions of $m_{\tilde{\chi}^\pm_{1,2}}$ and $m_{\tilde{t}_{1,2}}$ and
can be taken from the Appendix of \citere{Hofer:2009xb}. If we assume $m_{\tilde{g}}$ as well
as $m_{\tilde{b}_{1,2}}$ and $m_{\tilde{t}_2}$ to be at a high scale $\MSUSY$ and small mixing in
the chargino sector, $\Delta_b$ scales like
\begin{align}
\label{eq:approxdeltab}
 \Delta_b = F_1\frac{2}{3\pi}\alpha_s\frac{\mu\tan\beta }{\MSUSY}
 +F_2\frac{y_t^2}{16\pi^2}\frac{A_t\mu\tan\beta}{\text{max}(\MSUSY^2,\mu^2)}-F_3\frac{g^2}{16\pi^2}\frac{M_2\mu\tan\beta}{\text{max}(\MSUSY^2,\mu^2)}\,.
\end{align}
The functions $F_1,F_2$ and $F_3$ are of order $\mathcal{O}(1)$ and dependent on the exact values
of the masses at the electroweak and the high scale. $\Delta_b$ is thus in particular negative for
negative values of $\mu$, even though for no mixing in the stop sector, i.e. $A_t=\mu/\tan\beta$, the second term
and the third term in \eqn{eq:approxdeltab} yield a positive contribution. For negative $\Delta_b$ the radiative corrections lead to an enhancement of the light Higgs-fermion Yukawa coupling.
Similar to the bottom-quark Yukawa coupling $\tan\beta$-enhanced corrections can also be included in the $\tau$-lepton Yukawa coupling expressed in terms of $\Delta_\tau$.

We scan over the following five parameters, in the indicated ranges:
\begin{align}
\text{sgn}(\SMU{3}^2)\SMU{3} &\in [-150, 500]\gev\,, \nonumber\\
\SML{3}\equiv \SME{3} &\in [70, 300]\gev\,,\nonumber\\
M_A &\in [150, 1000]\gev\,,\nonumber\\
\mu&\in [-5, 5]\tev\,,\nonumber\\
\tan\beta &\in [1, 50]\,.
\end{align}
Additionally, we impose the stop mixing parameter to be zero, $X_{t} = 0\gev$. 

As in this scenario the remaining MSSM Higgs states --- the heavy $\CP$ even Higgs, $H$, the $\CP$ odd Higgs, $A$, and the charged Higgs $H^\pm$ --- can be relatively light, constraints from direct Higgs searches at the Tevatron and LHC experiments are important. We include these constraints with the public code \HB\ (version \text{4.2.1})~\cite{Bechtle:2008jh,Bechtle:2011sb,Bechtle:2013gu,Bechtle:2013wla,Bechtle:2015pma}, which determines for each parameter point whether it is allowed or excluded at the $\CL{95}$ by Tevatron or LHC Higgs searches, using the latest results from the experiments. Recently, a \HB\ extension was released~\cite{Bechtle:2015pma} that incorporates the results from the CMS search for non-standard Higgs bosons decaying into $\tau$ lepton pairs~\cite{Khachatryan:2014wca} in terms of an exclusion likelihood. In the following, we employ this exclusion likelihood from the CMS $h/H/A\to \tau^+\tau^-$ search and add it to the $\chi^2$ from the Higgs signal rates, resulting in a global $\chi^2$ function. 

\begin{figure*}[t]
 \subfigure[~Correlations between $M_A$ and $m_{\sstop_1}$.]{\includegraphics[width=0.44\textwidth]{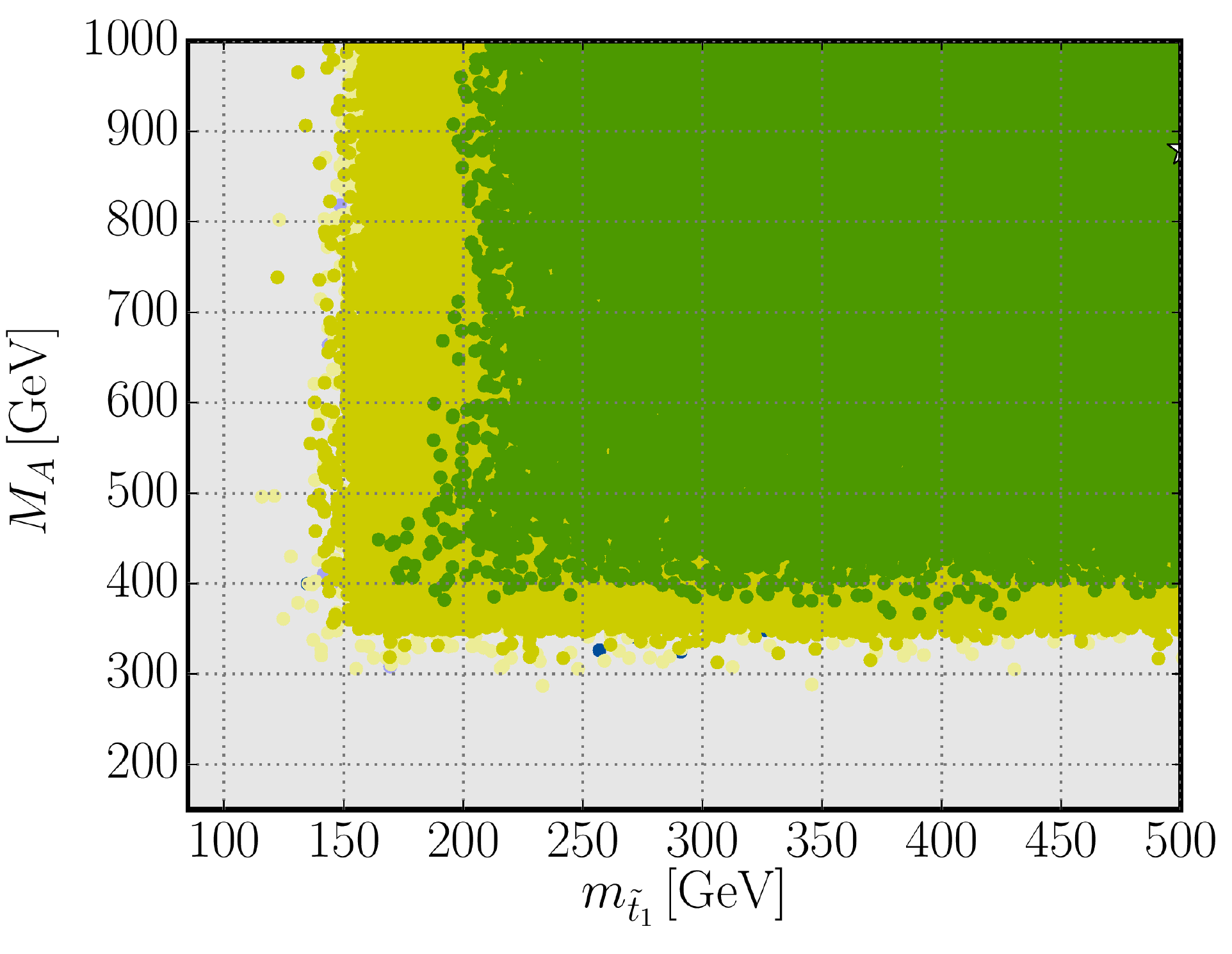}}\hfill
  \subfigure[~Correlations between $m_{\sstop_1}$ and $m_{\sstau_1}$.]{\includegraphics[width=0.44\textwidth]{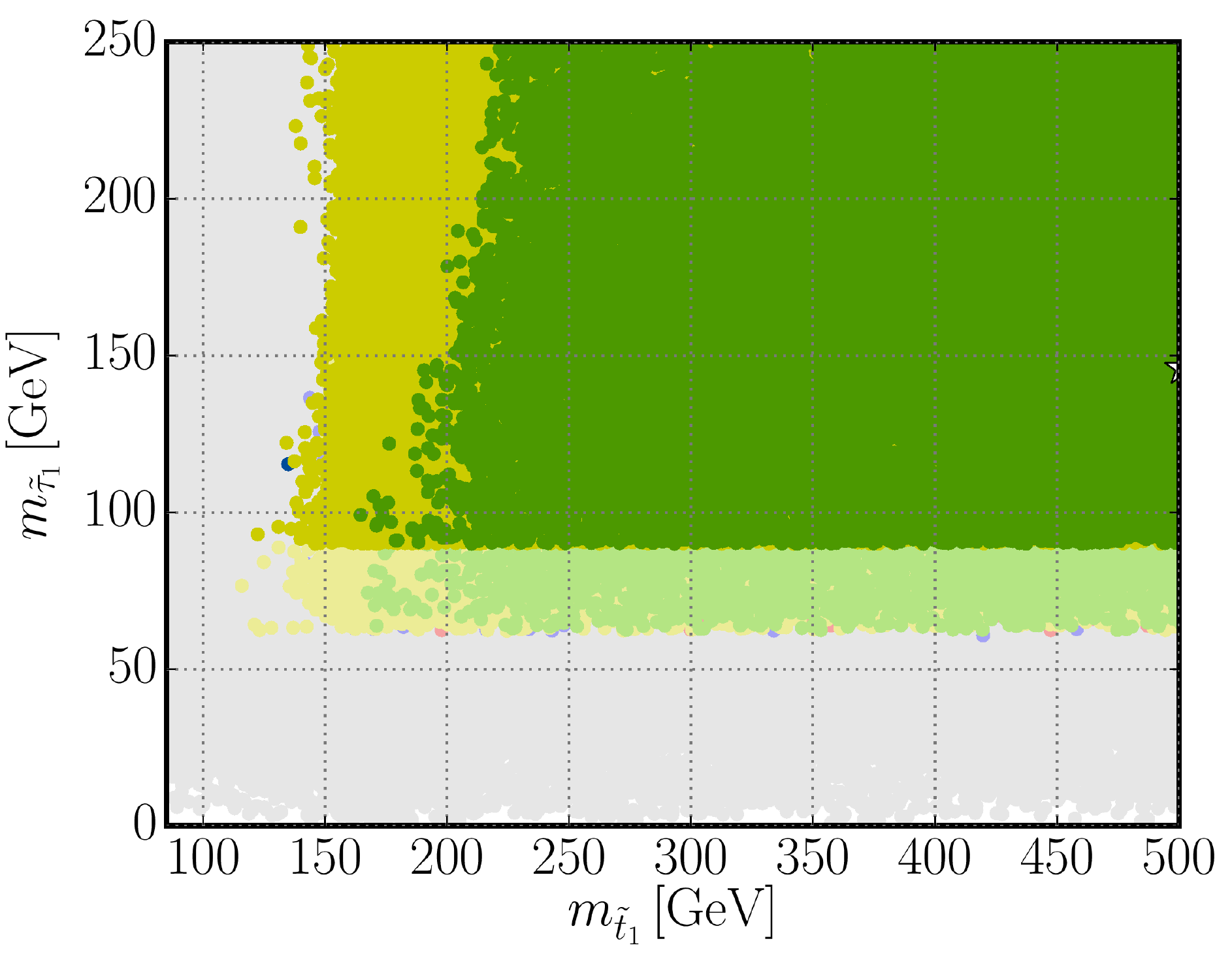}}\\
 \subfigure[~Correlations between $m_{\sstop_1}$ and $\tan\beta$.]{\includegraphics[width=0.44\textwidth]{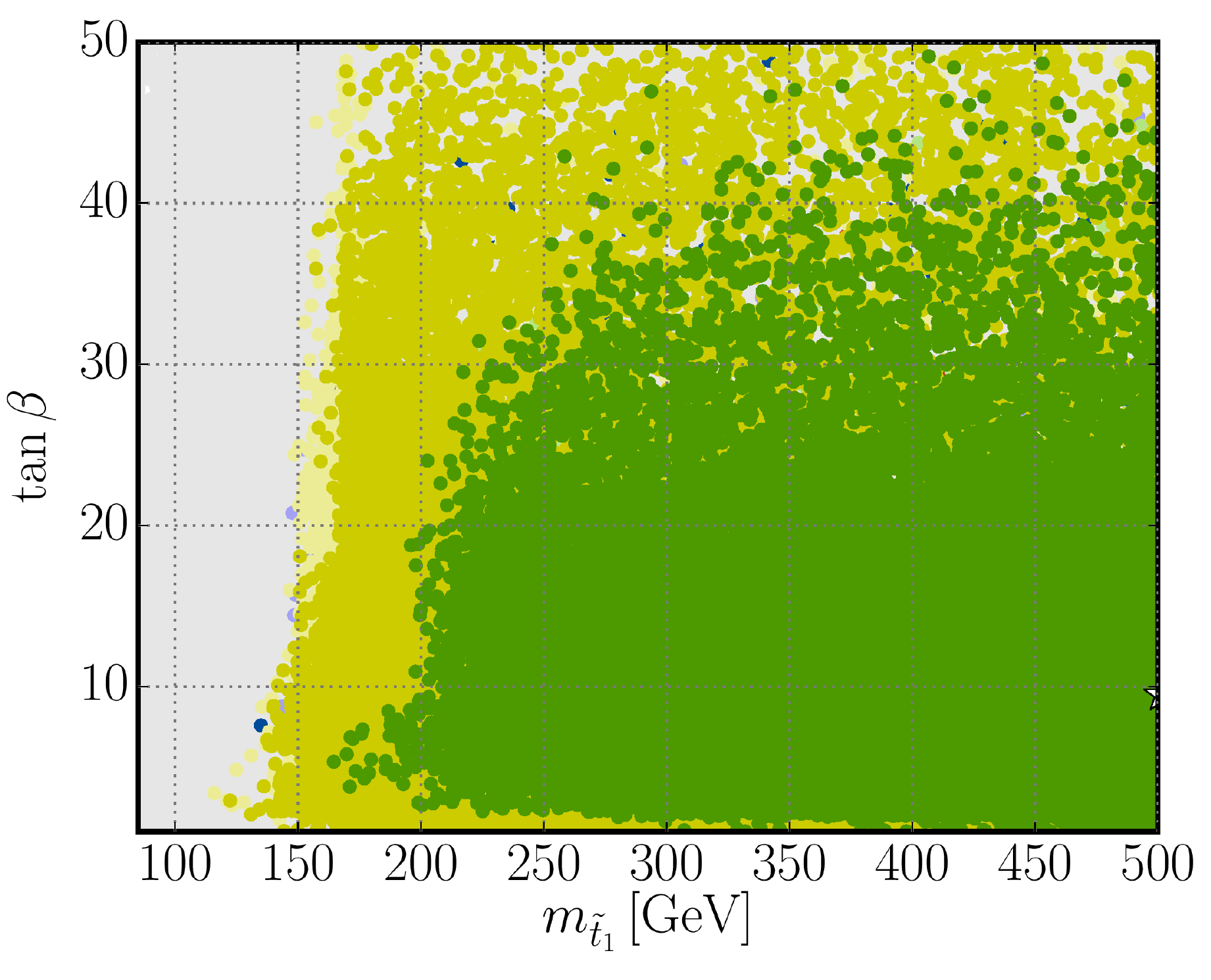}}\hfill
 \subfigure[~Correlation between $m_{\sstop_1}$ and $\mu$.]{\includegraphics[width=0.44\textwidth]{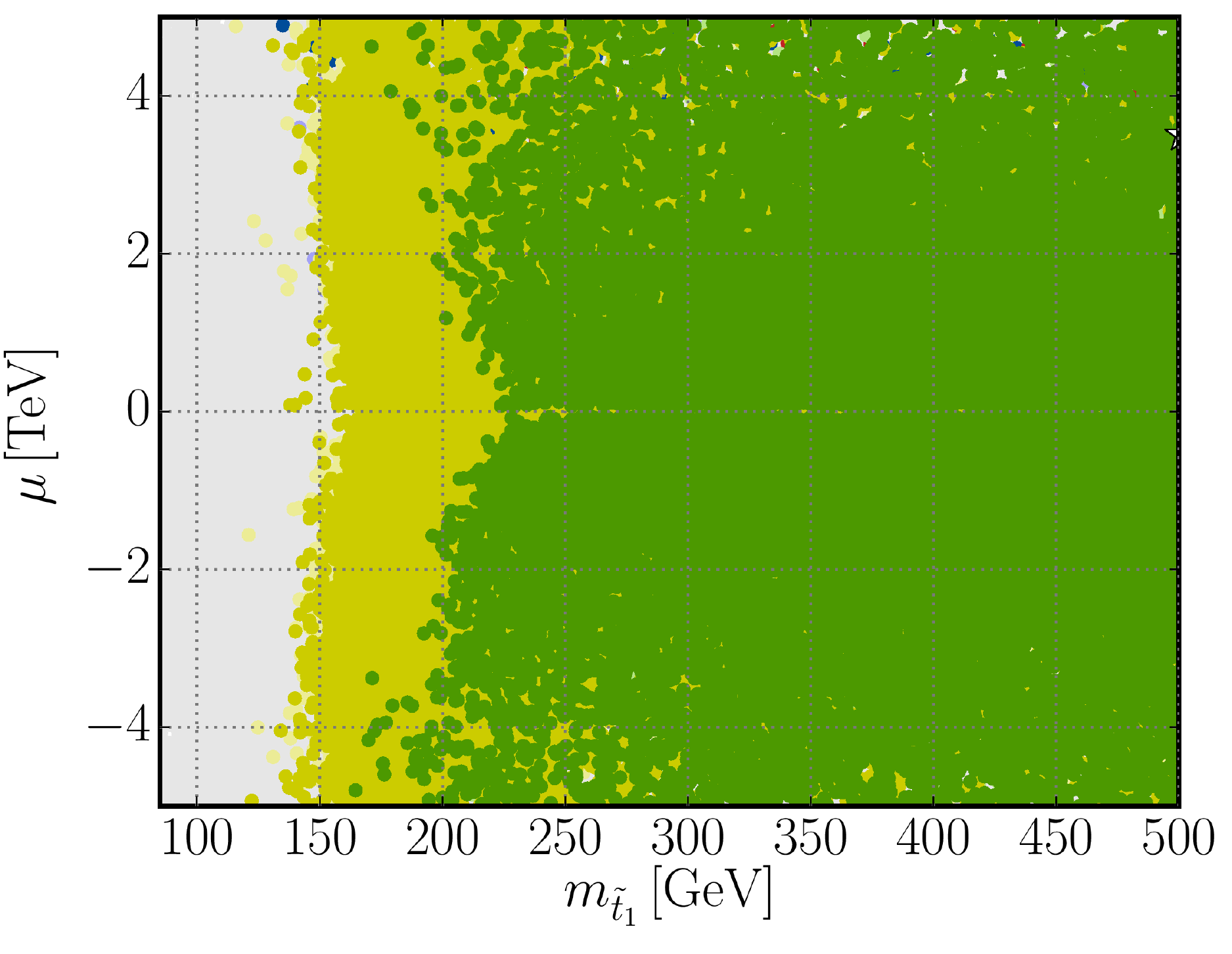}}
 \caption{Results for scenario~\textbf{D} `\emph{Non-decoupling effects}' for the parameter correlations of the light stop mass. Green [yellow] points indicate the $\CL{68}$ [$\CL{95}$] favored parameter points. Paler colors indicate points that are excluded at $\CL{95}$ by LHC Higgs searches, evaluated with \HBv{4.2.1}, or by the LEP stau mass limit, cf.~Section~\ref{Sect:Case2}. The blue [red] points (if visible) indicate the $\CL{68}$ [$\CL{95}$] favored points that do \emph{not} fulfill the vacuum metastability constraint, \eqn{Eq:CCBcondition}.}
 \label{Fig:Case4mstop}
\end{figure*}

The BF point is found at
\begin{align}
(m_{\sstop_1}, m_{\sstau_1}, M_A, \mu, \tan\beta) = (501\gev, 145\gev, 880\gev, 3.5\tev, 9.5)\,,
\end{align}
with a fit quality of $\chi^2/\text{ndf} = 68.1/81$, which is very similar to what we obtained in the previous fits where we allowed for a new physics Higgs decay mode. We show the correlation of the light stop mass, $m_{\sstop_1}$, with the other four scan parameters --- the pseudoscalar Higgs mass, $M_A$, the light stau mass, $m_{\sstau_1}$, the Higgsino mass parameter, $\mu$, and $\tan\beta$ --- in \fig{Fig:Case4mstop}. There is a slight tendency towards lower values of the pseudoscalar mass, $M_A\sim (400-500)\gev$, in the distribution of the most favored points featuring light stop masses, cf.~\fig{Fig:Case4mstop}(a). There is also a clear correlation with the light stau mass, shown in \fig{Fig:Case4mstop}(b): The lowest $\CL{95}$ allowed values of the stop mass, $m_{\sstop_1} \gtrsim 122\gev$, are obtained for small stau masses near the LEP limit. Disregarding the model-dependent stau mass LEP limit (cf.~Section~\ref{Sect:Case2}) weakens the lower stop mass limit to $116\gev$, while disregarding the vacuum metastability constraint does not impact the limit. Furthermore, the scan points with the lightest allowed stop mass values strongly favor low values of $\tan\beta \sim 2-5$ and tend to feature large $|\mu|$ values, as can be seen in \figs{Fig:Case4mstop}(c) and~\ref{Fig:Case4mstop}(d), respectively.

\begin{figure*}[t]
 \subfigure[~($M_A$, $\tan\beta$) plane.]{\includegraphics[width=0.44\textwidth]{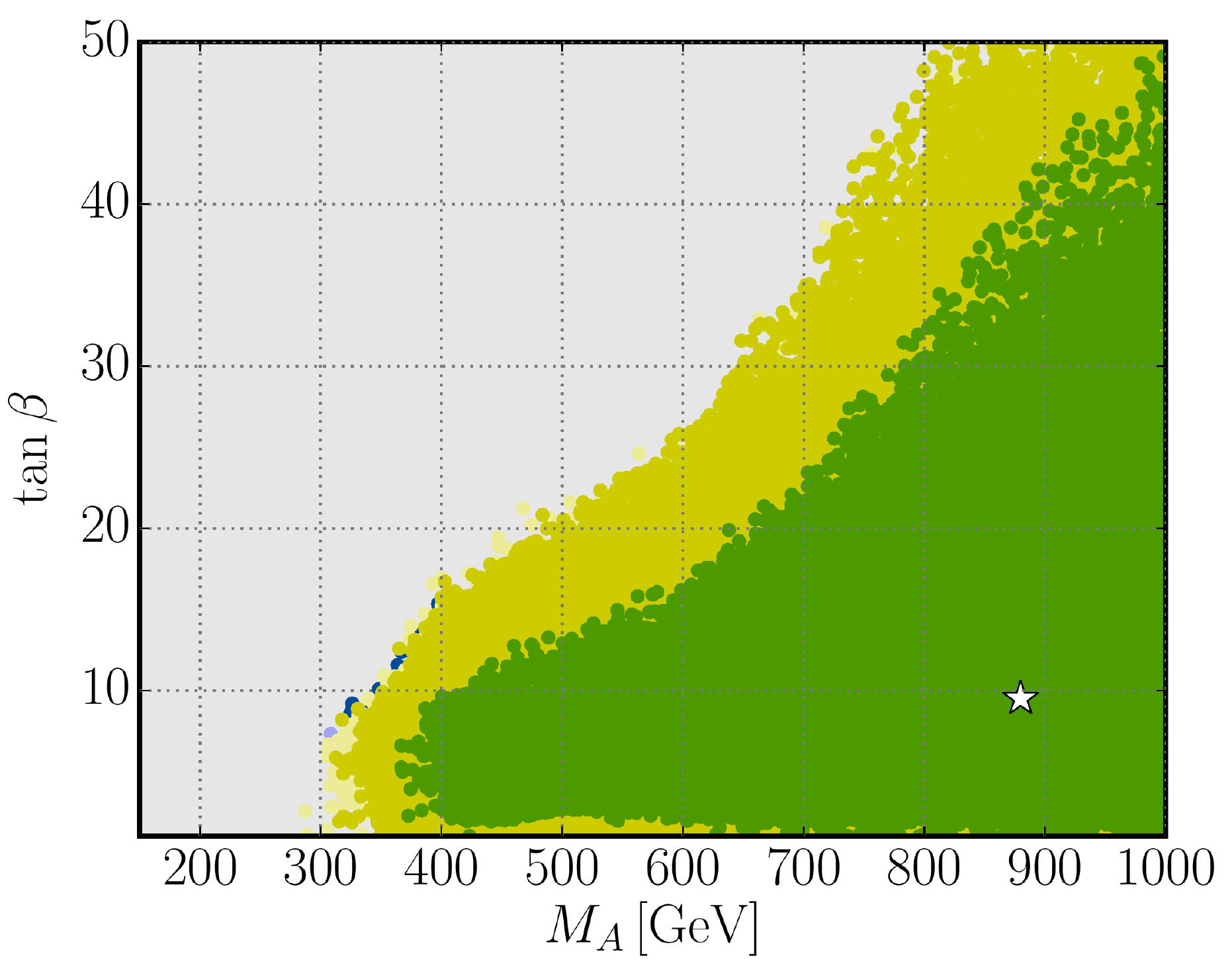}}\hfill
 \subfigure[~($M_A$, $m_{\sstau_1}$) plane.]{\includegraphics[width=0.44\textwidth]{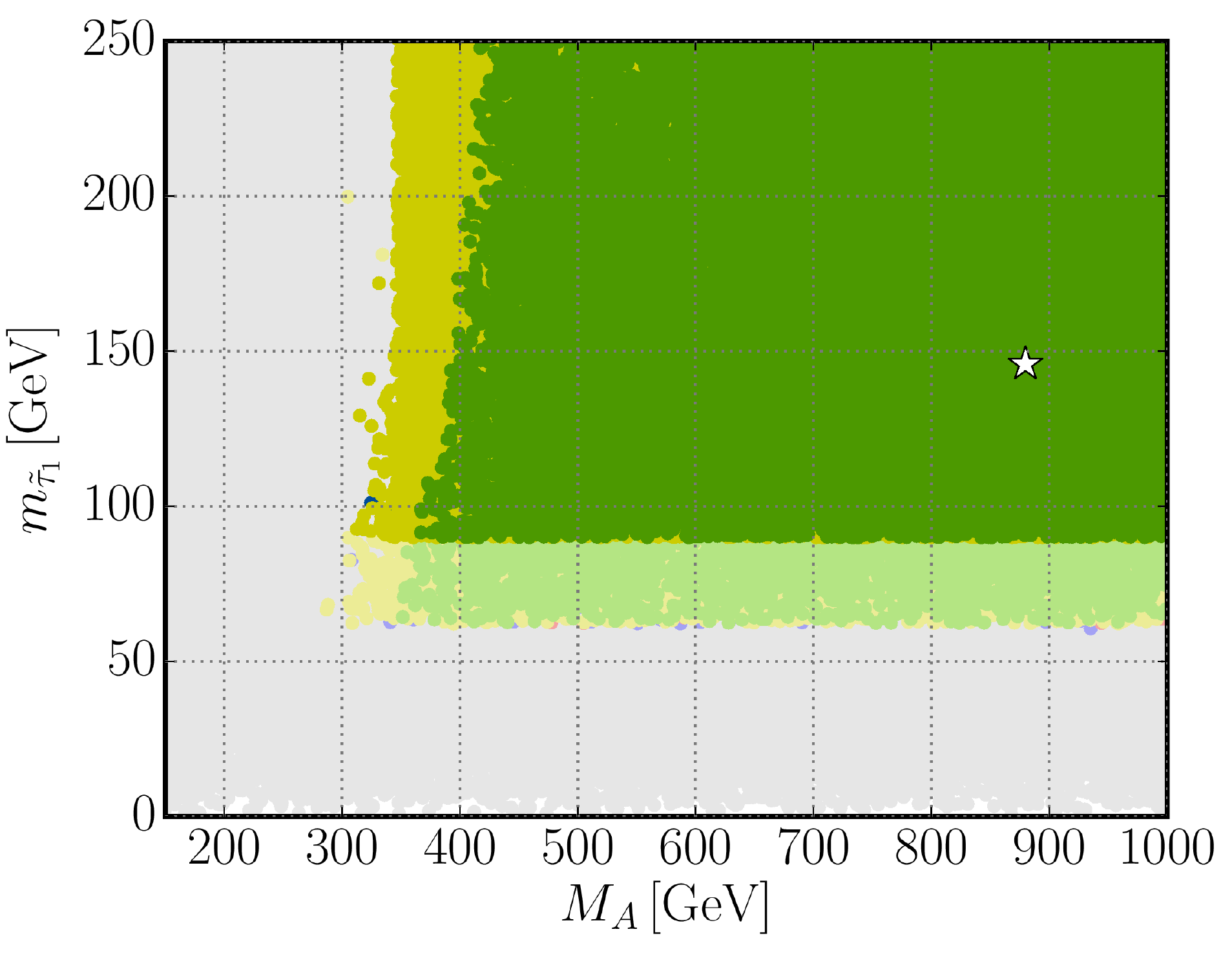}}\hfill
 \caption{Results for scenario~\textbf{D} `\emph{Non-decoupling effects}' for the correlations of the pseudoscalar Higgs mass, $M_A$, with $\tan\beta$ and with the light stau mass, $m_{\sstau_1}$. Color coding is the same as in \fig{Fig:Case4mstop}. The white star indicates the BF point.}
\label{Fig:Case4MA}
 \end{figure*}

In \fig{Fig:Case4MA} we show the results in the $(M_A, \tan\beta)$ plane, \fig{Fig:Case4MA}(a), and the $(M_A, m_{\sstau_1})$ plane, \fig{Fig:Case4MA}(b). For not too large $M_A$ values the high $\tan\beta$ region is excluded by the CMS $H/A\to \tau^+\tau^-$ search~\cite{Khachatryan:2014wca}. For $M_A$ values lower than $\sim 300\gev$ the light Higgs coupling to bottom quarks and $\tau$ leptons becomes too large to yield an acceptable fit. $M_A$ values around $\sim (300-350)\gev$ are only allowed if simultaneously the lighter stau state has a low mass, $m_{\sstau_1}\lesssim 150\gev$, as can be seen in \fig{Fig:Case4MA}(b).

\begin{figure*}[t]
 \subfigure[~$gg\to h\to VV$ rate as a function of $m_{\sstop_1}$.]{\includegraphics[width=0.44\textwidth]{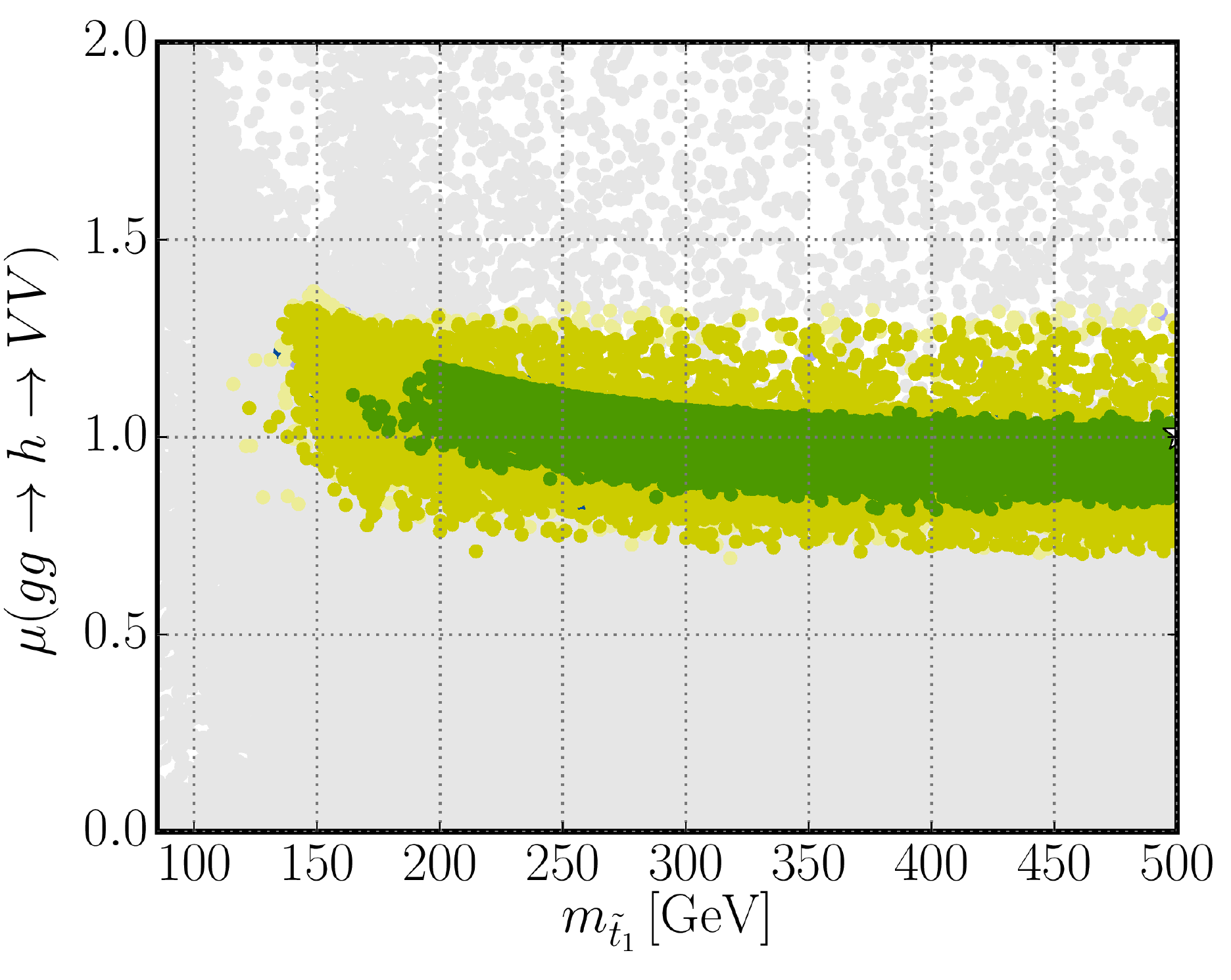}}\hfill
 \subfigure[~$Vh/\text{VBF}, h\to VV$ rate as a function of $m_{\sstop_1}$.]{\includegraphics[width=0.44\textwidth]{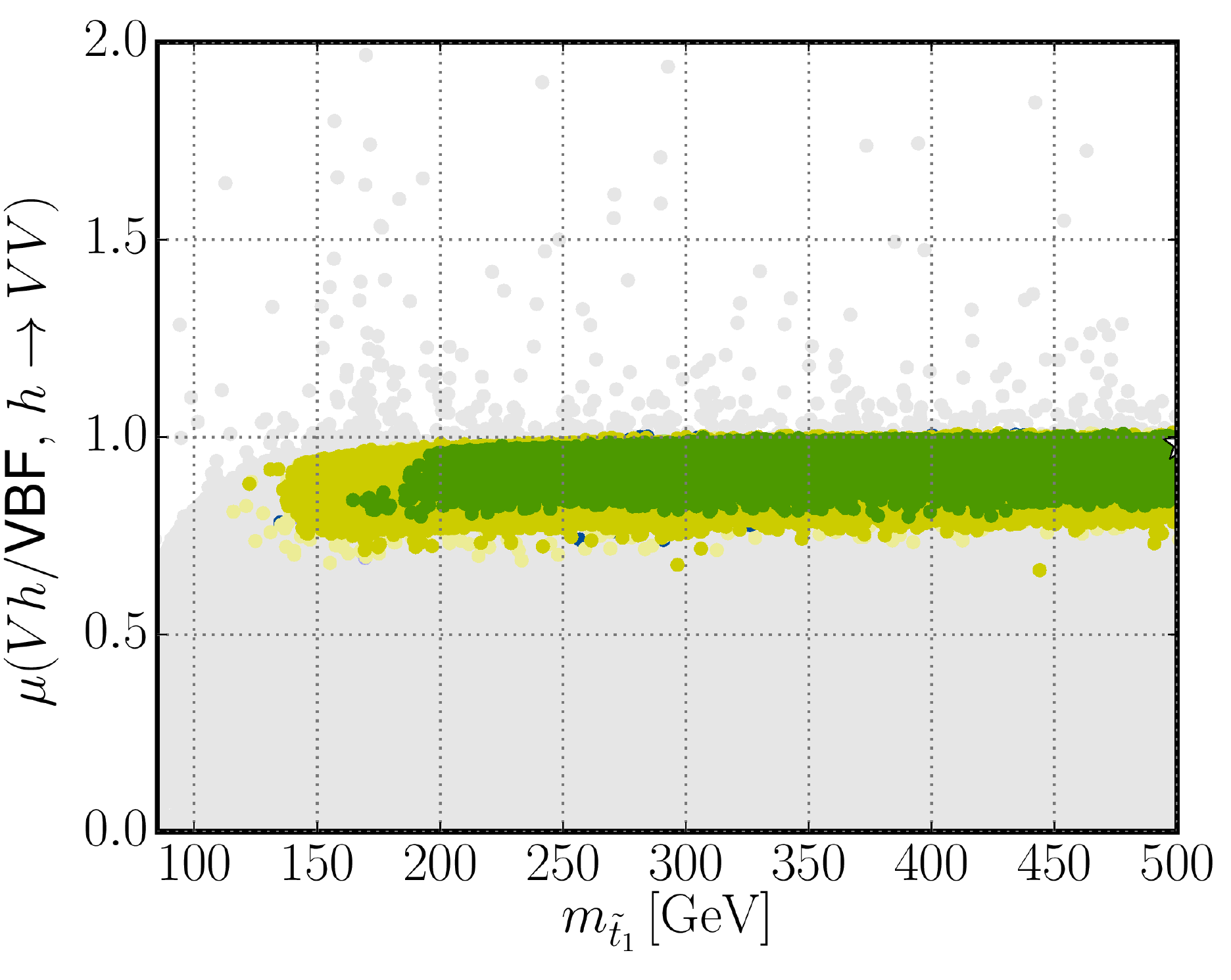}}
 \subfigure[~$gg\to h\to b\bar{b}$ rate as a function of $m_{\sstop_1}$.]{\includegraphics[width=0.44\textwidth]{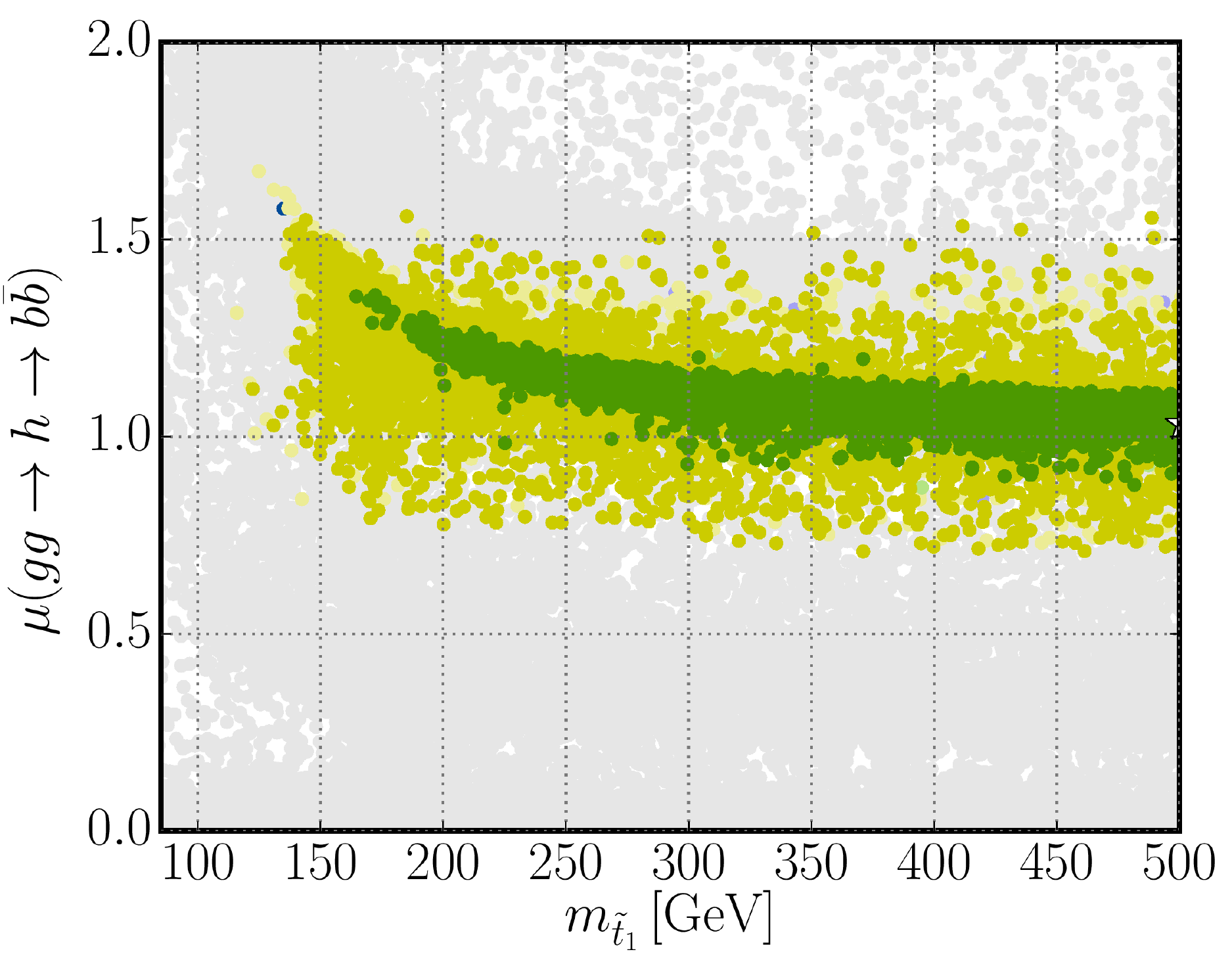}}\hfill
 \subfigure[~$Vh/\text{VBF}, h\to b\bar{b}$ rate as a function of $m_{\sstop_1}$.]{\includegraphics[width=0.44\textwidth]{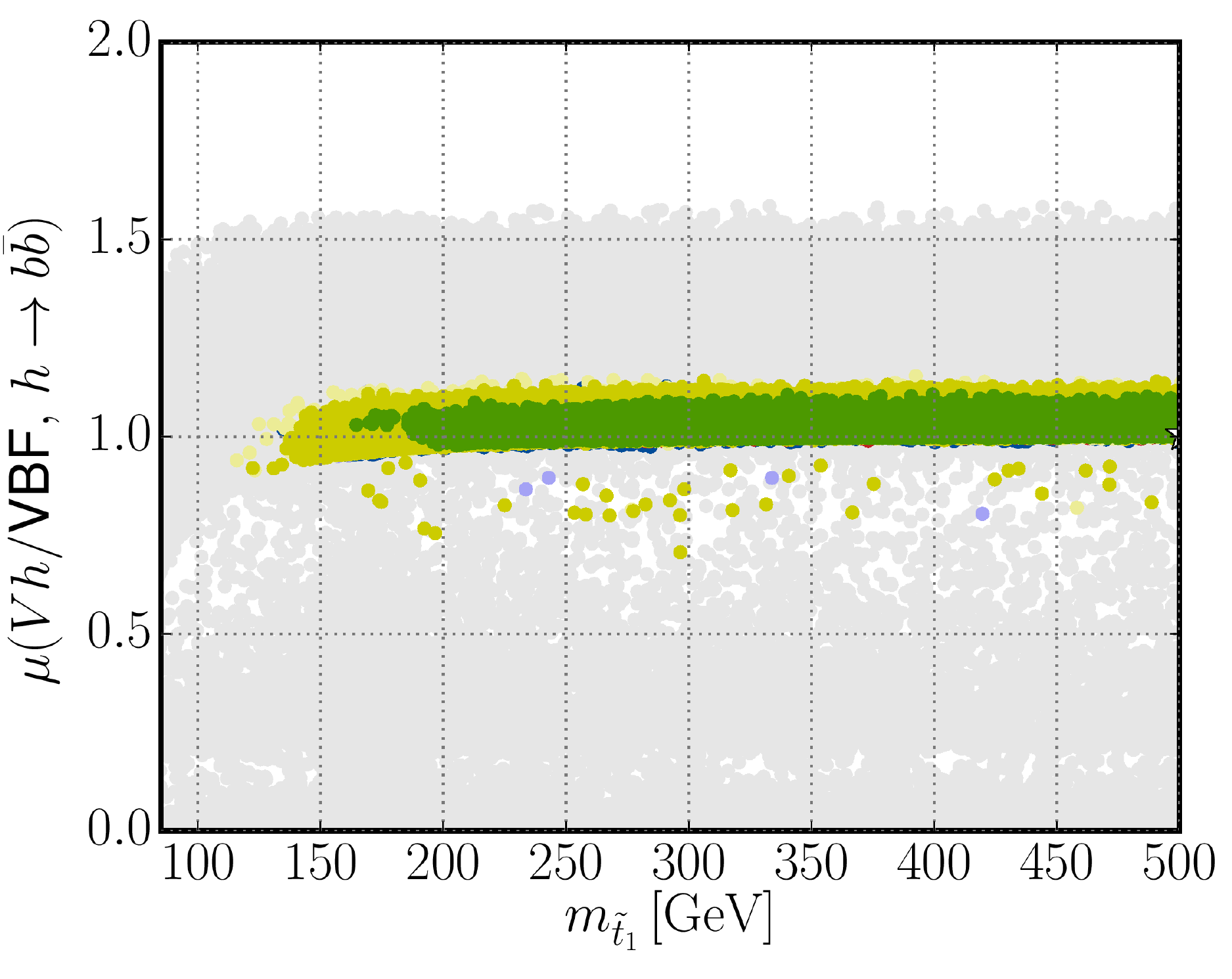}}
  \subfigure[~$gg\to h\to \gamma\gamma$ rate as a function of $m_{\sstop_1}$.]{\includegraphics[width=0.44\textwidth]{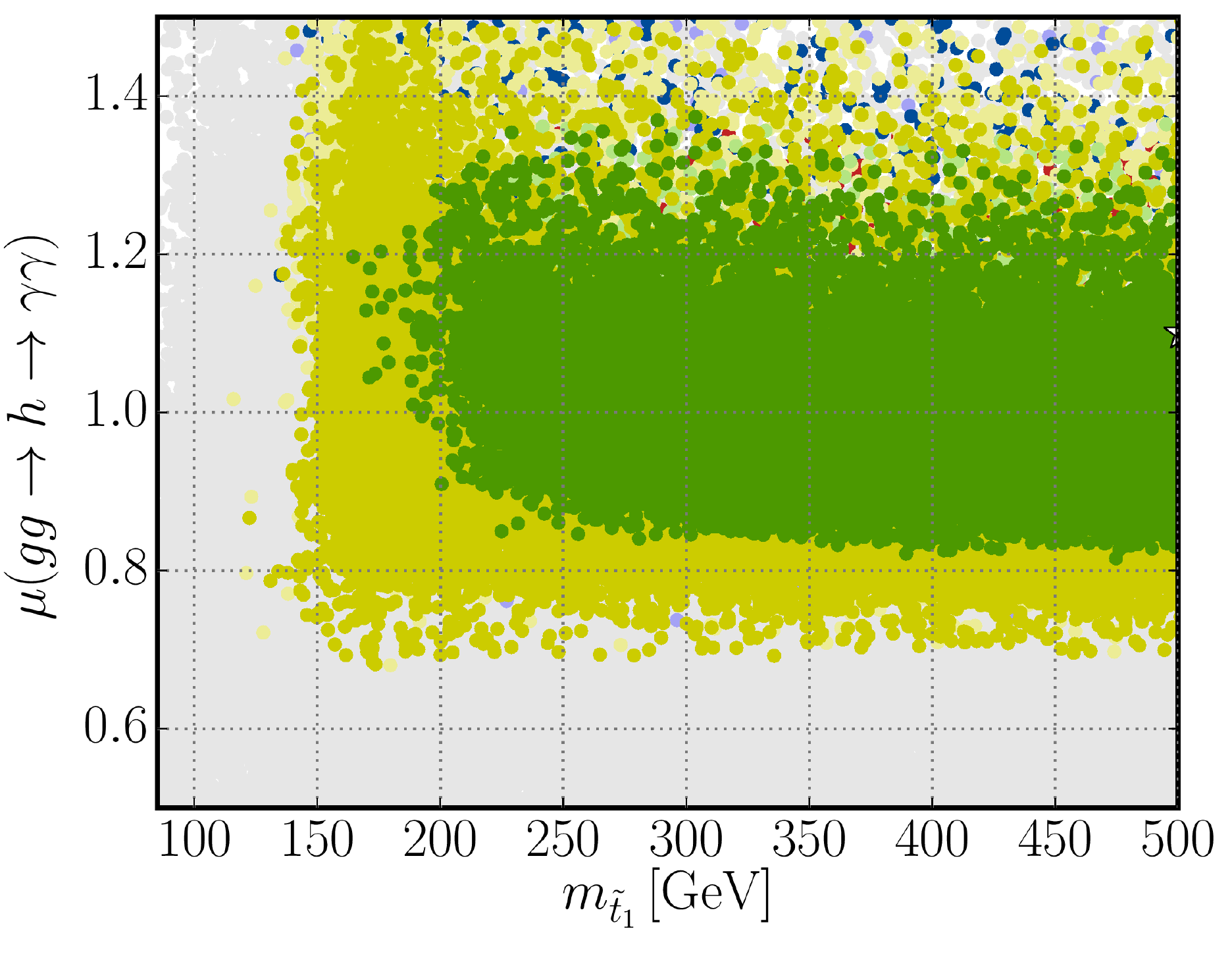}}\hfill
  \subfigure[~$Vh/\text{VBF}, h\to \gamma\gamma$ rate as a function of $m_{\sstop_1}$.]{\includegraphics[width=0.44\textwidth]{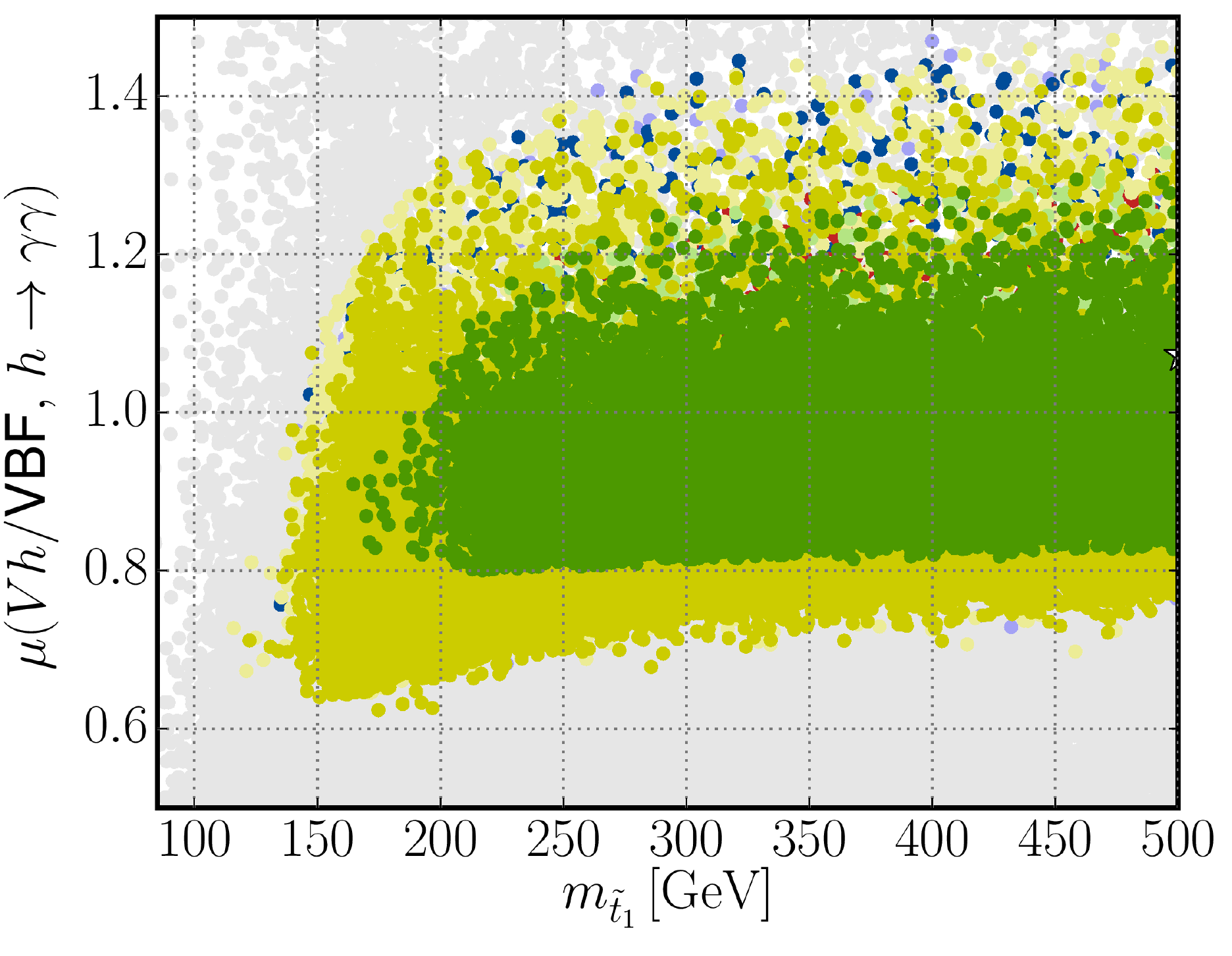}}
 \caption{SM normalized Higgs signal rates as a function of the light stop mass, $m_{\sstop_1}$, for the scenario~\textbf{D} `\emph{Non-decoupling effects}'. Color coding is the same as in \fig{Fig:Case4mstop}.}
\label{Fig:Case4_rates_mstop}
 \end{figure*}

\begin{figure*}[t]
 \subfigure[~$gg\to h\to VV$ rate as a function of $M_A$.]{\includegraphics[width=0.44\textwidth]{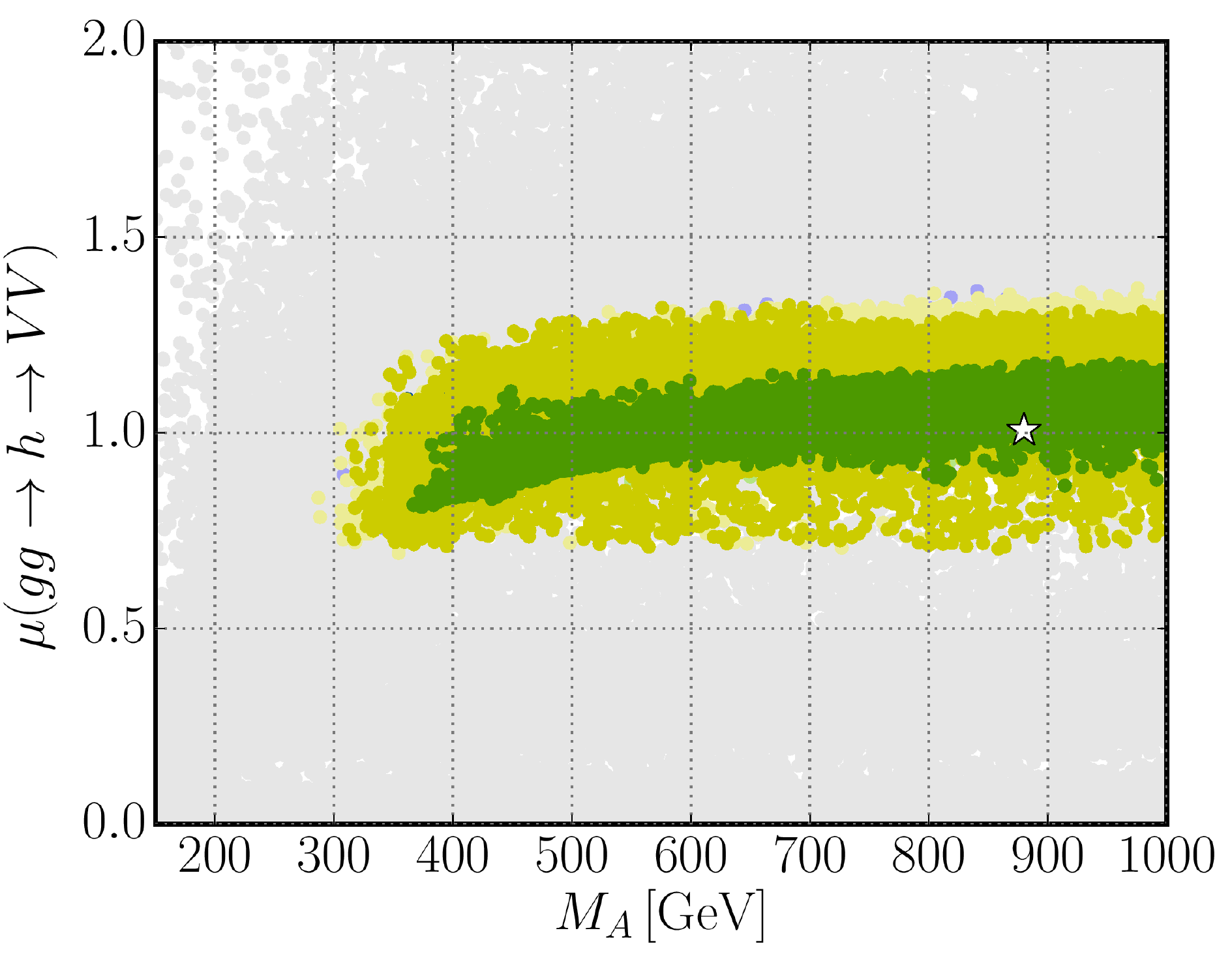}}\hfill
 \subfigure[~$Vh/\text{VBF}, h\to VV$ rate as a function of $M_A$.]{\includegraphics[width=0.44\textwidth]{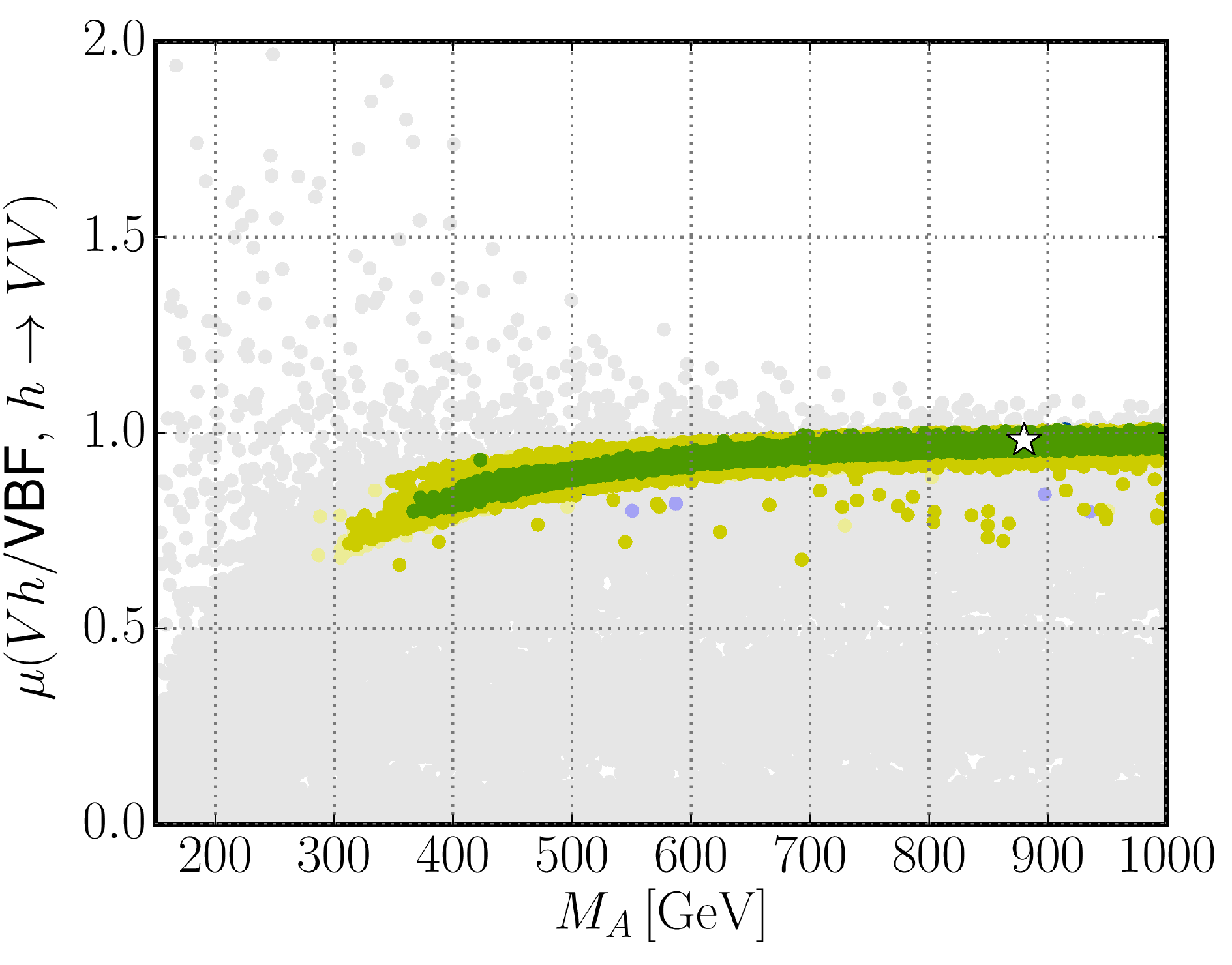}}
 \subfigure[~$gg\to h\to b\bar{b}$ rate as a function of $M_A$.]{\includegraphics[width=0.44\textwidth]{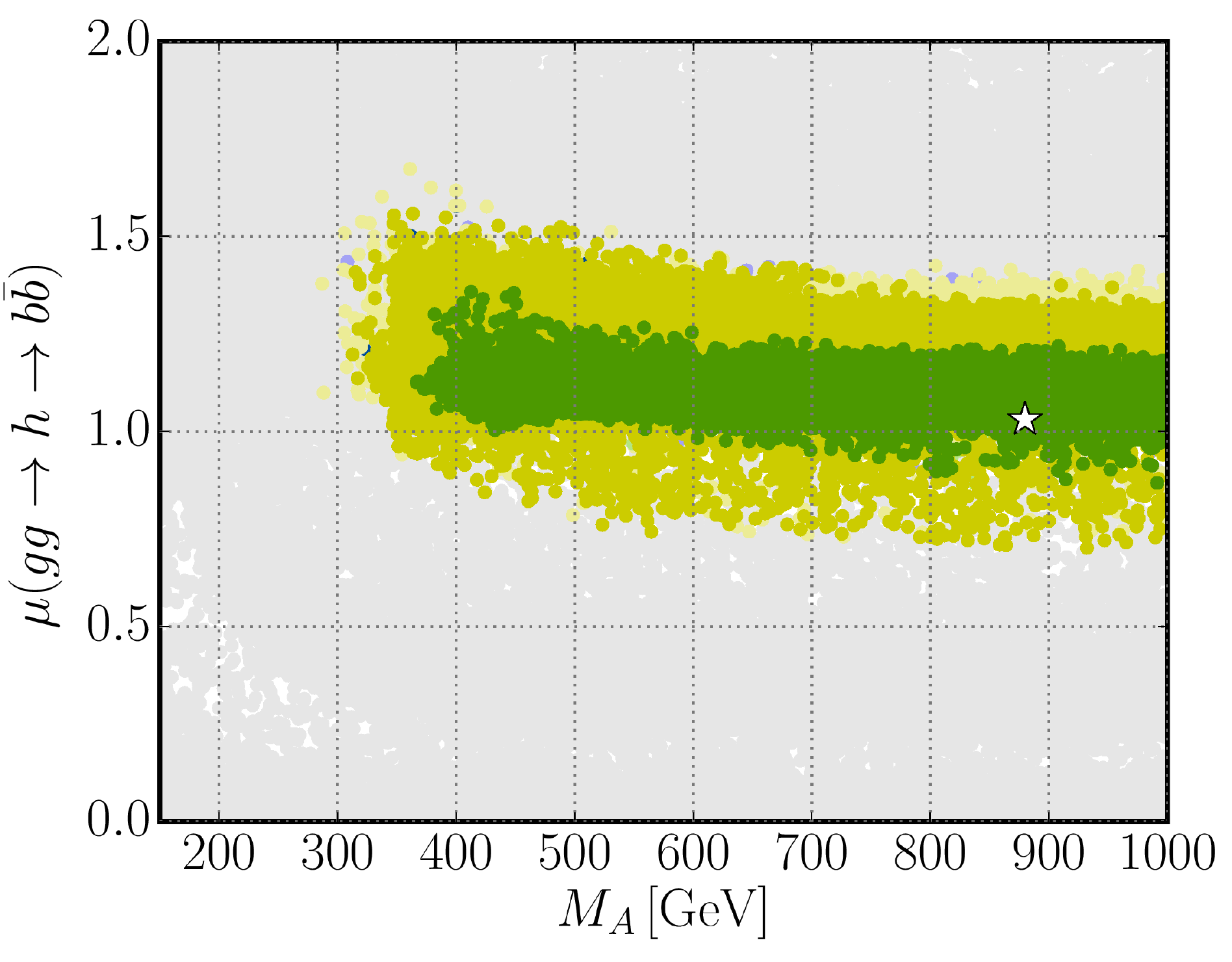}}\hfill
 \subfigure[~$Vh/\text{VBF}, h\to b\bar{b}$ rate as a function of $M_A$.]{\includegraphics[width=0.44\textwidth]{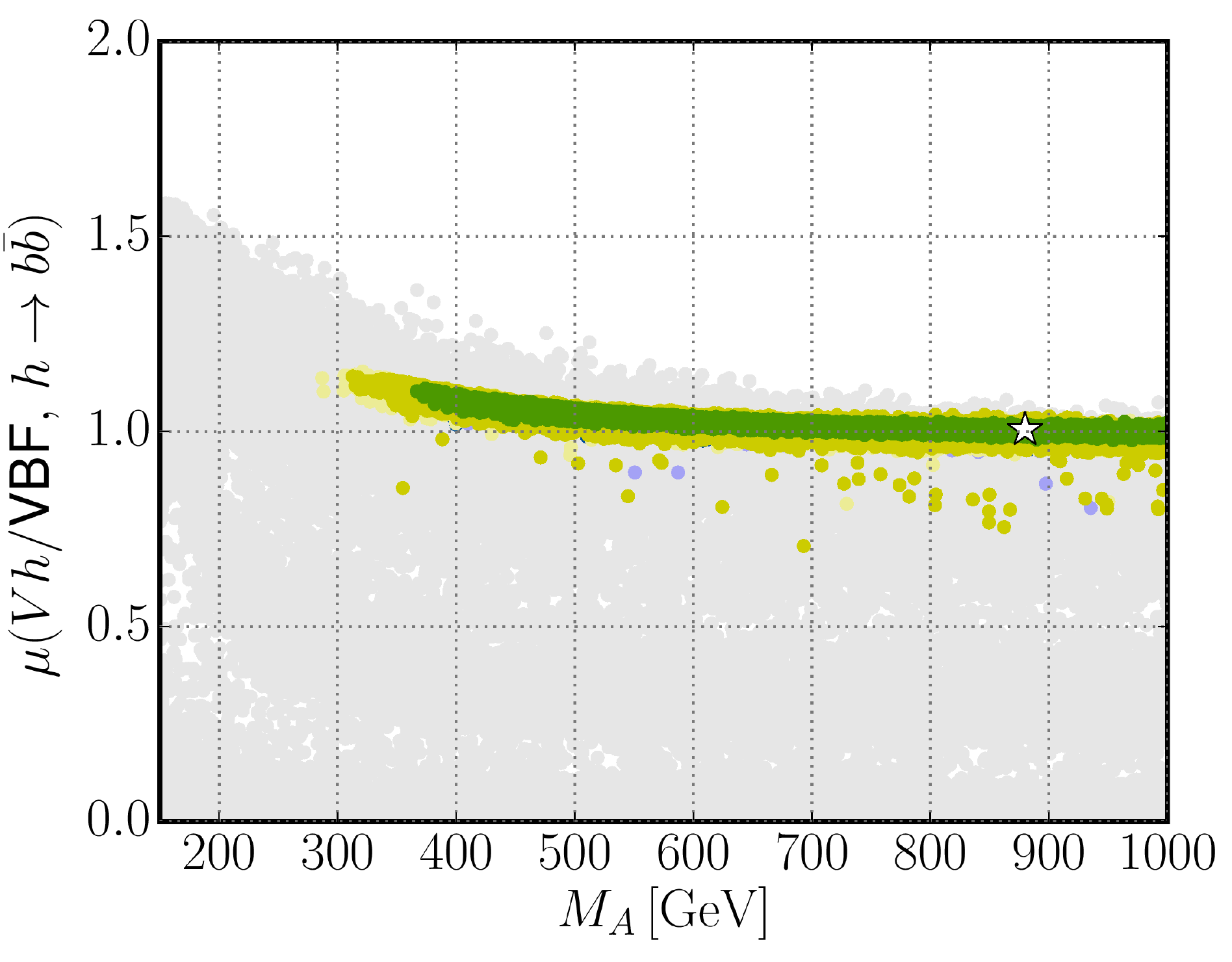}}
  \subfigure[~$gg\to h\to \gamma\gamma$ rate as a function of $M_A$.]{\includegraphics[width=0.44\textwidth]{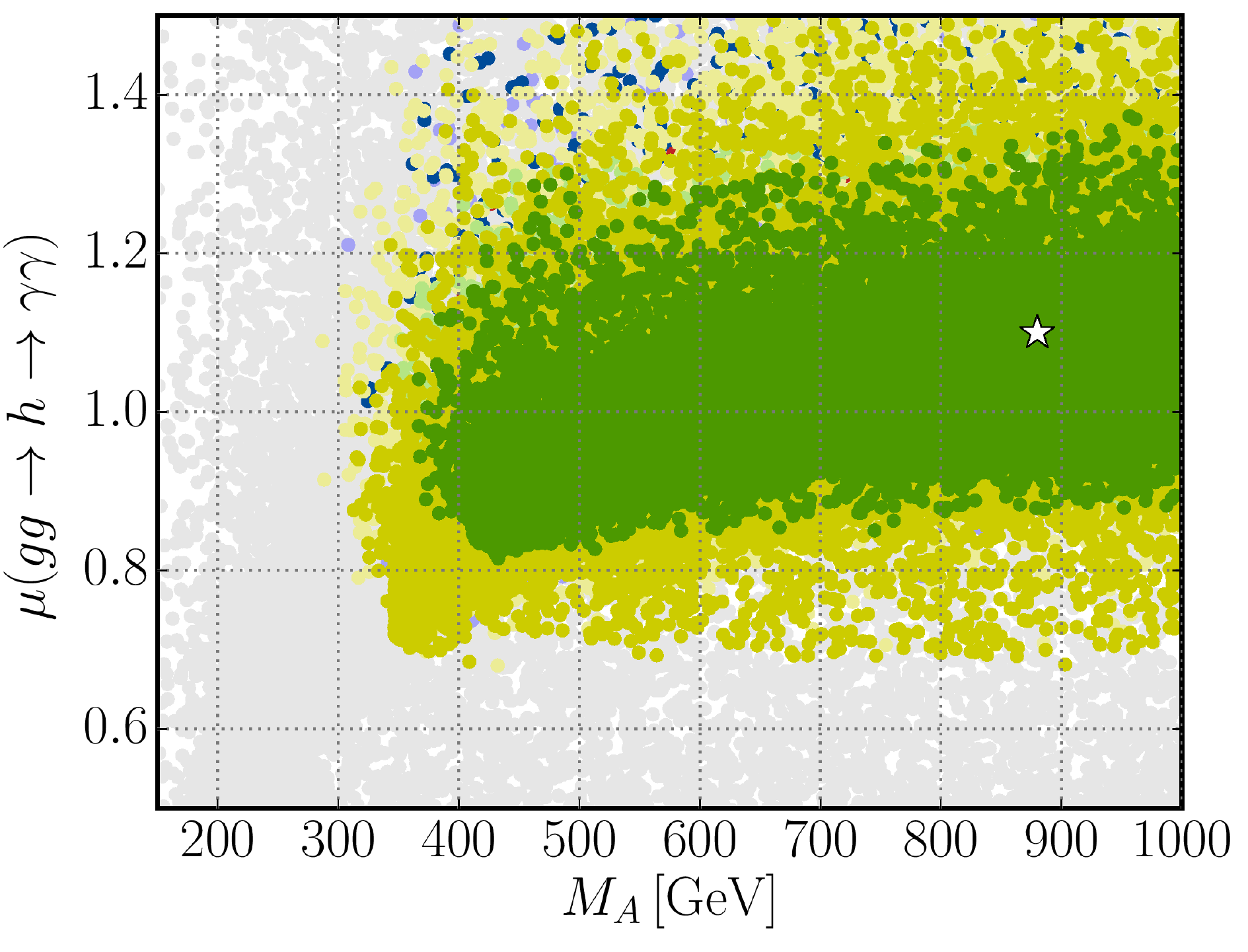}}\hfill
  \subfigure[~$Vh/\text{VBF}, h\to \gamma\gamma$ rate as a function of $M_A$.]{\includegraphics[width=0.44\textwidth]{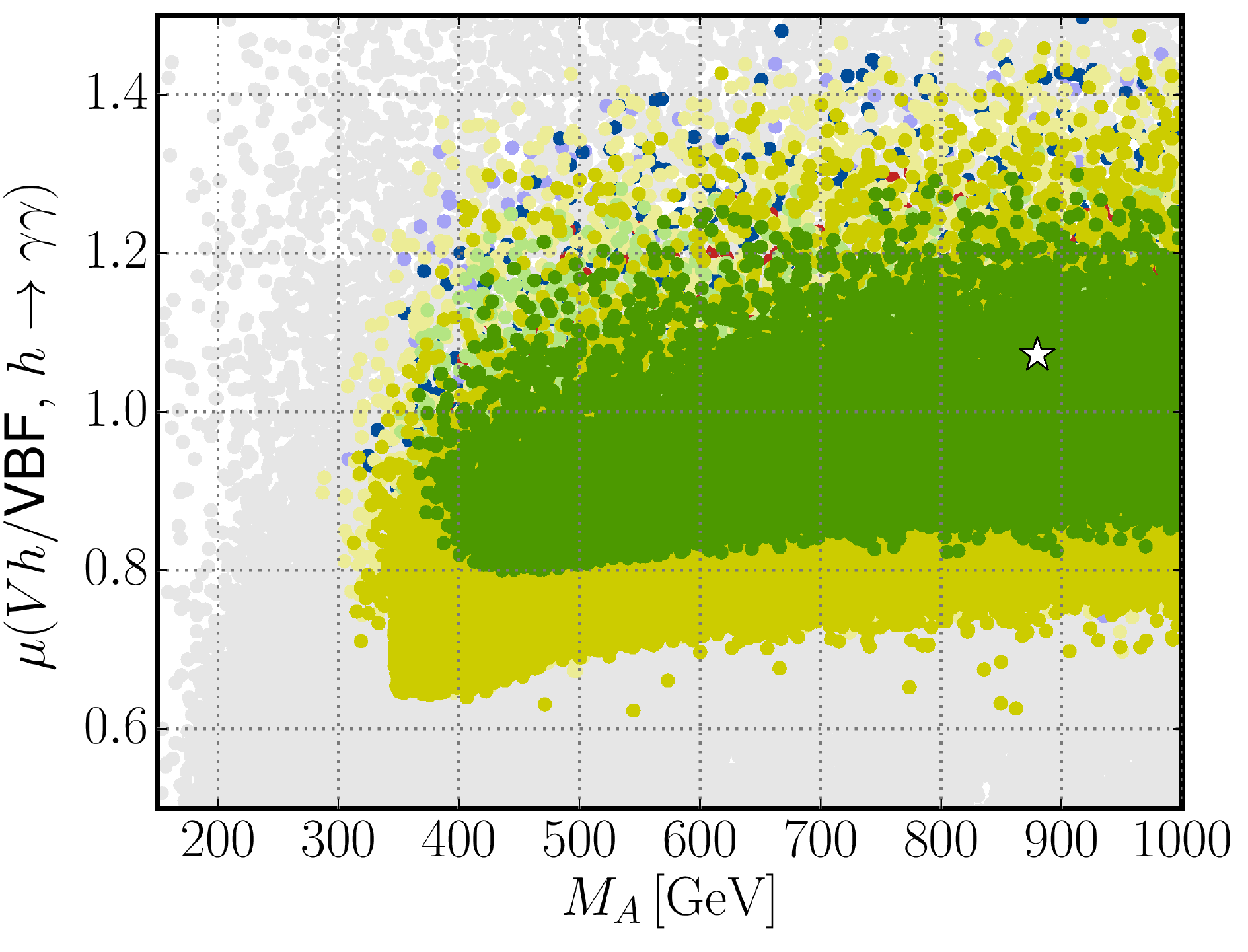}}
 \caption{SM normalized Higgs signal rates as a function of the pseudoscalar Higgs mass, $M_A$, for the scenario~\textbf{D} `\emph{Non-decoupling effects}'. Color coding is the same as in \fig{Fig:Case4mstop}.}
\label{Fig:Case4_rates_MA}
 \end{figure*}

The dependence of the predicted Higgs rates on the light stop mass, $m_{\sstop_1}$, and the pseudoscalar Higgs mass, $M_A$, is illustrated in \figs{Fig:Case4_rates_mstop} and~\ref{Fig:Case4_rates_MA}, respectively. The (idealized) SM normalized Higgs rates are defined according to \eqn{Eq:mu_definition}.
The figures show the rates for the channels for Higgs production in gluon fusion, $gg\to h$, or vector boson fusion (VBF) and vector boson associated production ($Vh$), with subsequent decay of the Higgs boson into $VV$, $b\bar{b}$ and $\gamma\gamma$.

At low stop masses the rates for the gluon fusion initiated signal channels increase. However, due to the freedom in the remaining scan parameters, SM-like rates, $\mu\approx 1$, can still be obtained at low stop masses. As the cross sections for the $\text{VBF}$ and $Vh$ Higgs production processes are unaffected by the light stop mass, the rates $\mu(Vh/\text{VBF}, h\to VV)$ and $\mu(Vh/\text{VBF}, h\to b\bar{b})$ reflect the $m_{\sstop_1}$ dependence of the branching fractions for $h\to VV$ and $h\to b\bar{b}$. While the former decay mode becomes suppressed by the increasing $h\to gg$ partial widths, the latter can be compensated by a slightly enhanced $hb\bar{b}$ coupling. Moreover this enhancement in the $hb\bar{b}$ coupling reduces further $\mathrm{BR}(h\to VV)$ (and all other branching fractions except the ones for $h\to b\bar{b}$ and potentially $h\to\tau^+\tau^-$, as the $h\tau^+\tau^-$ coupling is identical to the $hb\bar{b}$ coupling at tree-level).

The $M_A$ dependence of the Higgs rates exhibits an enhancement of the $h\to b\bar{b}$ channels and a reduction of the $h\to VV$ channels towards lower $M_A$ values. This is clearly correlated with the non-decoupling behavior of the light Higgs couplings at lower $M_A$ values, where its Yukawa couplings to bottom-quarks and $\tau$-leptons become enhanced (given $\tan\beta>1$) and the coupling to vector bosons becomes reduced. As these effects compensate each other in the rate for the $\text{VBF}/Vh, h\to b\bar{b}$ channel, which is experimentally observable at the LHC, this rate stays rather SM-like, see \fig{Fig:Case4_rates_MA}(d). The $h\to\gamma\gamma$ rates also tend to decrease at lower $M_A$ due to the reduced light Higgs coupling to the $W$ boson, however, the additional contribution from a light stau to $\Gamma(h\to \gamma\gamma)$ can lift this rate to a SM-like value. As a result, a low stau mass and large $|\mu|$ values are required to allow for low $M_A$ values, as pointed out above. This is particularly needed if, simultaneously, the stop mass is very low, since the $\sstop_1$ contributions tend to lower the $h\to \gamma\gamma$ partial width.

The suppression of the light Higgs decay rates due to the increase of the $h\to b\bar{b}$ partial width at lower $M_A$ values generally allows for a larger enhancement of the cross section in the dominant Higgs production mode -- the gluon fusion channel. Therefore, it is easier to obtain a viable parameter point with a very light stop at smaller $M_A$ values, as observed in \fig{Fig:Case4mstop}(a). 

\begin{figure*}
\centering
\includegraphics[width=0.44\textwidth]{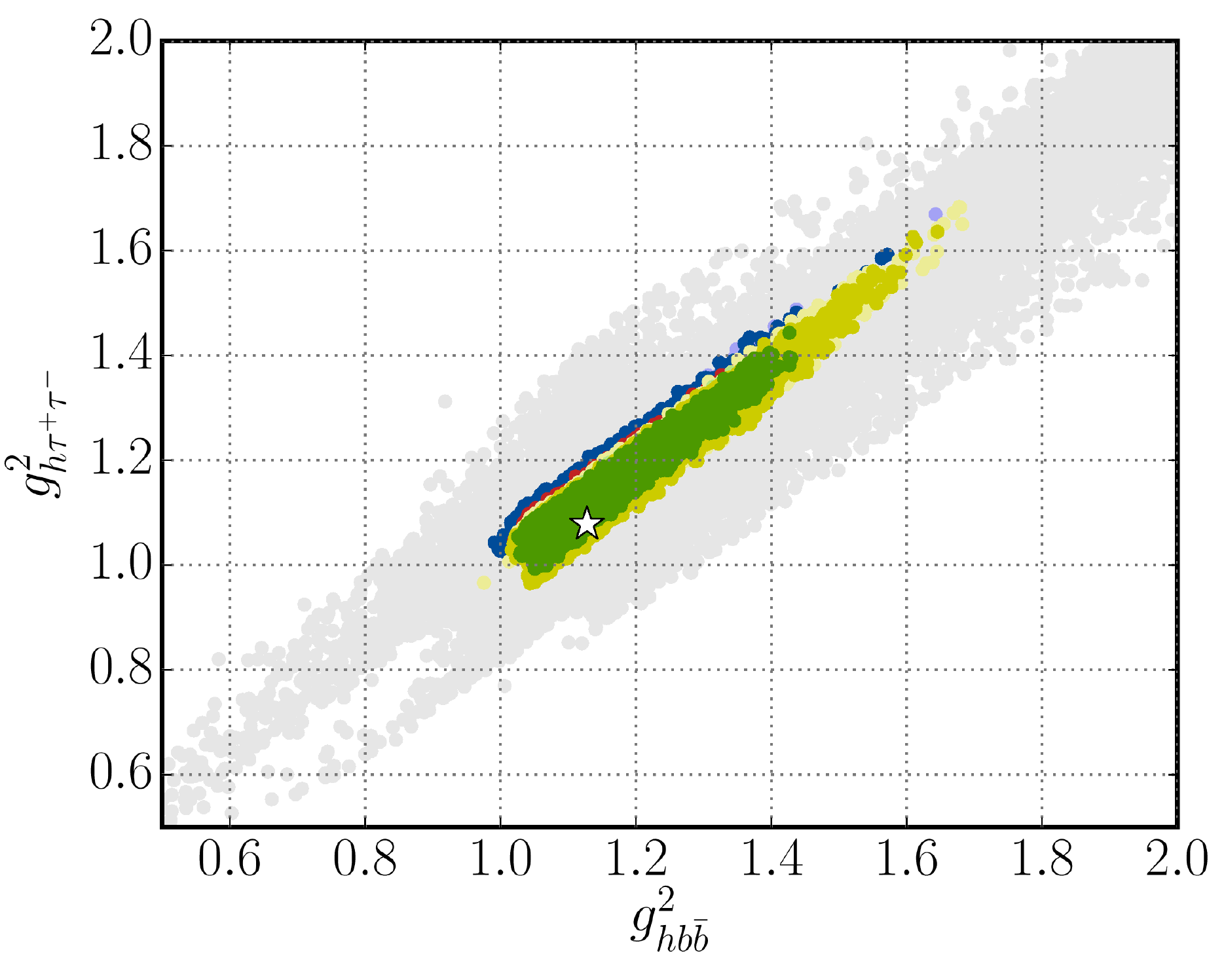}
 \caption{Correlation between the squared (SM normalized) light Higgs coupling to bottom-quarks, $g_{hb\bar{b}}^2$, and $\tau$-leptons, $g_{h\tau^+\tau^-}^2$, for the scenario~\textbf{D} `\emph{Non-decoupling effects}'. Color coding is the same as in \fig{Fig:Case4mstop}.}
\label{Fig:Case4_couplings}
\end{figure*}

We show the correlation between the squared SM-normalized light Higgs couplings to bottom-quarks, $g^2_{hb\bar{b}}$, and $\tau$-leptons, $g^2_{h\tau^+\tau^-}$, in \fig{Fig:Case4_couplings}. At tree-level, these SM normalized couplings are identical and enhanced for large $\tan\beta$. However, radiative corrections --- the aforementioned $\Delta_b$ and $\Delta_\tau$ corrections -- lead to a splitting between these couplings and can in principle lead to values $< 1$. It is remarkable to see that for most of the allowed points both couplings are above the corresponding SM value, with possible enhancements of the squared SM normalized couplings ranging up to $\sim 60\%$. Note also, that the SM normalized $h\tau^+\tau^-$ coupling tends to be somewhat smaller than the SM normalized $hb\bar{b}$ coupling. The (SM normalized) light Higgs coupling to vector bosons, $g_{hVV} = \sin(\alpha - \beta)$, deviates at most by $\sim 1\%$ from the SM prediction in the $\CL{95}$ region.

\begin{figure*}
 \subfigure[~Branching fraction for the decay $H\to\sstop_1 \sstop_1^*$ as a function of $m_{\sstop_1}$.]{\includegraphics[width=0.44\textwidth]{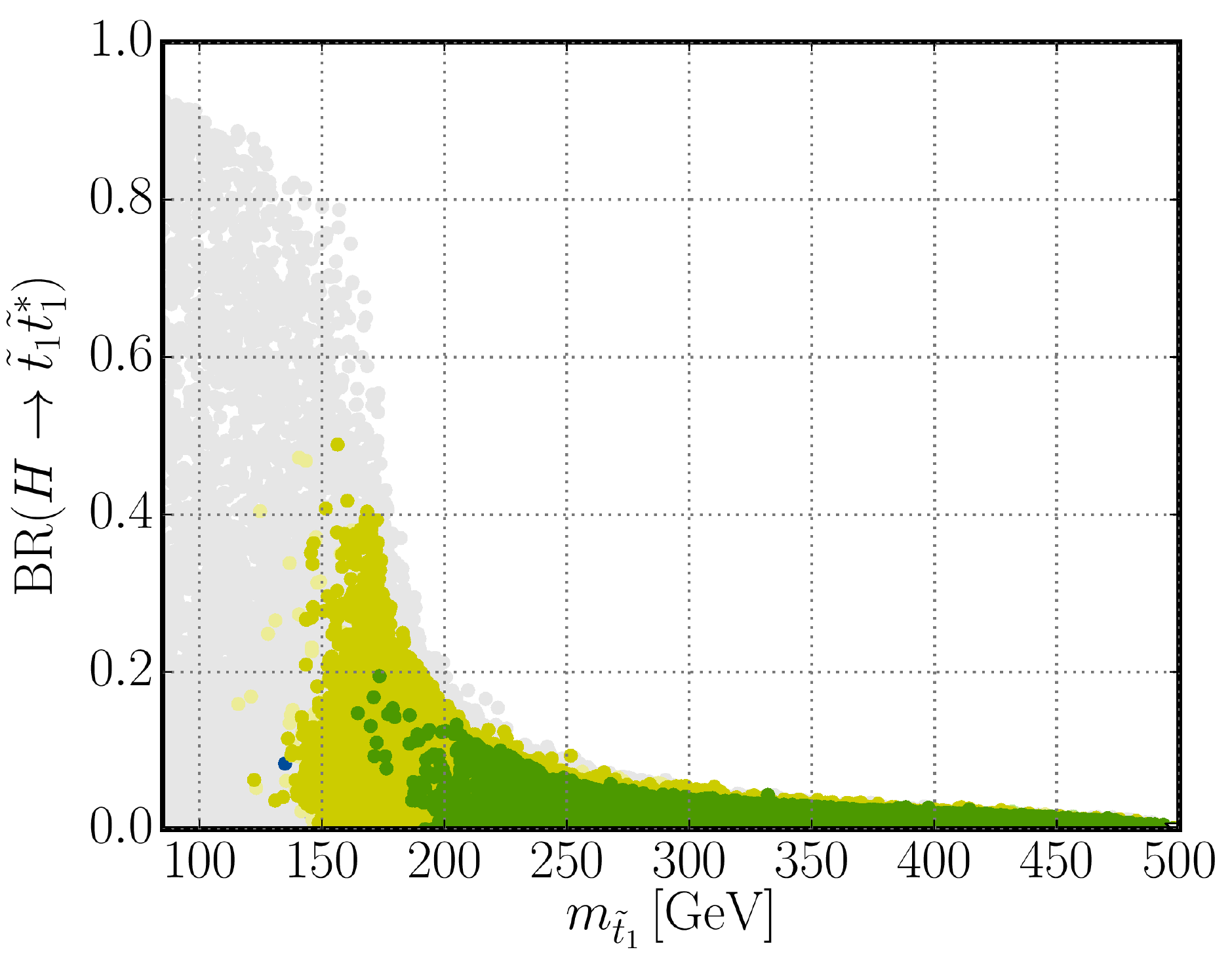}}\hfill
 \subfigure[~Branching fraction for the decay $H\to\sstau_1^+ {\sstau}_1^-$ as a function of $m_{\sstop_1}$.]{\includegraphics[width=0.44\textwidth]{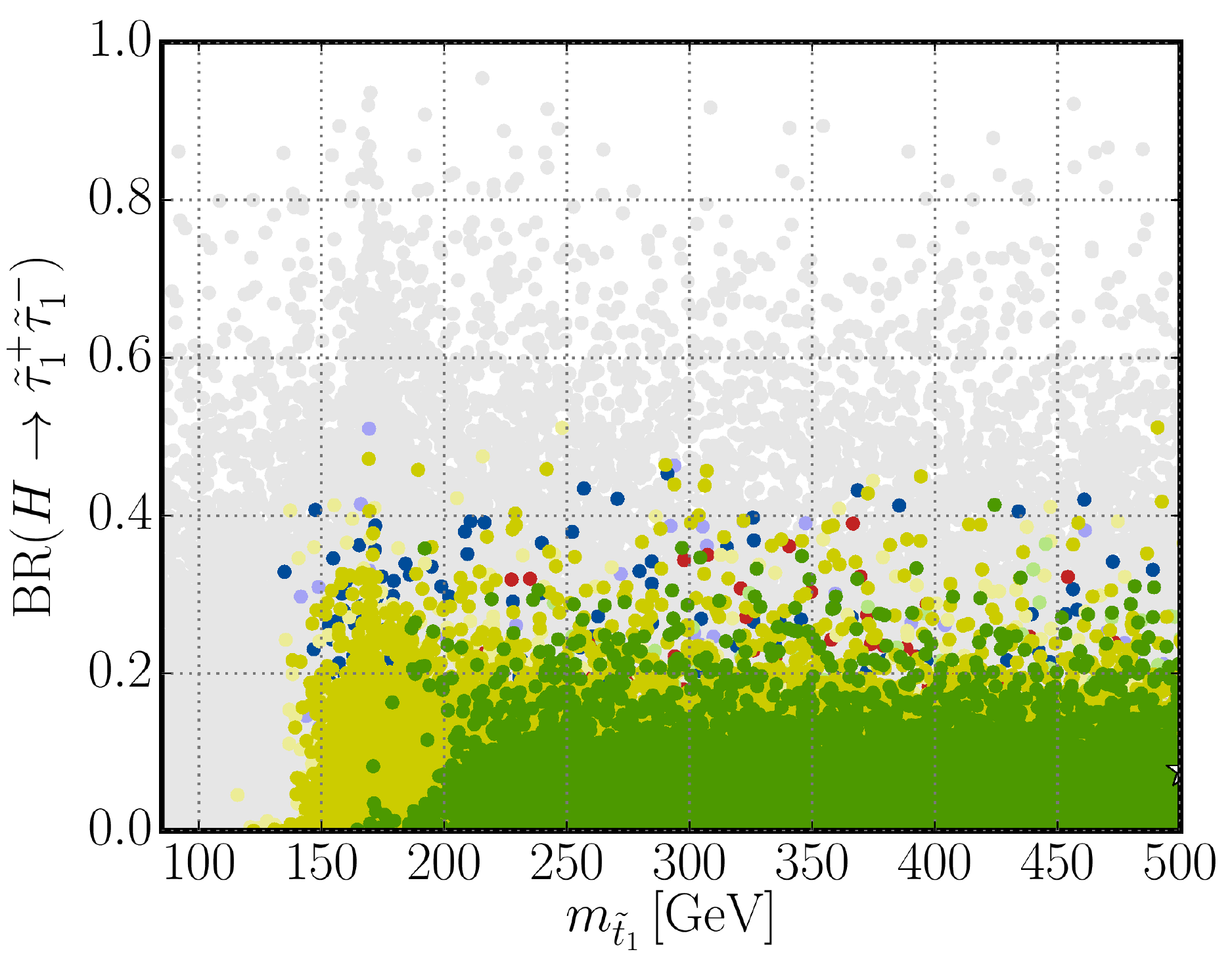}}\\
 \subfigure[~Branching fraction for the decay $H\to\sstop_1 \sstop_1^*$ as a function of $M_A$.]{\includegraphics[width=0.44\textwidth]{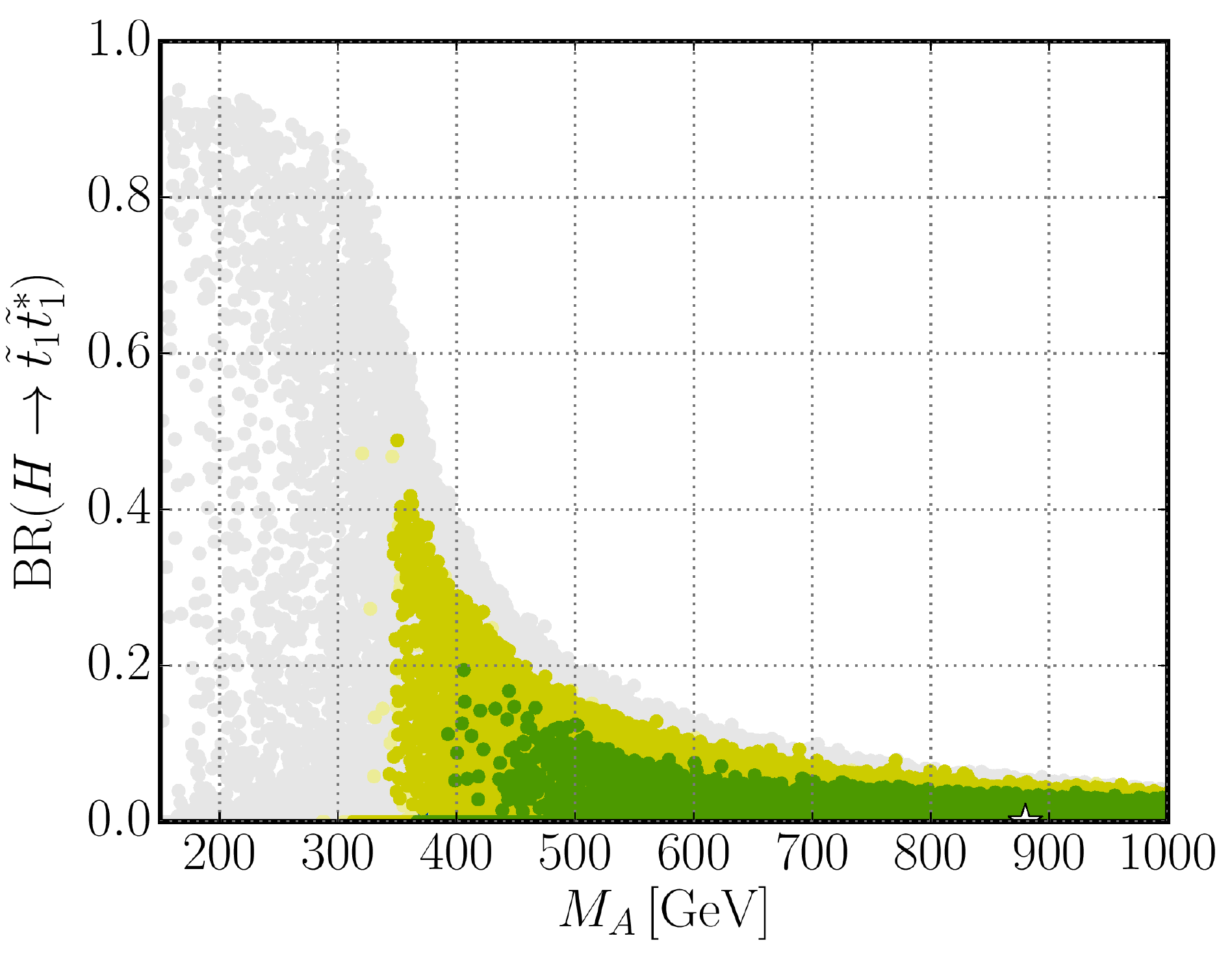}}\hfill
 \subfigure[~Branching fraction for the decay $H\to\sstau_1^+ {\sstau}_1^-$ as a function of $M_A$.]{\includegraphics[width=0.44\textwidth]{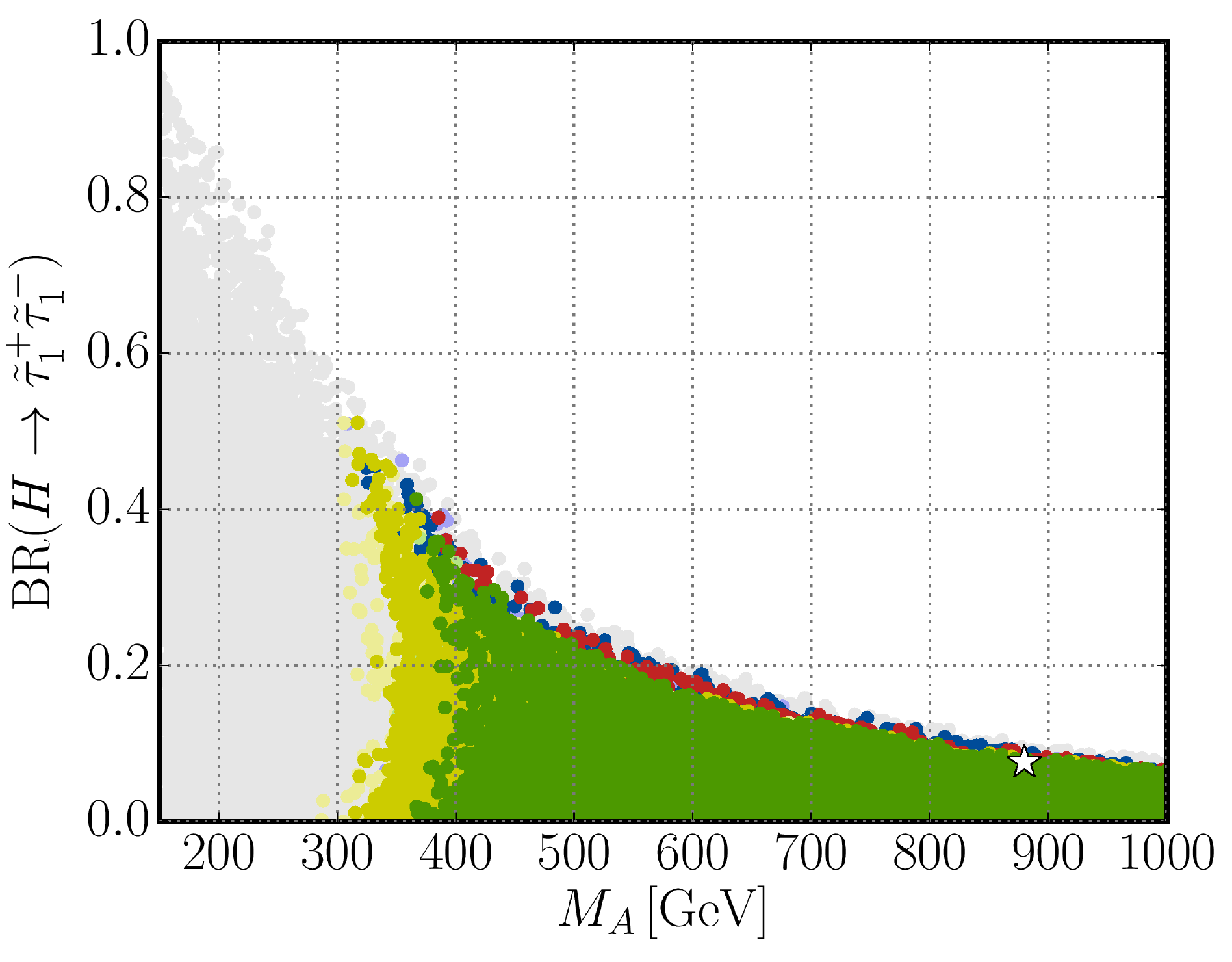}}\\
 \caption{Branching fractions for the heavy Higgs decays to light stops or light staus as a function of the light stop mass, $m_{\sstop_1}$, and pseudoscalar Higgs mass, $M_A$, for the scenario~\textbf{D} `\emph{Non-decoupling effects}'. Color coding is the same as in \fig{Fig:Case4mstop}.}
\label{Fig:Case4_BRs}
 \end{figure*}

Finally, we briefly want to discuss the possible phenomenology of the heavy $\CP$-even Higgs boson $H$ in this scenario. We show in \fig{Fig:Case4_BRs} the branching fractions for the heavy Higgs decays to light stops, $\mathrm{BR}(H\to \sstop_1 \sstop_1^*)$, and to light staus, $\mathrm{BR}(H\to \sstau_1^+ \sstau_1^-)$, as a function of the light stop mass, $m_{\sstop_1}$, and pseudoscalar Higgs mass, $M_A$. The latter is roughly equal to the mass of the heavy Higgs boson, $m_H \approx M_A$. Both branching fractions can become quite sizable, potentially reaching values up to $\sim 40\%$ for small $M_A \sim (300 - 400)\gev$. Evidently, the stop and stau masses have to be below $m_H/2 \approx M_A/2$ for this decay to be kinematically allowed, thus these high values can only be reached for $m_{\sstau_1}$ or $m_{\sstop_1}$ below around $\sim (150-200)\gev$. Although the presence of these decays is not a clear prediction of this light stop scenario, it still offers a genuine possible signature that should be searched for in the upcoming LHC Run~2 program.

\begin{figure}
 \centering
 \includegraphics[width=0.6\textwidth]{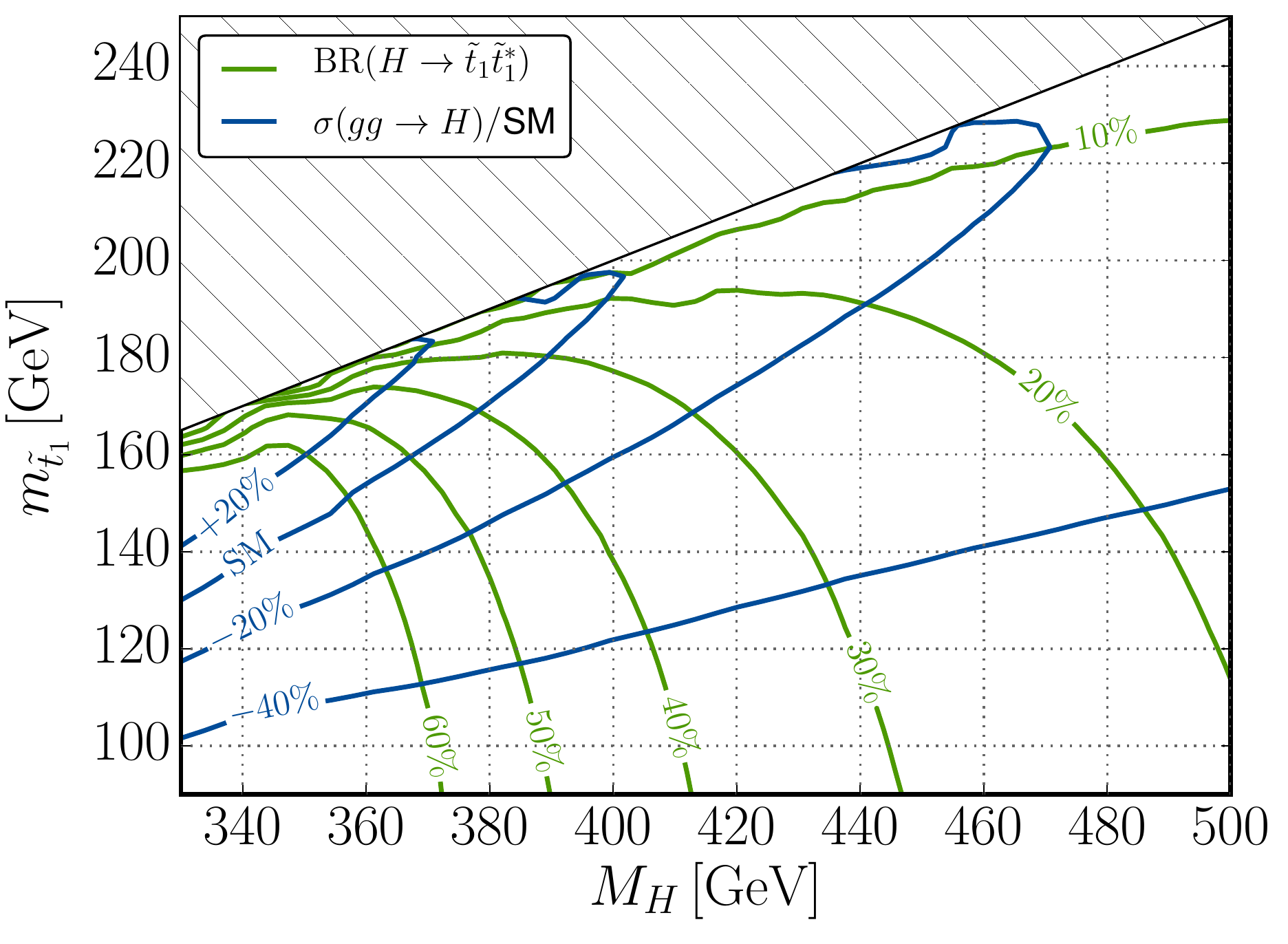}
 \caption{Branching ratio $\text{BR}(H\to \sstop_1 \sstop_1^*)$ and SM normalized gluon-fusion production cross section, $\sigma(gg\to H)/\text{SM}$ for the benchmark scenario `\texttt{HHLS}', given by green and blue contours, respectively, with values as indicated by the labels (the blue labels give the deviation from the SM prediction). In the hatched region the decay $H\to \sstop_1\sstop_1^*$ is kinematically inaccessible.}
 \label{Fig:HHLS}
 \end{figure}

As a benchmark scenario for future LHC searches we present in \fig{Fig:HHLS} the (SM normalized) gluon fusion cross section, $\sigma(gg\to H)/\text{SM}$, and $\text{BR}(H\to\sstop_1\sstop_1^*)$ predictions for a simplified model that we denote `Heavy Higgs to Light Stop' (\texttt{HHLS}) scenario here. It is defined by the following parameters:
\begin{align}
M_2 = \mu = 1\tev,\; \tan\beta = 1.5,\; X_t =0,\; A_t = A_b\,, \nonumber \\
 \SML{3}=\SME{3} = 150\gev,\; A_\tau = 1\tev\,, \nonumber \\
 \text{sgn}(\SMU{3}^2)\SMU{3} \in [-150, 250]\gev,\; M_A \in [300, 500]\gev\,.
 \label{Eq:HHLS}
\end{align}
The bino mass is fixed by the GUT relation, $M_1 = \tfrac{5}{3} \tfrac{\sin^2 \theta_w}{\cos^2 \theta_w} M_2$. The remaining squark and slepton masses as well as the gluino mass are decoupled according to the prescription described in Section~\ref{Sect:numerics} in order to obtain the correct light Higgs mass. Typical values of $\MSUSY$ are $\sim \mathcal{O}(10^3\tev)$ for $\tan\beta = 1.5$.

Note that at face value this benchmark scenario does not feature light Higgs signal rates consistent with current LHC observations. However, as has been demonstrated in our fit results above, the light Higgs rates can be made consistent even for low values of roughly $m_{\sstop_1}\gtrsim 120\gev$ and $M_A \gtrsim 300\gev$ by adjusting the light stau contribution to the $h\to \gamma \gamma$ rate via the $\sstau$ mass and $\mu$ parameters. These adjustments would only marginally affect the phenomenology of the proposed benchmark model, e.g.~by slightly lowering the branching fraction for the $H\to \sstop_1\sstop_1^*$ decay due to the  competing $H\to \sstau_1^+\sstau_1^-$ decay.

Another note is in order here: Our prediction for the gluon fusion production cross section, $\sigma(gg\to H)$, as displayed in \fig{Fig:HHLS}, includes the squark contributions only at leading-order (LO), while the higher-order corrections from SM loops are implemented at NNLO. This is because the approximations employed in the higher-order calculations of the squark contributions (see Section~\ref{Sect:uncertainties}) do clearly not hold if $m_H > 2 m_{\sstop_1}$.

For these reasons, the `\texttt{HHLS}' scenario defined by \eqn{Eq:HHLS} should only motivate \textit{model-independent} searches for the signal topology $gg\to H \to \sstop_1\sstop_1$ at the LHC, where the results are presented as limits on (or measurements of) the signal rate $\sigma(gg\to H)\times \text{BR}(H\to\sstop_1\sstop_1$).

An important ingredient for these searches is the assumption on the $\sstop_1$ decay. As discussed earlier, the vanilla $R$-parity conserving scenario with pure decays $\sstop_1\to \neut_1\, t$ or $\sstop_1\to \neut_1\, c$ with a stable lightest neutralino, $\neut_1$, is already highly constrained by LHC searches for stop pair production from Run~1. In fact, as long as the stop decay signature includes missing energy a search for direct stop pair production seems more appropriate due to a generally larger production cross section. However, in the absence of missing energy in the final state, a full reconstruction of the resonance mass of the intermediate heavy Higgs in the $gg\to H\to \sstop_1 \sstop_1$ channel is possible and might be instrumental to improve the signal-to-background discrimination. We therefore suggest to probe the following decay signatures for a stop LSP,
\begin{align}
\sstop_1 &\xrightarrow{\lambda^{''}_{3jk}} 2j \qquad \text{with~0~or~1~$b$-jet}\,, \label{Eq:stop2j}\\
\sstop_1 &\xrightarrow{\lambda^{''}_{ijk}} 4j \qquad \text{with~0,~1~or~2~$b$-jets} \qquad (i\ne 3)\,. \label{Eq:stop4j}
\end{align}
These decays are mediated by the baryon number violating (BNV) operator in the superpotential, $\mathcal{W}\supset \tfrac{1}{2}\lambda^{''}_{ijk} \bar{U}_i \bar{D}_j \bar{D}_k$, where $\bar{U}_i$ and $\bar{D}_i$ denote the up-type and down-type quark $SU(2)_L$ singlet superfields and $i,j,k \in \{1,2,3\}$ are generation indices ($j\ne k$) (see \citere{Dreiner:1997uz} for a review on SUSY with $R$-parity violation (RPV)). In the first case, \eqn{Eq:stop2j}, the BNV operator couples directly to the light stop (see \citere{Csaki:2015uza} for LHC constraints on this signature). In the second case, \eqn{Eq:stop4j}, the stop does not couple directly to the BNV operator and thus undergoes a four-body decay.

Since large BNV operators would wash out any baryon asymmetry generated at the EW scale or above, very small $\lambda^{''}$ couplings and thus larger $\sstop_1$ lifetimes are warranted, potentially leading to detached vertices~\cite{Barry:2013nva} and hadronized stops (so-called \emph{stoponium}~\cite{Batell:2015zla}). The four-momenta and mass of the decaying stops, and consequently the resonance mass of the heavy Higgs boson, could in principle be reconstructed from the four-momenta of the final state jets. Furthermore, for large splittings between the $H$ and $\sstop_1$ masses, the $\sstop_1$'s can be boosted. These features provide important handles to substantially reduce the QCD background.

A simple modification of the benchmark model `\texttt{HHLS}' towards much lower $M_2$ and/or $M_1$ values furthermore enables scenarios with a neutralino LSP.\footnote{Note that lowering also the parameter $\mu$ will give rise to $H\to \neut_1\neut_1$ decays.}
In such a case, a search for the cascade decay
\begin{align}
\sstop_1 \rightarrow c\, &\neut_1\nonumber\\
 &\overset{\,\lambda_{ijk}^{''}}{\hookrightarrow} 3j \qquad \text{with~0~or~1~$b$-jet}
\end{align}
seems feasible, where the charm jets from the initial $\sstop_1$ decay origin from a primary vertex, and the $\neut_1$ potentially decays in a detached secondary vertex. Similar decays of the lightest neutralino $\neut_1$ into leptonic final states are also possible in the context of lepton number violating (LNV) operators.
Detailed Monte-Carlo studies of the suggested collider signatures and their discovery potential at the LHC are unfortunately beyond the scope of this work.

\subsection{Summary of lower stop mass limits}

We present a summary of the derived $\CL{95}$ light stop mass limits for the four scenarios discussed in the previous sections in Tab.~\ref{Tab:summary}. We give the limits for the different cases of including or excluding additional constraints from LEP sparticle mass limits or vacuum metastability constraints (as discussed in the previous sections), and for the presence or absence of an additional unconstrained Higgs decay mode to new physics, $h\to\mathrm{NP}$.

\begin{table}
\scalebox{0.83}{
\begin{tabular}{l | c | c | c | c}
\toprule
\textbf{Lower $\mathbf{95\%}$~C.L.~stop mass limit} & \multicolumn{2}{c|}{\emph{(i)} $\brhnp$ free} & 
\multicolumn{2}{c}{\emph{(ii)} $\brhnp\equiv0$} \\
 \midrule
Scenario & \emph{all} constraints & \emph{no} constraints  & \emph{all} constraints & \emph{no} constraints  \\ 
 \midrule
 \textbf{A}: $\SMU{3}$ & $144\gev$ & N/A & $ 154\gev$ & N/A \\ 
 \textbf{B}: $\SMU{3},~\tan\beta,~\SML{3}\equiv\SME{3}$ &$123\gev$ & $119\gev$& $146\gev$ & $146\gev$\\%
 \textbf{C}: $\SMU{3},~\tan\beta,~\mu=M_2$ & $123\gev$ & $123\gev$ &$123\gev$ & $123\gev$ \\
 \textbf{D}: $\SMU{3},~\tan\beta,~M_A,~\mu,~\SML{3}\equiv\SME{3}$ & N/A & N/A & $122\gev$& $116\gev$\\%
 \bottomrule
 \end{tabular}
 }
 \caption{Summary of light stop mass limits at $\CL{95}$ for all considered scenarios and the following assumptions on the additional Higgs decay mode to ``new physics'': \emph{(i)} decay to undetectable final states, \emph{(ii)} no new decay mode. We  list both the limits obtained including \emph{all} and \emph{no} additional constraints (such as e.g.~LEP sparticle mass limits, vacuum metastability constraints, etc., see description in Section~\ref{Sect:Results} for details.)}.
 \label{Tab:summary}
\end{table}

Concerning the fit quality of the BF points in the four scenarios, we find very similar minimal $\chi^2$ values, while the number of statistical degrees of freedom (ndf) varies according to the number of fit parameters and included observables:
\begin{align}
\chi^2/\text{ndf} = \left\{ \begin{array}{ll} 
68.3/83, & \text{scenario~{\bf A},~with}\,\, \brhnp~\text{free}; \\
67.8/81, & \text{scenario~{\bf B},~with}\,\, \brhnp~\text{free};\\
68.0/81, & \text{scenario~{\bf C},~with}\,\, \brhnp~\text{free};\\
68.1/81, & \text{scenario~{\bf D}}. 
\end{array}
\right.
\end{align}
We can compare this with the $\chi^2$ outcome of the SM Higgs boson with a mass of $m_h=125.1\gev$:
\begin{align}
\chi^2/\text{ndf} = 68.3/85, \quad (\text{SM~Higgs~boson}).
\end{align}
We conclude that, regarding the Higgs data, none of the MSSM scenarios actually improves the goodness-of-fit with respect to the SM. On the other hand, unlike the SM, these MSSM scenarios potentially enable the successful generation of a baryon-antibaryon asymmetry at the electroweak phase transition, as will be discussed in the next section.

\section{Implications for electroweak baryogenesis}
\label{Sect:EWBG}

Electroweak baryogenesis is a compelling framework for the generation of the baryon asymmetry at a relatively low scale, namely energies corresponding to the nucleation temperature of the electroweak phase transition (for a recent comprehensive review and for references see \cite{Morrissey:2012db}). The nucleation temperature corresponds to the temperature $T_n$ at which the bubble formation and growth rate starts to exceed the Hubble rate. Such temperature is slightly lower than the critical temperature $T_c$ at which the broken and unbroken electroweak phase minima are at the same effective potential value. A convenient ``order parameter'' for the phase transition is
$\langle H^0\rangle\equiv v/\sqrt{2},$
i.e.~the vacuum expectation value of the real component of $H^0$, where the latter is the SM Higgs field.\footnote{One can always choose a basis where it is only the real component of $H^0$ which develops a vacuum expectation value.}

Schematically, electroweak baryogenesis models require a first-order electroweak phase transition, i.e.~one that proceeds through nucleation of bubbles of broken electroweak phase ($v\neq0$). The expanding bubble walls provide out-of-equilibirum regions which particles scatter off of. In the presence of $B$-violation via sphaleron transitions (which are large and unsuppressed in the unbroken electroweak phase), and as long as large enough $\CP$ violation exists amongst the particles scattering off of the expanding bubble walls, a net baryon number can be produced, and some will diffuse within the regions of broken electroweak phase (i.e.~inside the bubbles). As long as the sphaleron rate within the expanding bubbles is suppressed enough to limit the washout of the produced baryon number, a net baryon number is frozen in, and could explain the observed baryon asymmetry. An indicative criterion that has been historically used to estimate at which level sphalerons transitions are ``suppressed enough'' in the broken electroweak phase is $v(T_c)/T_c\gtrsim1$, a condition that would indicate that the electroweak phase transition is ``strongly enough'' first order.\footnote{The quantity $v(T_c)/T_c\gtrsim1$ is obviously not gauge invariant, which is of course problematic. A gauge-invariant baryon number preservation condition has been constructed and outlined in detail in \citere{Patel:2011th}.}

Electroweak baryogenesis requires elements of physics beyond the SM, as in the SM the electroweak phase transition, for the observed value of the Higgs mass, is not first order, and the relevant $\CP$-violating currents are too small even if the phase transition actually were first order. A natural framework that potentially accommodates all needed ingredients for successful electroweak baryogenesis is the MSSM, or minimal extensions thereof. Much theoretical work has been devoted in recent years to the question of whether the MSSM can feature a strongly enough first order phase transition. This question essentially depends, in the absence of additional Higgs sector superfields, on the mass of the stops, an SU$(3)$ triplet with a large Yukawa coupling to the Higgs, which strongly affects the Higgs effective potential. While a detailed discussion of the relevant calculations of the strength of the electroweak phase transition as a function of the lightest, right-handed stop mass\footnote{The second stop must of course be heavy to obtain a Higgs mass in agreement with experiments, as well as to avoid electroweak precision constraints.} is beyond the scope of the present discussion, we briefly summarize here the relevant state-of-the-art results:

\begin{enumerate} 
\item {\em Perturbative calculations.}  These calculations rely on a finite-temperature effective potential. For large left-handed stop masses, a one-loop analysis is not reliable due to large logarithmic corrections in the ratio of the heavy stop scale to the weak scale, and a renormalization group improved Higgs and stop effective potentials, including dominant two-loop effects, must be employed (see \citere{Carena:2008vj} for details). In the context of this treatment, and assuming very large values for the left-handed stop soft breaking mass, \citere{Carena:2012np} finds that a strongly-enough first order electroweak phase transition ($v(T_c)/T_c\gtrsim1$) can be obtained, for $m_h\simeq 125\gev$, for stop masses as large as $m_{\sstop}\lesssim 105\gev$ at most.
\item {\em Lattice calculations.}  The accuracy of a perturbative analysis of the electroweak phase transition is  intrinsically limited by infrared singularities in the thermal field theory for momentum scales $p\sim g^2T/\pi$, which thus warrant the use of non-perturbative techniques such as numerical lattice simulations. In the simple context of the SM, where perturbative calculations indicate a weaker but persistently first-order phase transition with increasing Higgs mass, the predictions for the nature of the electroweak phase transition from lattice studies indicate no first-order transition at all for $m_h\gtrsim 72\gev$, illustrate clearly the limitations of perturbative results~\cite{Kajantie:1996mn,Kajantie:1996qd,Gurtler:1997hr,Csikor:1998eu}. 

\citere{Laine:2012jy} studied the nature of the electroweak phase transition with lattice simulations for an effective theory where the dynamical degrees of freedom which have not been integrated out are restricted to two SU$(2)_L$ Higgs doublets and one scalar  SU$(3)_c$ triplet and  SU$(2)_L$ singlet, with parameters fixed so that the lightest $\CP$-even Higgs has a mass of around $126\gev$, and the light right-handed ``stop'' of $155\gev$. It is important to note that no resummation was carried out for the large logarithms arising from the other heavy squarks (assumed to be at a scale larger than $7\tev$), with a potential impact on the uncertainty on, for example, the lightest stop mass of several GeV. This notwithstanding, the key result of the analysis of \citere{Laine:2012jy} is that for a $155\gev$ stop the electroweak phase transition has a $v(T_c)/T_c\simeq1.1$, and is thus sufficiently strongly first-order to suppress sphaleron washout processes enough in the broken electroweak phase. Also, it confirms that the strength of the phase transition is systematically under-estimated by perturbative calculations compared to lattice results. 
\end{enumerate}

While a precise value for the largest possible mass of the lightest stop compatible with $v(T_c)/T_c\gtrsim1$ and its dependence on other supersymmetric parameters is unclear, we can thus conclude that masses as large as $150\gev$ could, potentially, lead to a strongly enough first-order electroweak phase transition in the MSSM. An independent confirmation of the lattice simulation results of Ref.~\cite{Laine:2012jy}, that furthermore includes a resummation of large logarithms arising from heavy squarks, would be very useful and may resolve the remaining puzzles.

A somewhat decoupled question from the strength of the phase transition, in the context of MSSM electroweak baryogenesis, is the origin of the $\CP$-violating currents needed to produce a large enough baryon asymmetry. We will not review here the array of possible relevant particle/operator contents, nor the state-of-the-art of the technical aspects involved in the calculations, nor the associated uncertainties. However, we will point out that within the MSSM three ``sectors'' are of relevance:
\begin{enumerate}
\item A generic possibility for MSSM $\CP$-violating currents is that of resonant sfermion sources \cite{Kozaczuk:2012xv}, i.e.~with $m_{\tilde f_L}\simeq m_{\tilde f_R}$. While quasi degenerate stops are highly problematic for obvious reasons, and constraints from chromo-electric dipole moments make the possibility of sbottom-induced electroweak baryogenesis problematic, the possibility of light and quasi-degenerate staus is open. In this stau-induced electroweak baryogenesis electric dipole moments are highly suppressed and limited to Barr-Zee type two-loop contributions with a stau loop. As long as staus are lighter than about one TeV, and $\tan\beta$ is large enough, this is a very interesting possibility for MSSM electroweak baryogenesis.
\item A second class of well-known $\CP$-violating currents is associated with the electroweak-ino sector, and especially with the soft breaking bino and wino masses, the $\mu$ parameter and the physical relative $\CP$-violating phases. Resonant contributions are induced for non-vanishing Higgsino-gaugino phases for $M_{1,2}\sim\mu$, where a resonant behavior arises in the VEV insertion approximation. In order for this $\CP$-violating source to be significant, the relevant particle species must be close to thermal equilibrium, and thus at least one chargino needs to be light (i.e.~with a mass comparable to the nucleation temperature of the electroweak phase transition)~\cite{Cirigliano:2006dg, Cirigliano:2009yd, Kozaczuk:2011vr}. Constraints on the size of the $\CP$-violating phase from null searches for the electric dipole moment of the electron, neutron and atoms might favor a resonant bino-Higgsino scenario, with two light and almost degenerate neutralinos and one light chargino, although the resonant wino-Higgsino scenario, featuring two almost degenerate, light charginos, is not excluded~\cite{Li:2008ez}.
\item The heavy Higgs sector scale $M_A$, which controls the overall normalization of a class of resonant sources calculable in the context of the VEV-insertion approximation, and which enters also as an important parameter in other frameworks for the calculation of the $\CP$-violating currents \cite{Morrissey:2012db}. In general, the generated baryon asymmetry in the Universe is enhanced by lower values of $M_A$, although values in excess of a TeV are still viable.
\end{enumerate}

In summary, electroweak baryogenesis motivates scenarios with (i) a light stau sector (corresponding to our benchmark scenario {\bf B} above), (ii) one or two quasi-degenerate and light charginos ({\bf C}), and (iii) a light ``heavy'' Higgs sector, i.e.~the non-decoupling regime of relatively low $M_A$ ({\bf D}). These choices are all reflected in our choices of benchmark scenarios in the present study, and in all cases significantly lower light stop masses than in the scenario with only a light stop and the remaining SUSY spectrum being decoupled ({\bf A}) are allowed.

\section{Conclusions}
\label{Sect:conclusions}

In this study we derived indirect limits on the light stop mass in the MSSM from the Higgs rate measurements performed at Run~1 of the LHC. These constraints are complementary to limits obtained in direct collider searches for light stops, and are of particular importance in cases where the underlying assumptions of these collider searches are not fulfilled.

We used the public code \FH{} for the prediction of Higgs masses and branching ratios, and the \SH{} package for
the calculation of the gluon fusion cross section including light stop contributions
up to \nnlo{} \qcd{}. 
We carefully analyzed the theoretical uncertainties for the gluon fusion cross section from the approximations we adopted in the higher order contributions of the light stop to gluon fusion, which we
also applied to the light Higgs boson partial width into gluons.

Within this setup we considered four distinct MSSM scenarios.
We used \HS\ to perform a $\chi^2$ analysis of the MSSM parameter space in the
four scenarios in order to derive lower limits on the light stop mass from Higgs rate measurements.
Our MSSM scenarios are motivated by considerations of possible scenarios for successful electroweak baryogenesis within the MSSM, wherein a light stop is required to achieve a strongly-enough first order electroweak phase transition (scenario~{\bf A}). Scenarios with a light stop and a light stau (scenario~{\bf B}) or a light chargino (scenario~{\bf C}) are suggested by potential resonant $\CP$-violating sources critical to produce a large-enough baryon asymmetry. Finally, low values for  $M_A$, i.e.~a non-decoupled heavy Higgs sector (scenario~{\bf D}) is also generically well-motivated, and in some cases required, to have sufficiently strong $\CP$ sources. For completeness we add that we assume $\CP$ conservation in our study. However, given that our analysis is based on a large mass splitting in the stop sector with a reduced influence of the stop mixing parameter, $\CP$ conservation in the stop sector is well-motivated.

For all four scenarios, we evaluated the lower limits on the possible value of the light (right-handed) stop  mass, as well as the correlation among relevant masses, parameters and rates for Higgs decay modes. We included the possible existence of a generic, possibly invisible new physics  decay mode for the SM-like Higgs boson, and we also considered the case where such additional decay mode is not allowed.
Our analysis also takes into account other theoretical as well as experimental constraints including vacuum stability, model-dependent LEP sparticle mass limits, and bounds on other Higgs states (through \HB{}).

We find that in all cases our 95\% C.L. limits on the light stop mass are below $155\gev$, a value that according to lattice studies might be compatible with a strongly-enough first order electroweak phase transition in the MSSM. Specifically, we find that allowing for a new physics decay mode and only a light stop the lightest possible value for the stop mass is $144\gev$ (scenario~{\bf A}), while with the addition of a possible light chargino  or light staus (scenarios {\bf B} and {\bf C}) such lower limit is as low as $123\gev$. Allowing for a non-decoupled heavy Higgs sector instead of a new physics Higgs decay mode, with the addition of light staus, (scenario~{\bf D}) provides comparable lower stop mass limits, of around $122\gev$. Relaxing constraints from model-dependent LEP sparticle searches or vacuum metastability requirements additionally lowers the lightest possible stop mass in this latter scenario to around $116\gev$.

In conclusion, we find that a light stop $m_{\sstop}\ll m_t$ is still a generic possibilty in the MSSM in light of currently available data on the Higgs sector. Under the least stringent possible assumptions, masses as low as $116\gev$ are viable. Low stop masses are possible in particular for corners of the MSSM parameter space which are independently motivated by considerations in electroweak baryogenesis such as the strength of specific $\CP$-violating sources. Our results keep the window for successful electroweak baryogenesis in the MSSM still open, and the search for a light stop in the data of the current LHC run an exciting possibility.

\acknowledgments

We thank Sven Heinemeyer and Thomas Hahn for helpful discussions about the numerical stability of \FH. We are also grateful to Robert Harlander, Pietro Slavich and Michael Spira for comments on the heavy gluino mass dependence of the gluon fusion cross section. We also like to thank Lisa Zeune for helpful discussions. SP and TS are partly supported by the U.S. Department of Energy grant number DE-SC0010107. TS is furthermore supported by a Feodor-Lynen research fellowship sponsored by the Alexander von Humboldt foundation. SL acknowledges support through the SFB 676 ``Particles, Strings and the Early Universe'' funded by ``Deutsche Forschungsgemeinschaft''.

\appendix

\section{Experimental Higgs data from Tevatron and LHC}
\label{App:measurements}

We list the Higgs signal strength measurements from ATLAS, CMS and the Tevatron experiments CDF and D\O\ in Tabs.~\ref{Tab:HSobs1} and~\ref{Tab:HSobs2}. These observables are implemented in \HSv{1.4.0} and used in our numerical analysis. In total, we have 85 observables. Besides the measured signal strength value, $\hat{\mu}$, and its $1\sigma$ uncertainty, $\Delta\hat{\mu}$, the tables list for each observable the signal composition for the Higgs production/decay modes expected for a SM Higgs with mass $\sim 125.1\gev$.

\begin{table}
\scalebox{0.86}{
\renewcommand{\arraystretch}{1.0}
\begin{threeparttable}[b]
\footnotesize
 \begin{tabular}{lccrrrrr}
\toprule
 Analysis & \quad energy $\sqrt{s}$\quad & \quad$\hat{\mu} \pm \Delta \hat{\mu}$\quad & \multicolumn{5}{c}{\quad SM signal contamination [in \%]\quad} \\
 & & & \quad ggH \quad &  \quad VBF \quad & \quad WH \quad &   \quad ZH \quad & \quad $t\bar{t}H$\quad \\
\midrule
ATLAS $h\to WW\to \ell\nu\ell\nu~\mathrm{(VBF)}$~\cite{ATLAS:2014aga} & $7/8\tev$ & $  1.27\substack{+  0.53\\ -  0.45}$ & $  24.1$ & $  75.9$ & $   0.0$ & $   0.0$ & $   0.0$\\ 
ATLAS $h\to WW\to \ell\nu\ell\nu~\mathrm{(ggH)}$~\cite{ATLAS:2014aga} & $7/8\tev$ & $  1.01\substack{+  0.27\\ -  0.25}$ & $  97.8$ & $   1.2$ & $   0.6$ & $   0.3$ & $   0.1$\\ 
ATLAS $h\to ZZ\to 4\ell~\mathrm{(VBF/VH)}$~\cite{Aad:2014eva} & $7/8\tev$ & $  0.26\substack{+  1.64\\ -  0.94}$ & $  37.8$ & $  35.7$ & $  16.8$ & $   9.7$ & $   0.0$\\ 
ATLAS $h\to ZZ\to 4\ell~\mathrm{(ggH)}$~\cite{Aad:2014eva} & $7/8\tev$ & $  1.66\substack{+  0.51\\ -  0.44}$ & $  91.6$ & $   4.6$ & $   2.2$ & $   1.3$ & $   0.4$\\ 
ATLAS $h\to \gamma\gamma~\mathrm{(VBF, loose)}$~\cite{Aad:2014eha} & $7/8\tev$ & $  1.33\substack{+  0.92\\ -  0.77}$ & $  39.0$ & $  60.0$ & $   0.6$ & $   0.3$ & $   0.1$\\ 
ATLAS $h\to \gamma\gamma~\mathrm{(VBF, tight)}$~\cite{Aad:2014eha} & $7/8\tev$ & $  0.68\substack{+  0.67\\ -  0.51}$ & $  18.2$ & $  81.5$ & $   0.1$ & $   0.1$ & $   0.1$\\ 
ATLAS $h\to \gamma\gamma~(Vh,E_T^\text{miss})$~\cite{Aad:2014eha} & $7/8\tev$ & $  3.51\substack{+  3.30\\ -  2.42}$ & $   8.7$ & $   3.7$ & $  35.8$ & $  44.8$ & $   7.1$\\ 
ATLAS $h\to \gamma\gamma~(Vh,2j)$~\cite{Aad:2014eha} & $7/8\tev$ & $  0.23\substack{+  1.67\\ -  1.39}$ & $  45.0$ & $   3.3$ & $  31.9$ & $  19.8$ & $   0.1$\\ 
ATLAS $h\to \gamma\gamma~(Vh,1\ell)$~\cite{Aad:2014eha} & $7/8\tev$ & $  0.41\substack{+  1.43\\ -  1.06}$ & $   0.7$ & $   0.2$ & $  91.4$ & $   5.9$ & $   1.8$\\ 
ATLAS  $h\to \gamma\gamma~\mathrm{(central,high}~p_{Tt})$~\cite{Aad:2014eha} & $7/8\tev$ & $  1.62\substack{+  1.00\\ -  0.83}$ & $  72.6$ & $  16.4$ & $   6.1$ & $   3.7$ & $   1.2$\\ 
ATLAS $h\to \gamma\gamma~\mathrm{(central,low}~p_{Tt})$~\cite{Aad:2014eha} & $7/8\tev$ & $  0.62\substack{+  0.42\\ -  0.40}$ & $  93.2$ & $   4.1$ & $   1.6$ & $   1.0$ & $   0.1$\\ 
ATLAS $h\to \gamma\gamma~\mathrm{(forward,high}~p_{Tt})$~\cite{Aad:2014eha} & $7/8\tev$ & $  1.73\substack{+  1.34\\ -  1.18}$ & $  71.4$ & $  16.7$ & $   6.9$ & $   4.1$ & $   0.9$\\ 
ATLAS $h\to \gamma\gamma~\mathrm{(forward,low}~p_{Tt})$~\cite{Aad:2014eha} & $7/8\tev$ & $  2.03\substack{+  0.57\\ -  0.53}$ & $  92.5$ & $   4.2$ & $   2.0$ & $   1.2$ & $   0.1$\\ 
ATLAS $h\to \gamma\gamma~(tth,\mathrm{hadr.)}$~\cite{Aad:2014eha} & $7/8\tev$ & $ -0.84\substack{+  3.23\\ -  1.25}$ & $  15.0$ & $   1.3$ & $   1.3$ & $   1.4$ & $  81.0$\\ 
ATLAS $h\to \gamma\gamma~(tth,\mathrm{lep.)}$~\cite{Aad:2014eha} & $7/8\tev$ & $  2.42\substack{+  3.21\\ -  2.07}$ & $   8.4$ & $   0.1$ & $  14.9$ & $   4.0$ & $  72.6$\\ 
ATLAS $h\to \tau\tau~\mathrm{(VBF,hadr.hadr.)}$~\cite{Aad:2015vsa} & $7/8\tev$ & $  1.40\substack{+  0.90\\ -  0.70}$ & $  30.1$ & $  69.9$ & $   0.0$ & $   0.0$ & $   0.0$\\ 
ATLAS $h\to \tau\tau~\mathrm{(boosted,hadr.hadr.)}$~\cite{Aad:2015vsa} & $7/8\tev$ & $  3.60\substack{+  2.00\\ -  1.60}$ & $  69.5$ & $  13.3$ & $  11.3$ & $   5.8$ & $   0.0$\\ 
ATLAS $h\to \tau\tau~\mathrm{(VBF,lep.hadr.)}$~\cite{Aad:2015vsa} & $7/8\tev$ & $  1.00\substack{+  0.60\\ -  0.50}$ & $  17.2$ & $  82.8$ & $   0.0$ & $   0.0$ & $   0.0$\\ 
ATLAS $h\to \tau\tau~\mathrm{(boosted,lep.hadr.)}$~\cite{Aad:2015vsa} & $7/8\tev$ & $  0.90\substack{+  1.00\\ -  0.90}$ & $  73.0$ & $  13.3$ & $   9.1$ & $   4.6$ & $   0.0$\\ 
ATLAS $h\to \tau\tau~\mathrm{(VBF,lep.lep.)}$~\cite{Aad:2015vsa} & $7/8\tev$ & $  1.80\substack{+  1.10\\ -  0.90}$ & $  15.4$ & $  84.6$ & $   0.0$ & $   0.0$ & $   0.0$\\ 
ATLAS $h\to \tau\tau~\mathrm{(boosted,lep.lep.)}$~\cite{Aad:2015vsa} & $7/8\tev$ & $  3.00\substack{+  1.90\\ -  1.70}$ & $  70.9$ & $  21.4$ & $   5.7$ & $   2.1$ & $   0.0$\\ 
ATLAS $Vh\to V(bb)~(0\ell)$~\cite{Aad:2014xzb} &  $7/8\tev$ &$ -0.35\substack{+  0.55\\ -  0.52}$ & $   0.0$ & $   0.0$ & $  20.8$ & $  79.2$ & $   0.0$\\ 
ATLAS $Vh\to V(bb)~(1\ell)$~\cite{Aad:2014xzb} & $7/8\tev$ & $  1.17\substack{+  0.66\\ -  0.60}$ & $   0.0$ & $   0.0$ & $  96.7$ & $   3.3$ & $   0.0$\\ 
ATLAS $Vh\to V(bb)~(2\ell)$~\cite{Aad:2014xzb} & $7/8\tev$ & $  0.94\substack{+  0.88\\ -  0.79}$ & $   0.0$ & $   0.0$ & $   0.0$ & $ 100.0$ & $   0.0$\\ 
ATLAS $Vh\to V(WW)~(2\ell)$~\cite{Aad:2015ona} & $7/8\tev$ & $  3.70\substack{+  1.90\\ -  1.80}$ & $   0.0$ & $   0.0$ & $  83.3$ & $  16.7$ & $   0.0$\\ 
ATLAS $Vh\to V(WW)~(3\ell)$~\cite{Aad:2015ona} & $7/8\tev$ & $  0.72\substack{+  1.30\\ -  1.10}$ & $   0.0$ & $   0.0$ & $  86.5$ & $  13.5$ & $   0.0$\\ 
ATLAS $Vh\to V(WW)~(4\ell)$~\cite{Aad:2015ona} & $7/8\tev$ & $  4.90\substack{+  4.60\\ -  3.10}$ & $   0.0$ & $   0.0$ & $   0.0$ & $ 100.0$ & $   0.0$\\ 
ATLAS $tth\to \mathrm{multilepton}~(1\ell, 2\tau_h)$~\cite{Aad:2015iha} & $7/8\tev$ & $ -9.60\substack{+  9.60\\ -  9.70}$ & $   0.0$ & $   0.0$ & $   0.0$ & $   0.0$ & $   100.0$\tnote{1}\\ 
ATLAS $tth\to \mathrm{multilepton}~(2\ell, 0\tau_h)$~\cite{Aad:2015iha} & $7/8\tev$ & $  2.80\substack{+  2.10\\ -  1.90}$ & $   0.0$ & $   0.0$ & $   0.0$ & $   0.0$ & $   100.0$\tnote{2}\\ 
ATLAS $tth\to \mathrm{multilepton}~(2\ell, 1\tau_h)$~\cite{Aad:2015iha} & $7/8\tev$ & $ -0.90\substack{+  3.10\\ -  2.00}$ & $   0.0$ & $   0.0$ & $   0.0$ & $   0.0$ & $   100.0$\tnote{3}\\ 
ATLAS $tth\to \mathrm{multilepton}~(3\ell)$~\cite{Aad:2015iha} &  $7/8\tev$ &$  2.80\substack{+  2.20\\ -  1.80}$ & $   0.0$ & $   0.0$ & $   0.0$ & $   0.0$ & $   100.0$\tnote{4}\\ 
ATLAS $tth\to \mathrm{multilepton}~(4\ell)$~\cite{Aad:2015iha} & $7/8\tev$ & $  1.80\substack{+  6.90\\ -  6.90}$ & $   0.0$ & $   0.0$ & $   0.0$ & $   0.0$ & $   100.0$\tnote{5}\\ 
ATLAS $tth\to tt(bb)$~\cite{Aad:2015gra} & $7/8\tev$ & $  1.50\substack{+  1.10\\ -  1.10}$ & $   0.0$ & $   0.0$ & $   0.0$ & $   0.0$ & $ 100.0$\\ 
\midrule
CDF $h\to WW$~\cite{Aaltonen:2013ipa} & $1.96\tev$& $  0.00\substack{+  1.78\\ -  1.78}$ & $  77.5$ & $   5.4$ & $  10.6$ & $   6.5$ & $   0.0$\\ 
CDF $h\to \gamma\gamma$~\cite{Aaltonen:2013ipa} & $1.96\tev$& $  7.81\substack{+  4.61\\ -  4.42}$ & $  77.5$ & $   5.4$ & $  10.6$ & $   6.5$ & $   0.0$\\ 
CDF $h\to \tau\tau$~\cite{Aaltonen:2013ipa} & $1.96\tev$& $  0.00\substack{+  8.44\\ -  8.44}$ & $  77.5$ & $   5.4$ & $  10.6$ & $   6.5$ & $   0.0$\\ 
CDF $Vh\to V(bb)$~\cite{Aaltonen:2013ipa} & $1.96\tev$& $  1.72\substack{+  0.92\\ -  0.87}$ & $   0.0$ & $   0.0$ & $  62.0$ & $  38.0$ & $   0.0$\\ 
CDF $tth\to tt(bb)$~\cite{Aaltonen:2013ipa} & $1.96\tev$& $  9.49\substack{+  6.60\\ -  6.28}$ & $   0.0$ & $   0.0$ & $   0.0$ & $   0.0$ & $ 100.0$\\ 
\midrule
D\O\ $h\to WW$~\cite{Abazov:2013gmz} & $1.96\tev$& $  1.90\substack{+  1.63\\ -  1.52}$ & $  77.5$ & $   5.4$ & $  10.6$ & $   6.5$ & $   0.0$\\ 
D\O\  $h\to bb$~\cite{Abazov:2013gmz} & $1.96\tev$& $  1.23\substack{+  1.24\\ -  1.17}$ & $   0.0$ & $   0.0$ & $  62.0$ & $  38.0$ & $   0.0$\\ 
D\O\  $h\to \gamma\gamma$~\cite{Abazov:2013gmz} & $1.96\tev$& $  4.20\substack{+  4.60\\ -  4.20}$ & $  77.5$ & $   5.4$ & $  10.6$ & $   6.5$ & $   0.0$\\ 
D\O\  $h\to \tau\tau$~\cite{Abazov:2013gmz} & $1.96\tev$& $  3.96\substack{+  4.11\\ -  3.38}$ & $  77.5$ & $   5.4$ & $  10.6$ & $   6.5$ & $   0.0$\\ 
\bottomrule
 \end{tabular}
   \begin{tablenotes}
 \footnotesize
 \item[1] The SM Higgs signal composition is $h\to \tau\tau$ ($93.0\%$), $h\to WW$ ($4.0\%$), $h\to bb$ ($3.0\%$).
 \item[2] The SM Higgs signal composition is $h\to WW$ ($80.1\%$), $h\to \tau\tau$ ($14.9\%$), $h\to ZZ$ ($3.0\%$), $h\to bb$ ($2.0\%$).
 \item[3] The SM Higgs signal composition is $h\to \tau\tau$ ($61.8\%$), $h\to WW$ ($35.2\%$), $h\to ZZ$ ($2.0\%$), $h\to bb$ ($1.0\%$).
 \item[4] The SM Higgs signal composition is $h\to WW$ ($74.1\%$), $h\to \tau\tau$ ($14.9\%$), $h\to ZZ$ ($7.0\%$), $h\to bb$ ($3.9\%$).
 \item[5] The SM Higgs signal composition is $h\to WW$ ($68.1\%$), $h\to \tau\tau$ ($13.9\%$), $h\to ZZ$ ($14.0\%$), $h\to bb$ ($4.0\%$).
 \end{tablenotes}
 \end{threeparttable}
}
 \caption{Higgs signal strengths measurements from the ATLAS collaboration at the LHC, and the CDF and D\O\ collaborations at the Tevatron.}
 \label{Tab:HSobs1}
\end{table}

\begin{table}
\scalebox{0.86}{
\renewcommand{\arraystretch}{1.0}
\begin{threeparttable}[b]
\footnotesize
 \begin{tabular}{lccrrrrr}
\toprule
 Analysis & \quad energy $\sqrt{s}$\quad & \quad$\hat{\mu} \pm \Delta \hat{\mu}$\quad & \multicolumn{5}{c}{\quad SM signal contamination [in \%]\quad} \\
 & & & \quad ggH \quad &  \quad VBF \quad & \quad WH \quad &   \quad ZH \quad & \quad $t\bar{t}H$\quad \\
\midrule
CMS $h\to WW\to 2\ell2\nu~(0/1j)$~\cite{Chatrchyan:2013iaa} &  $7/8\tev$ & $  0.74\substack{+  0.22\\ -  0.20}$ & $  85.8$ & $   8.9$ & $   3.3$ & $   1.9$ & $   0.0$\\ 
CMS $h\to WW\to 2\ell2\nu~\mathrm{(VBF)}$~\cite{Chatrchyan:2013iaa} &  $7/8\tev$ & $  0.60\substack{+  0.57\\ -  0.46}$ & $  24.1$ & $  75.9$ & $   0.0$ & $   0.0$ & $   0.0$\\ 
CMS $h\to ZZ\to 4\ell~(0/1j)$~\cite{Chatrchyan:2013mxa}  &  $7/8\tev$& $  0.88\substack{+  0.34\\ -  0.27}$ & $  91.9$ & $   8.1$ & $   0.0$ & $   0.0$ & $   0.0$\\ 
CMS $h\to ZZ\to 4\ell~(2j)$~\cite{Chatrchyan:2013mxa} &  $7/8\tev$ & $  1.55\substack{+  0.95\\ -  0.66}$ & $  76.1$ & $  23.9$ & $   0.0$ & $   0.0$ & $   0.0$\\ 
CMS $h\to \gamma\gamma~(\text{untagged}~0)$~\cite{Khachatryan:2014ira}&  $7\tev$  & $  1.97\substack{+  1.51\\ -  1.25}$ & $  80.8$ & $   9.7$ & $   5.8$ & $   3.2$ & $   0.6$\\ 
CMS $h\to \gamma\gamma~(\text{untagged}~1)$~\cite{Khachatryan:2014ira} &  $7\tev$ & $  1.23\substack{+  0.98\\ -  0.88}$ & $  92.3$ & $   4.1$ & $   2.3$ & $   1.2$ & $   0.1$\\ 
CMS $h\to \gamma\gamma~(\text{untagged}~2)$~\cite{Khachatryan:2014ira} &  $7\tev$ & $  1.60\substack{+  1.25\\ -  1.17}$ & $  92.3$ & $   4.0$ & $   2.3$ & $   1.3$ & $   0.1$\\ 
CMS $h\to \gamma\gamma~(\text{untagged}~3)$~\cite{Khachatryan:2014ira} &  $7\tev$ & $  2.61\substack{+  1.74\\ -  1.65}$ & $  92.5$ & $   3.9$ & $   2.3$ & $   1.2$ & $   0.1$\\ 
CMS $h\to \gamma\gamma~(\mathrm{VBF,dijet}~0)$~\cite{Khachatryan:2014ira} &  $7\tev$ & $  4.85\substack{+  2.17\\ -  1.76}$ & $  19.9$ & $  79.6$ & $   0.3$ & $   0.2$ & $   0.1$\\ 
CMS $h\to \gamma\gamma~(\mathrm{VBF,dijet}~1)$~\cite{Khachatryan:2014ira} &  $7\tev$ & $  2.60\substack{+  2.16\\ -  1.76}$ & $  39.0$ & $  58.9$ & $   1.2$ & $   0.7$ & $   0.3$\\ 
CMS $h\to \gamma\gamma~(Vh,E_T^\text{miss})$~\cite{Khachatryan:2014ira} &  $7\tev$ & $  4.32\substack{+  6.72\\ -  4.15}$ & $   4.9$ & $   1.2$ & $  43.2$ & $  44.4$ & $   6.3$\\ 
CMS $h\to \gamma\gamma~(Vh,\text{dijet})$~\cite{Khachatryan:2014ira} &  $7\tev$ & $  7.86\substack{+  8.86\\ -  6.40}$ & $  28.6$ & $   2.9$ & $  43.8$ & $  23.3$ & $   1.5$\\ 
CMS $h\to \gamma\gamma~(Vh,\text{loose})$~\cite{Khachatryan:2014ira} &  $7\tev$ & $  3.10\substack{+  8.29\\ -  5.34}$ & $   3.8$ & $   1.1$ & $  79.7$ & $  14.6$ & $   0.7$\\ 
CMS $h\to \gamma\gamma~(tth, \text{tags})$~\cite{Khachatryan:2014ira} &  $7\tev$ & $  0.71\substack{+  6.20\\ -  3.56}$ & $   4.3$ & $   1.5$ & $   2.9$ & $   1.6$ & $  89.7$\\ 
CMS $h\to \gamma\gamma~(\text{untagged}~0)$~\cite{Khachatryan:2014ira} &  $8\tev$& $  0.13\substack{+  1.09\\ -  0.74}$ & $  75.7$ & $  11.9$ & $   6.9$ & $   3.6$ & $   1.9$\\ 
CMS $h\to \gamma\gamma~(\text{untagged}~1)$~\cite{Khachatryan:2014ira} &  $8\tev$& $  0.92\substack{+  0.57\\ -  0.49}$ & $  85.1$ & $   7.9$ & $   4.0$ & $   2.4$ & $   0.6$\\ 
CMS $h\to \gamma\gamma~(\text{untagged}~2)$~\cite{Khachatryan:2014ira}&  $8\tev$ & $  1.10\substack{+  0.48\\ -  0.44}$ & $  91.1$ & $   4.7$ & $   2.5$ & $   1.4$ & $   0.3$\\ 
CMS $h\to \gamma\gamma~(\text{untagged}~3)$~\cite{Khachatryan:2014ira}&  $8\tev$ & $  0.65\substack{+  0.65\\ -  0.89}$ & $  91.5$ & $   4.4$ & $   2.4$ & $   1.4$ & $   0.3$\\ 
CMS $h\to \gamma\gamma~(\text{untagged}~4)$~\cite{Khachatryan:2014ira}&  $8\tev$ & $  1.46\substack{+  1.29\\ -  1.24}$ & $  93.1$ & $   3.6$ & $   2.0$ & $   1.1$ & $   0.2$\\ 
CMS $h\to \gamma\gamma~(\mathrm{VBF,dijet}~0)$~\cite{Khachatryan:2014ira} &  $8\tev$ & $  0.82\substack{+  0.75\\ -  0.58}$ & $  17.8$ & $  81.8$ & $   0.2$ & $   0.1$ & $   0.1$\\ 
CMS $h\to \gamma\gamma~(\mathrm{VBF,dijet}~1)$~\cite{Khachatryan:2014ira} &  $8\tev$& $ -0.21\substack{+  0.75\\ -  0.69}$ & $  28.4$ & $  70.6$ & $   0.6$ & $   0.2$ & $   0.2$\\ 
CMS $h\to \gamma\gamma~(\mathrm{VBF,dijet}~2)$~\cite{Khachatryan:2014ira} &  $8\tev$& $  2.60\substack{+  1.33\\ -  0.99}$ & $  43.7$ & $  53.3$ & $   1.4$ & $   0.8$ & $   0.8$\\ 
CMS $h\to \gamma\gamma~(Vh,E_T^\text{miss})$~\cite{Khachatryan:2014ira} &  $8\tev$& $  0.08\substack{+  1.86\\ -  1.28}$ & $  16.5$ & $   2.7$ & $  34.4$ & $  35.3$ & $  11.1$\\ 
CMS $h\to \gamma\gamma~(Vh,\text{dijet})$~\cite{Khachatryan:2014ira} &  $8\tev$& $  0.39\substack{+  2.16\\ -  1.48}$ & $  30.4$ & $   3.1$ & $  40.5$ & $  23.3$ & $   2.6$\\ 
CMS $h\to \gamma\gamma~(Vh,\text{loose})$~\cite{Khachatryan:2014ira} &  $8\tev$& $  1.24\substack{+  3.69\\ -  2.62}$ & $   2.7$ & $   1.1$ & $  77.9$ & $  16.8$ & $   1.5$\\ 
CMS $h\to \gamma\gamma~(Vh,\text{tight})$~\cite{Khachatryan:2014ira} &  $8\tev$& $ -0.34\substack{+  1.30\\ -  0.63}$ & $   0.2$ & $   0.2$ & $  76.9$ & $  19.0$ & $   3.7$\\ 
CMS $h\to \gamma\gamma~(tth, \text{multijet})$~\cite{Khachatryan:2014ira} &  $8\tev$& $  1.24\substack{+  4.23\\ -  2.70}$ & $   4.1$ & $   0.9$ & $   0.8$ & $   0.9$ & $  93.3$\\ 
CMS $h\to \gamma\gamma~(tth, \text{lepton})$~\cite{Khachatryan:2014ira} &  $8\tev$& $  3.52\substack{+  3.89\\ -  2.45}$ & $   0.0$ & $   0.0$ & $   1.9$ & $   1.9$ & $  96.1$\\ 
CMS $h\to \mu\mu$~\cite{CMS:2013aga} &  $7/8\tev$& $  2.90\substack{+  2.80\\ -  2.70}$ & $  94.1$ & $   5.9$ & $   0.0$ & $   0.0$ & $   0.0$\\ 
CMS $h\to \tau\tau~(0j)$~\cite{Chatrchyan:2014nva}&  $7/8\tev$ & $  0.40\substack{+  0.73\\ -  1.13}$ & $  98.5$ & $   0.8$ & $   0.4$ & $   0.3$ & $   0.0$\\ 
CMS $h\to \tau\tau~(1j)$~\cite{Chatrchyan:2014nva} &  $7/8\tev$& $  1.06\substack{+  0.47\\ -  0.47}$ & $  79.7$ & $  12.1$ & $   5.2$ & $   3.0$ & $   0.0$\\ 
CMS $h\to \tau\tau~\mathrm{(VBF)}$~\cite{Chatrchyan:2014nva} &  $7/8\tev$& $  0.93\substack{+  0.41\\ -  0.41}$ & $  20.9$ & $  79.1$ & $   0.0$ & $   0.0$ & $   0.0$\\ 
CMS $Vh\to V(\tau\tau)$~\cite{Chatrchyan:2014nva}&  $7/8\tev$ & $  0.98\substack{+  1.68\\ -  1.50}$ & $   0.0$ & $   0.0$ & $  47.1$\tnote{1} & $  27.3$\tnote{1} & $   0.0$\\ 
CMS $Vh\to V(bb)$~\cite{CMS:2013dda}&  $7/8\tev$ & $  1.00\substack{+  0.51\\ -  0.49}$ & $   0.0$ & $   0.0$ & $  63.3$ & $  36.7$ & $   0.0$\\ 
CMS $Vh\to V(WW)\to 2\ell2\nu$~\cite{Chatrchyan:2013iaa} &  $7/8\tev$& $  0.39\substack{+  1.97\\ -  1.87}$ & $  60.2$ & $   3.8$ & $  22.8$ & $  13.2$ & $   0.0$\\ 
CMS $Vh\to V(WW)~(\text{hadr.})$~\cite{CMS:2013xda} &  $7/8\tev$& $  1.00\substack{+  2.00\\ -  2.00}$ & $  63.7$ & $   3.3$ & $  21.9$ & $  11.1$ & $   0.0$\\ 
CMS $Wh\to W(WW)\to 3\ell3\nu$~\cite{Chatrchyan:2013iaa} &  $7/8\tev$& $  0.56\substack{+  1.27\\ -  0.95}$ & $   0.0$ & $   0.0$ & $ 100.0$ & $   0.0$ & $   0.0$\\ 
CMS $tth\to 2\ell~\text{(same-sign)}$~\cite{Khachatryan:2014qaa} &  $7/8\tev$& $  5.30\substack{+  2.10\\ -  1.80}$ & $   0.0$ & $   0.0$ & $   0.0$ & $   0.0$ & $  100.0$\tnote{2}\\ 
CMS $tth\to 3\ell$~\cite{Khachatryan:2014qaa}&  $7/8\tev$ & $  3.10\substack{+  2.40\\ -  2.00}$ & $   0.0$ & $   0.0$ & $   0.0$ & $   0.0$ & $  100.0$\tnote{3}\\ 
CMS $tth\to 4\ell$~\cite{Khachatryan:2014qaa} &  $7/8\tev$& $ -4.70\substack{+  5.00\\ -  1.30}$ & $   0.0$ & $   0.0$ & $   0.0$ & $   0.0$ & $  100.0$\tnote{4}\\ 
CMS $tth\to tt(bb)$~\cite{Khachatryan:2014qaa}&  $7/8\tev$ & $  0.70\substack{+  1.90\\ -  1.90}$ & $   0.0$ & $   0.0$ & $   0.0$ & $   0.0$ & $ 100.0$\\ 
CMS $tth\to tt(\gamma\gamma)$~\cite{Khachatryan:2014qaa}&  $7/8\tev$ & $  2.70\substack{+  2.60\\ -  1.80}$ & $   0.0$ & $   0.0$ & $   0.0$ & $   0.0$ & $ 100.0$\\ 
CMS $tth\to tt(\tau\tau)$~\cite{Khachatryan:2014qaa}&  $7/8\tev$ & $ -1.30\substack{+  6.30\\ -  5.50}$ & $   0.0$ & $   0.0$ & $   0.0$ & $   0.0$ & $ 100.0$\\ 
\bottomrule
 \end{tabular}
   \begin{tablenotes}
 \footnotesize
 \item[1] The signal is contaminated to $16.2\%$ [$9.4\%$] by $WH\to WWW$ [$ZH\to ZWW$] in the SM.
 \item[2] The SM Higgs signal composition is $h\to WW$ ($73.3\%$), $h\to \tau\tau$ ($23.1\%$), $h\to ZZ$ ($3.6\%$)
 \item[3] The SM Higgs signal composition is $h\to WW$ ($71.8\%$), $h\to \tau\tau$ ($23.8\%$), $h\to ZZ$ ($4.4\%$).
 \item[4] The SM Higgs signal composition is $h\to WW$ ($53.0\%$), $h\to \tau\tau$ ($30.1\%$), $h\to ZZ$ ($16.9\%$).
 \end{tablenotes}
 \end{threeparttable}
 }
 \caption{Higgs signal strengths measurements from CMS collaboration at the LHC.} 
  \label{Tab:HSobs2}
\end{table}

\newpage
\bibliographystyle{JHEP}
\bibliography{main}

\end{document}